\documentclass{article}[fulpage]
\usepackage[a4paper, total={6.3in, 9.7in}]{geometry}
\usepackage[utf8]{inputenc}
\usepackage[english]{babel}
\usepackage[autostyle]{csquotes}
\usepackage{graphicx}
\usepackage{amsmath,amsthm}
\usepackage{mathtools}
\usepackage{verbatim}
\usepackage{cite}
\usepackage{centernot}
\usepackage{amssymb}
\usepackage{tikz}
\usepackage{tikz-qtree}
\usepackage{pdfpages}
\usepackage{thmtools}
\usepackage{thm-restate}
\usepackage{hyperref}
\usepackage{cleveref}

\usetikzlibrary{trees}
\usetikzlibrary{decorations.pathmorphing}
\usetikzlibrary{decorations.markings}
\usetikzlibrary{decorations.pathmorphing,shapes}
\usetikzlibrary{calc,decorations.pathmorphing,shapes}
\usetikzlibrary{patterns}
\usetikzlibrary{snakes}
\newtheorem{theorem}{Theorem}
\newtheorem{definition}[theorem]{Definition}
\newtheorem{observation}[theorem]{Observation}
\newtheorem{example}[theorem]{Example}
\newtheorem{lemma}[theorem]{Lemma}
\newtheorem{fact}[theorem]{Fact}
\newtheorem{claim}[theorem]{Claim}
 
 \newtheoremstyle{break}
  {\topsep}{\topsep}%
  {\itshape}{}%
  {\bfseries}{}%
  {\newline}{}%
\theoremstyle{break}
 \newtheorem{algorithm}{Algorithm}
\newcommand{\QED}{\hfill$\qed$}
\newcommand{\cOtilde}{\tilde{\mathcal{O}}}

 \newcommand{\defproblema}[2]{
  \vspace{2mm}
\noindent\fbox{
  \begin{minipage}{0.96\textwidth}
  #1\\
  #2
  \end{minipage}
  }
  \vspace{2mm}
}

\def\infinity{\rotatebox{90}{8}}

\title{Dynamic Suffix Array with Sub-linear update time and Poly-logarithmic Lookup Time}

\author{
\begin{tabular}{cc}
Amihood Amir\thanks{ Department of Computer Science, Bar-Ilan
University, Ramat-Gan 52900, Israel, +972 3 531-8770; {\tt
amir@cs.biu.ac.il}} & 
Itai Boneh\thanks{Department of Computer Science, Bar-Ilan University, 
Ramat Gan, Israel, email: \texttt{itai.bone@biu.live.ac.il}.
}
\\
{\small Bar-Ilan University}& {\small Bar-Ilan University}\\
{\small and} \\
{\small Georgia Tech} \\
\end{tabular}
}

\date{}

\begin{document}

\maketitle

\begin{abstract}
The {\em suffix array} $SA_S[1\ldots n]$ of an $n$-length string $S$ is a lexicographically sorted array of the suffixes of $S$. The suffix array is one of the most well known and widely used data structures in string algorithms. We present a data structure for maintaining a representation of the suffix array of a dynamic string which undergoes symbol substitutions, deletions, and insertions. 

For every string manipulation, our data structure can be updated in $\cOtilde(n^{\frac{2}{3}})$ time with $n$ being the current length of the string. For an input query $i\in [1\ldots n]$, our data structure reports $SA_S[i]$ in $O(\log^5(n))$ time.

We also present a faster data structure, with $\cOtilde(\sqrt{n})$ update time, for maintaining the {\em Inverted Suffix Array} of a dynamic string undergoing symbol substitutions updates. For an input query $i\in [1\ldots n]$, our data structure reports the $i$'th entry in the inverted suffix array in $O(\log^4(n))$ time.

Our data structures can be used to obtain sub-linear dynamic algorithms for several classical string problems for which efficient dynamic solutions were not previously known.

\end{abstract}
\newpage
\section{Introduction}

Recently, there has been a growing interest in dynamic strings algorithms. Many classical stringology problems have been lately investigated in dynamic settings. Perhaps the most natural problem of stringology is the one of constructing an \textbf{index} for a given string. That is, preprocessing a string $S$ in order to obtain a data structure that given an input word $p$, outputs the set $occ$ of indexes in $S$ in which the word $p$ occurs. In the dynamic settings, we wish to maintain an index for a string that undergoes edit operations (substitution, deletion, or insertion of a single symbol in any location in the text). The current state of the art for dynamic indexing is the data structure of ~\cite{GKKL:18} that requires $O(\log^2(n))$ time per update and supports $O(p+ \log^2(n) + |occ|)$ time queries.

In the \textbf{Longest Common Factor} problem, we are given two strings $S$ and $T$. We are interested in finding the largest words that is a substring of both $S$ and $T$. Charalampopoulos et al. ~\cite{cgp:20} show how to maintain the LCF of two dynamic strings $S$ and $T$ that are going through edit operation updates with polylogarithmic time per update. In the \textbf{Longest Increasing Subsequence} problem, we are given an input string $S[1\ldots n]$ that consists of numbers. We are interested in finding the longest increasing sequence of indexes ${i_1,i_2 \ldots i_x} \subseteq [1\ldots n]$ such that $S[i_j] < S[i_{j+1}]$ for every $j\in [1\ldots x-1]$. Gawrychowski et al. ~\cite{DBLP:conf/stoc/GawrychowskiJ21} provided an algorithm for maintaining an approximation of the LIS of a dynamic string in polylogarithmic time per update. Kociumaka and Seddighin ~\cite{kociumaka2021improved} provided an algorithm for maintaining the exact LIS of a dynamic string with polynomial sublinear time per update. In addition to the above, the problems of finding the longest palindrome substring in a text ~\cite{DBLP:journals/corr/abs-1906-09732ab:19, fnibt:19}, finding all the runs in a text ~\cite{abck:19}, approximated pattern matching ~\cite{charalampopoulos2020faster}, and computing the edit distance between two strings ~\cite{charalampopoulos_et_al:LIPIcs:2020:12134} have all been recently investigated in dynamic settings.   

The natural interest in dynamic strings algorithms grows from the fact
that the biggest digital library in the world - the web - is
constantly changing, as well as from the fact that other big digital
libraries - genomes and astrophysical data, are also subject to
change through mutation and time, respectively.

The {\em suffix tree}~\cite{W-73} and {\em suffix array}~\cite{MM-90}
have been, arguably, the most powerful and heavily used tools in
Stringology. The {\em suffix tree} of string $S$ is a compressed trie of all suffixes of $S$, and the {\em suffix array} of $S$ corresponds to a pre-order traversal of all the leaves of the suffix tree of $S$. The natural application of the suffix tree is for indexing, but it
has been used for many purposes. An incomplete list includes
approximate matching~\cite{LV-86tcs,LV-89}, parameterized
matching~\cite{bak:96,bak:97,ael:07,hls:04,lnp:11}, efficient
compression~\cite{ZL-77,ZL-78,abf:sleeping:95,W-84,FT-95,s:dcc:98,nr:99,bk:00,apostolico2007fast,abm:08},
finding syntactic regularities in
strings~\cite{abg:92,LS-93,b:99,LSS01,hr:03,wykg:05,kps:06,dp:07,sbt:07,KS:16,LSB:17},
and much more.

It was not until recently that an algorithm for maintaining the suffix tree
or suffix array of a dynamically changing text had been sought. The
difficulty is that even a single change in the text may cause a change in linear
number of entries in the suffix array. Thus, although a dynamic suffix
array algorithm would be extremely useful to automatically adapt many
static pattern matching algorithms to a dynamic setting, other
techniques had to be sought.

The key to most recent efficient dynamic solutions that appeared in the Stringology literature is renaming. 
However, renaming is not a panacea for dynamic algorithms to all the
problems that the suffix tree or array solved in the static
setting. Perhaps the key property of the suffix array is that the
suffixes are sorted lexicographically. The powerful renaming and locally persistent parsing 
techniques developed thus far do not maintain lexicographic
ordering. It is, thus, no surprise that problems like maintaining the
Burrows-Wheeler transform, or finding the Lyndon root of a substring,
do not hitherto have an efficient dynamic version.

The current state of the art for maintaining the suffix array of a dynamic text is the flexible data structure of Amir and Boneh ~\cite{isaac_ab:20}. For every parameter $k\in [1\ldots n]$, they provide a data structure with $\cOtilde(k)$ update time supporting look up queries to $SA[i]$ in $\cOtilde(\frac{n}{k})$ time.

\subsection{Our Contribution}
We enhance the data structure of Amir and Boneh ~\cite{isaac_ab:20} to obtain  the following.
\begin{enumerate}
\item A data structure for maintaining the suffix array
  of a dynamic string in $\cOtilde(n^\frac{2}{3})$ time per text update 
  with $O(\log^5(n))$ time for a look up query $SA[i]$. This data structure supports symbol substitution, symbol insertion and symbol deletion updates. 
  \item A data structure for maintaining the \textit{inverted} suffix array
  of a dynamic string in $\cOtilde(\sqrt{n})$ time per text update
  with $O(\log^4(n))$ time for a look up query $SA[i]$. This data structure only supports symbol substitution updates.
\end{enumerate}

Both data structures use $\cOtilde(n)$ space. Note that our data structures maintain the sublinear polynomial update time of  ~\cite{isaac_ab:20}, and exponentially improve the query time from polynomial to polylogarithmic.
Additionally, our data structure for dynamic suffix array can be used as a black box, alongside previously existing results, to obtain the following.

\begin{enumerate}
    \item A dynamic text $S[1\ldots n]$ can be maintained with $\cOtilde(n^\frac{2}{3})$ time per edit operation update to support random access queries to $BWT[i]$, the burrows wheeler transform of $S$, in polylogarithmic time. 
    \item A dynamic text $S[1\ldots n]$ can be maintained with $\cOtilde(n^\frac{2}{3})$ time per edit operation update to enable random access queries to the $LCP$ array of $S$ in polylogarithmic time.
    \item A dynamic text $S[1\ldots n]$ can be maintained with $\cOtilde(n^\frac{2}{3})$ time per edit operation update to enable a simulation of the suffix tree of $S$ with polylogarithmic time for traversing a suffix tree edge.

\end{enumerate}

In ~\cite{isaac_ab:20}, summarized in Section 6, the above results are presented with a polynomial query time, rather then polylogarithmic. Replacing the data structure of ~\cite{isaac_ab:20} with ours yields the above. We refer the interested reader to ~\cite{isaac_ab:20} for an overview of how a dynamic suffix array is utilized to achieve these results.

\subsection{Technical Overview}
\textbf{Dynamic Inverse Suffix Array} Similarly to ~\cite{isaac_ab:20}, we partition the suffixes of the the dynamic string $S[1\ldots n]$ into clusters. Every cluster $C$ corresponds to a word $w_C$ with length $|w_C| = \sqrt{n}$. The cluster $C$ contains all the suffixes starting with the prefix $w_C$. We say that two suffixes are \textbf{close} if they are located within the same cluster, and \textbf{far} otherwise. It is shown in ~\cite{isaac_ab:20} that these clusters can be maintained in $\cOtilde(\sqrt{n})$ time when a substitution update is applied to $S$.

Given a query $i$ for $iSA[i]$, we are interested in finding the lexicographical rank of $S[i\ldots n]$ among the suffixes of $S$. Equivalently, we want to count the number of suffixes of $S[1\ldots n]$ that are lexicographically smaller than $S[i\ldots n]$. The inverse suffix array algorithm of ~\cite{isaac_ab:20} counts the number of suffixes \textbf{far} from $S[i\ldots n]$ that are smaller than $S[i\ldots n]$. The bottleneck that causes for the polynomial query time in ~\cite{isaac_ab:20} is counting the number of \textbf{close} suffixes that are smaller than $S[i\ldots n]$. This task can be equivalently described as evaluating the lexicographical rank of $S[i\ldots n]$ within the suffixes in its cluster.

In order to overcome this bottleneck, we wish to maintain an 'array' $R[1\ldots n]$. For every suffix $S[i\ldots n]$, $R[i]$ stores the rank of $S[i\ldots n]$ within the suffixes in its cluster. In order to update $R$ in sublinear time, we provide a deep analysis of the nature in which the ranks in $R$ change is as a result of a substitution update. The conclusion of our analysis is that while $R$ may require updates in $\Omega(n)$ entries as a result in a single substitution, these updates can be compactly described as a sequence of $O(\sqrt{n})$ special updates called \textbf{stairs updates}. 

\begin{definition}
A {\em decreasing stairs update} (See Figure \ref{fig:stairsexp2}) is denoted by $S = (i,j,p)$. Applying $S$ to $R$ increases the values stored in $R[i\ldots j]$ as follows: the rightmost $p$ indexes $R[j - p + 1 \ldots j]$ are increased by $1$, the $p$ indexes to their left are increased by $2$, and so on until we finally exceed $R[i\ldots j]$. 

Formally, for every $t \in [1 \ldots \lceil \frac{j - i + 1}{p} \rceil]$, the counters $R[a]$ with $a \in [max(j - t\cdot p + 1,i) \ldots j - (t-1) \cdot p]$ are increased by $t$. 
\end{definition}

We show how to obtain the sequence of $O(\sqrt{n})$ stairs updates that represent the changes to $R$ in $\cOtilde(\sqrt{n})$ time. We then provide a data structure that maintains an ordered set of integers that is applied a sequence of stairs updates. The utility of provided by our data structure for stairs updates is described by Theorem \ref{t:stairsp}.

\begin{restatable}{theorem}{stairsp}\label{t:stairsp}
There is a data structure for maintaining an indexed set of integers $R$ to support both applying a stairs update to $R$ and look-up queries $R[i]$ in $O(\log^3(U))$ time, provided that all the updates $(i,j,p)$ have the same value of $p$. $U$ is the overall amount of updates. The data structure consumes $O(U\log^2(U))$ space. Initializing the data structure (all counters are set to $0$) takes $O(1)$ time.
\end{restatable}

The restriction that every stair update that is applied to the ranks has the same width $p$ does not necessarily hold in our dynamic algorithm. It may be the case that the stairs updates within the sequence have various step widths, or that the stairs updates applied in the most recent update have different step width from the updates that were applied in previous update. This denies us from using our data structure for stairs update in a straight forward manner to maintain $R$.

In order to solve that problem, we identify a unique property of the stairs updates that need to be applied to $R$. 
\begin{observation}\label{o:stairsisrun}
Every stairs update $(i,j,p)$ that is applied to $R$ corresponds to a run $S[i\ldots j]$ in $S$ with period $p$. 
\end{observation}
Observation \ref{o:stairsisrun} allows us to exploit existing insights on the structure of runs within a string to bound the number of different values of $p$ for which a stairs update with step width $p$ was applied to an index $i$ in $R$. Specifically, we show that there are $O(\log(n))$ such values of $p$. With this property, our problem can be reduced to maintaining a sequence $\{St_1,St_2 \ldots St_{n} \}$ of restricted stairs updates data structure described in Theorem \ref{t:stairsp}. Every stairs update data structure $St_p$ is responsible for updates with a step width $p$. When we wish to evaluate $R[i]$, we sum the values stored in $St_p$ over the $p$ values such that $St_p[i]$ was affected by a stairs update. Since there are $O(\log(n))$ such values, the query time remains polylogarithmic.

\textbf{Dynamic Suffix Array} As in our solution for inverse suffix arrays, we build upon the data structure presented in ~\cite{isaac_ab:20}. For an integer parameter $k$, the dynamic data structure of ~\cite{isaac_ab:20} provides the following utility.
\begin{fact}\label{f:ab2020dsa}
A query $i$ for $SA_S[i]$ of a dynamic $S$ can be reduced to a Close Suffix Selection query (defined below) in $O(\log^2(n))$ time. The data structure for executing this reduction can be maintained in $O(k)$ time per edit operation update.
\end{fact}

{\defproblema{\textsc{Close Suffixes Select}}{Given a subword of size $k$ of $S$ represented by its starting index $i \in [1 \ldots n - k + 1]$ and a rank $r$, return the suffix $S[i \ldots n]$ with lexicographic rank $r$ among the suffixes starting with $S[i\ldots i + k - 1]$}}

In Section \ref{s:dcssq}, we prove the following: 
\begin{theorem}\label{t:selectClose}
Given a dynamic text $S[1\ldots n]$, there is a data structure that support close suffixes selection queries in $O(\log^5(n))$ time. The data structure can be maintained in $\cOtilde((\frac{n}{k})^2 + k)$ time when an edit operation is applied to $S$. The data structure uses $\cOtilde(n + (\frac{n}{k})^2)$ space.
\end{theorem}

By setting $k = n^{\frac{2}{3}}$ and applying the above reduction, we obtain the following:
\begin{theorem}\label{t:dsa}
For a dynamic string $S[1\ldots n]$, there is a data structure that supports look up queries to $SA_S[i]$ in $O(\log^5(n))$ time. When an edit operation update is applied to $S$, the data structure can be maintained in $\cOtilde(n^{\frac{2}{3}})$ time. The data structure uses $\cOtilde(n)$ space.
\end{theorem}

We present the main idea for proving Theorem \ref{t:selectClose}. 
Let $w$ be a word of size $t = \frac{k}{2}$. By exploiting periodic structures in strings, we can represent all the occurrences of a $w$ in $S[1\ldots n]$ with $O(\frac{n}{k})$ arithmetic progressions. We denote this compact set of arithmetic progressions representing the occurrences of $w$ as $POR_w$. The data structure of ~\cite{isaac_ab:20} allows the evaluation of $POR_w$ from an index $i$ representing a starting index of $w$ in $\cOtilde(|POR_w|)$ time. We denote the explicit set of occurrences of $w$ in $S$ as $Occ_w$. We denote as $A_w[1 \ldots |Occ_w|]$ the array consisting of the starting indexes of occurrences of $w$ sorted by lexicographic order of the suffixes starting in these indexes. While the size of $POR_w$ is bounded by $O(\frac{n}{k})$, the size of $A_w$ may be $\Omega(n)$. Note that a Close suffix selection query is equivalent to a random access query in $A_w$.

Let $C = (a,b,p) \in POR_w$ be an arithmetic progression in the representation of the occurrences of $w$. We denote the size of the cluster as $|C| = \frac{b-a}{p} + 1$. The arithmetic progression $C$ corresponds to occurrences of $w$ starting in $a, a+p , a+2\cdot p \ldots a+ |C| \cdot p =  b$. We say that $C$ \textit{implies} these occurrences of $w$. We define the following properties of clusters and of occurrences implied by clusters.

\begin{definition}
Let $w_i = S[a + i\cdot p \ldots a+ i\cdot p + t - 1]$ be an occurrence implied by $C$ of $w$. The \textit{period rank} of $w_t$ is defined to be $r_p(w_t) = |C| - i $.
\end{definition}

\begin{definition}[Cluster Head, Cluster Tail, Increasing and Decreasing Clusters]
Let $w_{|C|} = [b \ldots b + t -1]$ be the rightmost occurrence of $w$ represented by $C$ (with respect to starting index). Let $w_s = w[e - p + 1 \ldots e]$ be the suffix of size $p$ of $w$. We call the suffix $S[b + t \ldots n]$ the \textit{tail} of $C$ denoted as $Tail(C)$. We call $C$ an \textit{increasing} cluster if $Tail(C) >_L w_s$, or a \textit{decreasing} cluster if $Tail(C) <_L w_s$. 
\end{definition}

See Figure \ref{fig:clusterexample} for an example of a cluster and its properties. In Section \ref{s:ersq}, we prove that $A_w$ has the following structure (derived from Lemma \ref{l:decthaninc} and Lemma \ref{l:decRanks}):

$A_w$ can be described as a concatenation $A_w = D[1\ldots |D|] , I[1\ldots |I|]$ with $D$ being the sub array containing the set of occurrences implied by decreasing clusters, and $I$ being the set of occurrences implied by increasing clusters. 

$D$ can be written as the concatenation $D = D_0, D_1 \ldots D_{m-1},D_{m}$. $m$ is the maximal period rank of an occurrence in a decreasing cluster. For every $r\in [0\ldots m]$ , $D_{r}$ contains the occurrences $w'$ from decreasing clusters with $r_p(w') = r$ sorted by lexicographic order of the tails of their clusters.

Similarly, $I = I_t ,I_{t-1} \ldots I_1,I_0$. $t$ is the maximal period rank of an occurrence in an increasing cluster. For every $r\in [0\ldots t]$ , $I_{r}$ contains the occurrences $s$ from increasing clusters with $r_p(s) = r$ sorted by lexicographic order of the tails of their clusters (See Figure \ref{fig:awdemons}).

Let $w = S[s\ldots e]$ be a subword of $S$. Let $l$ and $r$ be two non-negative integers. 
\begin{definition}
An occurrence $w_1 = S[s_1\ldots e_1]$ of $w$ is an $(l,r)$-extendable occurrence of $w$ if $lcs(s-1,s_1-1) \ge l$ and $lcp(e+1,e_1+1) \ge r$. We define the array $A^{(l,r)}_w$ as the lexicographically sorted array of suffixes of $S$ starting with an $(l,r)$-extendable occurrence of $w$.
\end{definition}

It is easy to see that given a word $w^c = S[s - l \ldots e + r]$ containing $w$, an occurrence of $w^c$ in index $i$ is equivalent to an $(l,r)$-extendable occurrence of $w$ in $i + l$. Therefore, $A_{w^c}[i] = A^{(l,r)}_w[i] - l$ for every $i \in [1 \ldots |A_{w^c}|]$. We present the following problem:

{\defproblema{\textsc{Extension Restricting Selection}}
{
\textbf{Preprocess:} A word $w=S[s\ldots e]$ and $POR_w$
\\
\textbf{Query:} upon input $l,r,i$. Report $A^{(l,r)}_w[i]$ }
}

In Section \ref{s:ersq}, we prove the following:
\begin{theorem}\label{t:extSelect}
If $LCE$ queries can be evaluated in $O(\log(n))$ time, the periodic occurrences representation $POR_w$ of a word $w=S[s\ldots e]$ can be preprocessed in time $\cOtilde(|POR_w|)$ to answer Extension Restricting Selection queries in $O(\log^5(n))$ time. The data structure uses $\cOtilde(|POR_w|)$ space.
\end{theorem}

LCE queries are formally defined in Section \ref{s:prel}. Existing results in dynamic stringology can be used in a straight forward manner to enable $O(\log(n))$ LCE queries in dynamic settings.

Constructing the data structure described in Theorem \ref{t:extSelect} is the most technically involved part in the construction of our dynamic suffix array data structure. It is achieved by exploiting the characterization of the structure of $A_w$. Our insights on the structure of $A_w$ allow us to reduce an Extension Restricting Selection query to a small set of multidimensional range queries.

With the data structure of Theorem \ref{t:extSelect}, a dynamic data structure for Close Suffix selection queries can be constructed as follows.
We initialize a data structure for supporting dynamic $LCE$ queries on $S$ by employing Lemma \ref{lem:dynLCE}. We also initialize the dynamic data structure of Fact \ref{f:ab2020dsa} for obtaining periodic occurrences representations and reducing suffix array queries to close selection queries.
Let $t=\lfloor \frac{k}{2} \rfloor$. We partition $S$ into words of size $t$ starting in indexes $i = 0 \bmod t$. For every word $w_x=S[x\cdot t \ldots (x+1)\cdot t - 1]$ for $x \in [0 \ldots \lfloor \frac{n}{t} \rfloor]$, we obtain $POR_{w_x}$ in $\cOtilde(\frac{n}{t})$ time. We preprocess $POR_{w_x}$ for Extension Restricting Selection queries using Theorem \ref{t:extSelect}. We keep an array $ERSQ[1 \ldots  \lfloor \frac{n}{t} \rfloor]$ with $ERSQ[x]$ containing a pointer to the Extension Restricting Selection queries data structure of $w_x$.

Upon an edit operation update, we update the data structures of Fact \ref{f:ab2020dsa}. We then partition the updated $S$ in the same manner as in the initialization step. Note that the words in the partition may change, as the size of $S$ may change due to an insertion or a deletion update, and $k$ may be a function of $|S|=n$. For every word $w_x$ in the updated partition, we extract $POR_{w_x}$ and evaluate the extension restricting selection queries data structures for every word $w_x$ from scratch. We also construct the array $ERSQ$ from scratch as in the initialization step.

Upon a Close Suffix Selection query $(i,r)$, the word $w_{\lceil \frac{i}{t} \rceil}$ is completely contained within $w = S[i \ldots i + k -1]$. Let $L$ be the distance between the starting indexes of  $w_{\lceil \frac{i}{t} \rceil}$ and $w$ and let $R$ be the distance between the ending indexes. We execute an Extension Restricting Selection query $(L,R,r)$ on the Extension Restricting Selection queries data structure of $w_{\lceil \frac{i}{t} \rceil}$. Let the output of the query be $z$. We return $z  -L$ .

\textbf{Correctness:} A Close Suffix Selection query $(i,r)$ is equivalent to reporting $A_w[r]$. Denote the indexes of $w_{\lceil \frac{i}{t} \rceil}=S[s \ldots e]$. The word $w = S[i \ldots i + k - 1]$ can be written as $w = S[s - L , e + R]$. Therefore, $A_w[r] =  A^{(L,R)}_{w_{\lceil \frac{i}{t} \rceil}}[r] - L$. 

\textbf{Preprocessing time:} The preprocessing times of the $LCE$ data structure and the periodic occurrences data structure are both bounded by $\cOtilde(n)$. The remaining operations are, in fact, an initial update. The complexity analysis for these is provided below.

\textbf{Update time:} Evaluating the partition can be executed in $O(\frac{n}{k})$ time in a straightforward manner. There are $\frac{n}{t} \in O(\frac{n}{k})$ words in our partition. For every word $w$, we find $POR_w$ in $\cOtilde(\frac{n}{k})$ time. Since $|w| = \frac{k}{2}$, we have $|POR_w| \in O(\frac{n}{k})$. So constructing the Extension Restricting Selection data structure from $POR_w$ takes $\cOtilde(\frac{n}{k})$. The update time for the $LCE$ data structure is polylogarithmic and the update time for the periodic occurrences representation data structure is $\cOtilde(k)$. The overall update complexity is dominated by $\cOtilde((\frac{n}{k})^2 + k)$.

\textbf{Space:} The space consumed by the Extension Restricting Selection data structures is bounded by $\cOtilde((\frac{n}{k})^2)$. The space consumed by the array $ERSQ$ is bounded by $O(\frac{n}{k})$. The space consumed by the data structure for periodic occurrences representation is bounded by $O(n)$. The space consumed by the $LCE$ data structure is bounded by $\cOtilde(n)$. The overall space complexity is bounded by $\cOtilde(n + (\frac{n}{k})^2)$.

\textbf{Query time:} The query consists of a constant number of basic arithmetic operations, a single array lookup and a single Extension Restricting Selection query. The query time is dominated by $O(\log^5(n))$. \QED

\subsection{Roadmap}
This paper is organized as follows: Section~\ref{s:prel} gives basic definitions and terminology. Section \ref{s:prev} provides an overview of ~\cite{isaac_ab:20}, and presents the bottleneck in the complexity of the algorithms presented there. This identifies the difficulty that the current paper solves. In Section \ref{s:dcssq} we present a {\em static} data structure problem, and show that, surprisingly, an efficient data structure for solving this problem, enables us to construct an efficient {\em dynamic} suffix array data structure. We provide an efficient solution for this problem in Section \ref{s:ersq}. For dynamic inverted suffix array queries, our results appear in Section \ref{s:dcsrq}. Section \ref{s:comsrq} provides some complementary proofs for the result of Section \ref{s:dcsrq}. Section \ref{s:updates} provides two data structures that are required for the construction of Section \ref{s:dcsrq}.

\section{Preliminaries}\label{s:prel}
\subsection{Strings}
Let $S=S[1]S[2]\ldots S[n]$ be a \textit{string} (or a \textit{word}) of length $|S|=n$
over an ordered alphabet $\Sigma$ of size
$|\Sigma|=\sigma$.  By $\varepsilon$ we denote an empty string.
For two positions $i$ and $j$ in $S$, we denote by 
$S[i\ldots  j]=S[i]\ldots  S[j]$ the \textit{factor}  (sometimes called
\textit{substring}) of $S$ that starts at position 
$i$ and ends at position $j$ (it equals $\varepsilon$ if $j<i$).  
We recall that a {\em prefix} of $S$ is a factor that starts at
position $1$ ($S[1\ldots j]$) and a {\em suffix} is a factor that ends at
position $n$ ($S[i \ldots n]$). The prefix ending in index $i$ is denoted as 
$S_i$ and the suffix starting in index $i$ is denoted as $S^i$.

An integer $p$ is \textit{a period} of $S$ if $S[i] = S[i+p]$ for every $i \in [1 \ldots n - p]$. $p$ is called \textbf{the} period of $S$ if it is the minimal period of $S$. We denote the period of a string $S$ as $per(S)$. If $per(S)< \frac{|S|}{2}$, we say that $S$ is periodic. A substring $R = S[i\ldots j]$ of $S$ is called a \textit{run} if it is periodic with period $p$, and every substring of $S$ containing $R$ does not have a period $p$. Equivalently, $R$ is a run if it is periodic with period $p$, and $S[i-1] \neq S[i-1 + p]$ and $S[j + 1] \neq S[j + 1 - p]$. If $R$ is a prefix (resp. suffix) of $S$, the first (resp. second) constraint is not required.

Let $w[1\ldots m]$ be a string. We say that $w$ occurs in position $i \in [1 \ldots n-m+1]$ in $S$ if $S[i + j -1] = w[j]$ for every $j \in [1 \ldots m]$. Every substring $S[i \ldots i + m -1] = w$ is an \textit{occurrence} of $w$ starting in index $i$.

The \textit{longest common prefix} of two strings $S[1 \ldots n],T[1\ldots m]$ , denoted as $LCP(S,T)$, is the maximal integer $l \in [1 \ldots \min(n,m)]$ such that $S[1 \ldots l] = T[1 \ldots l]$. The \textit{longest common suffix} of $S$ and $T$, denoted as $LCS(S,T)$, is the maximal integer $l \in [1 \ldots \min(n,m)]$ such that $S[n-l + 1\ldots n] = T[m-l + 1 \ldots m]$.

Given a string $S$, a \textit{Longest Common Extension} (often shortened to $LCE$) data structure for $S$ supports the two following queries:
\begin{enumerate}
    \item $LCP(i,j)$ - return the longest common prefix of $S[i\ldots n]$ and $S[j\ldots n]$.
    \item $LCS(i,j)$ - return the longest common suffix of $S[1\ldots i]$ and $S[1\ldots j]$.
\end{enumerate}

The results of (\cite{GKKL:18}) or of (\cite{niibt:16}) can be applied to obtain the following:
\begin{lemma}\label{lem:dynLCE}
There is a deterministic (res. randomized) data structure for maintaining a dynamic string with $O(\log(n)\log^*(n))$ (resp. $O(\log(n))$ excepted) time per edit operation (deletion, insertion, or substitution of a symbol) so that an $LCE$ query can be answered in $O(\log(n)\log^*(n))$ (resp. $O(\log(n))$ expected) time. The data structure uses $\cOtilde(n)$ space.
\end{lemma}

\textbf{Remark:} Throughout the paper, we use the randomized variant of Lemma \ref{lem:dynLCE}. Thus, we refer to the application of an $LCE$ query as an $O(\log(n))$ time operation. Every such reference to Lemma \ref{lem:dynLCE} can be made deterministic by applying the deterministic variant of the lemma with the cost of an $O(\log^*(n))$ multiplicative slowdown.

We point out that other than $LCE$ queries, the algorithms presented in this paper do not use randomization. Therefore, our results can be derandomized by adding a multiplicative $\log^*(n)$ factor to the complexity of an $LCE$ query. 
\\
Let $S[1\ldots n]$ and $T[1\ldots m]$ be two strings over an ordered alphabet $\Sigma$. $S$ is lexicographically smaller than $T$, denoted as $S <_L T$, if $S[LCP(S,T) + 1] < T[LCP(S,T) + 1]$. In the case in which $LCP(S,T) = \min(|S|,|T|)$, $S <_L T$ if $|S| < |T|$. Note that given two indexes $i$ and $j$ representing two suffixes of a string $S$, Lemma \ref{lem:dynLCE} enables to  lexicographically compare $S^i$ and $S^j$ in $O(\log(n))$ time. 
\\
The \textit{suffix array} $SA_S$ is an array of size $n$ containing the suffixes of $S$ sorted lexicographically. 
A suffix $S[i\ldots n]$ is represented in $SA_S$ by its starting index $i$, so the size of $SA_S$ is linear. The \textit{Inverse Suffix Array} $iSA_S$ is the inverse permutation of the suffix array. Namely, $iSA_S[SA_S[x]] = x$ for every $x \in [1\ldots n]$.
\\
In this paper, we handle a dynamic string that is undergoing  substitution, deletion, and insertion updates. A substitution update is given as a pair $(i,\sigma)$. Applying the update $(i,\sigma)$ to $S$ sets $S[i] = \sigma$ and does not affect any other index in $S$. We always assume that the applied substitution update is not trivial. Namely, $S[i] \neq \sigma$ prior to the substitution. An insertion update is also given as a pair $(i,\sigma)$. Applying the insertion $(i,\sigma)$ to $S$ results in a new text $S' = S[1\ldots i] \sigma S[i+1 \ldots n]$. A deletion update is given as an index $i$, Applying a deletion in index $i$ to $S$ results in a new text $S' = S[1\ldots i-1] S[i+1 \ldots n]$. We collectively refer to substitution updates, insertion updates and deletion updates as {\em edit updates} or {\em edit operations}.

\subsection{Periodic Structures}
To achieve efficiency, we often need to simultaneously process the set of occurrences of a word $w$ in $S$. To that end we exploit the periodic structure of words with numerous occurrences within $S$. The periodicity lemma (\cite{10.2307/2034009}) states the following.
\begin{lemma}\label{l:periodicity}
If a string $S[1\ldots n]$ has a period $p$ and a period $q$ such that $n \ge p + q - gcd(p,q)$, then $gcd(p,q)$ is also a period of $S$. 
\end{lemma}

The following is a basic observation regarding the occurrences of a word within $S$.
\begin{observation}\label{o:consocdif}
Given a word $w$ with size $|w| = x$ and period $p$, starting in index $i$ in $S$, let $j$ be the index closest to $i$ that is also a starting index of an occurrence of $w$ in $S$. Either $|j-i|=p$ or $|j-i|\geq \frac{x}{2}$.
\end{observation}

With Observation \ref{o:consocdif}, we are ready to introduce the periodic occurrences representation.

\begin{fact} \label{f:perrep}
The set of occurrences of a word $w$ having length $x$ and with period $p$ in a string $S[1\ldots n]$ can be represented by $O(\frac{n}{x})$ arithmetic progressions (often referred to as {\em clusters}) of the form $C = (a,b,p)$. The cluster $C=(a,b,p)$ represents a set of occurrences of $w$ starting in indexes $i_t = a + p \cdot t$ with $t \in [0 .. \frac{b-a}{p}]$. The size of a cluster, denoted as $|C| = \frac{b-a}{p} + 1$ is the number of occurrences represented by $C$. Every cluster $C$ is associated with a run $R_C$ in $S$ with period $p$. The occurrences represented by $C$ are all the occurrences of $w$ contained within $R_C$. It follows that every cluster is locally maximal, in the sense that there are no occurrences of $w$ starting in $a-p$ or in $b+p$.
\end{fact}

We call the representation of the occurrences of $w$ described in Fact \ref{f:perrep} the \textit{periodic occurrences representation} of $w$.

\begin{lemma}\label{l:fastCluster}
For every parameter $x \in [1 \ldots n]$, there is a data structure that, given input integer $i$, outputs the periodic occurrences representation of $S[i\ldots i+ x - 1]$ in a dynamic text $S[1 \ldots n]$ in $\cOtilde(\frac{n}{x})$ time. This data structure can be maintained in $\cOtilde(x)$ time after an edit operation update is applied to $S[1\ldots n]$.
\end{lemma}

Lemma 3 of~\cite{isaac_ab:20} and the discussion following Observation 1 there, serve as a proof for Lemma \ref{l:fastCluster}. Specifically, the discussion following Observation 1 presents a method for obtaining the periodic occurrences representation of $w$ in $\cOtilde(\frac{n}{x})$ time using, what they called, the $k$-Words tree data structure. Lemma 3 ensures that the $k$-Words tree can be maintained over edit operations updates in $\cOtilde(x)$ time. While the lemma in \cite{isaac_ab:20} refers to a dynamic text undergoing substitution updates, it can be easily generalized to edit operations.

\subsection{Dynamic multidimensional range queries}
Dynamic multidimensional range queries is an algorithmic tool enabling the maintenance of a set of $d$-dimensional points. A data structure for $d$-dimensional range queries supports queries concerning the points within an input $d$-dimensional range. A $d$-dimensional point is given by the values of its $d$ coordinates. A $d$-dimensional range $R$ is given by a set of $d$ ranges $R = [X_1 \ldots Y_1] \times [X_2 \ldots Y_2] \times  \cdots \times [X_d \ldots Y_d]$. The point $p=(x_1,x_2 \ldots x_d)$ is in the range $R$ ($p \in R$) if $X_i \leq x_i \leq Y_i$ for every $i \in [1\ldots d]$.  Every point $p$ is assigned a numeric value $v(p)$.
The result below can be achieved by using the range tree data structure ~\cite{bentley1979data}.
\begin{fact}
There is a data structure for maintaining a set $D$ of $d$-dimensional points and supports the following queries:
\begin{enumerate}
    \item Add a new point $p = (x_1, x_2, \ldots x_d)$ with value $v(p)=v$ to $D$.
    \item Remove an existing point $p = (x_1, x_2, \ldots x_d)$ from $D$.
    \item \textbf{$COUNT(R)$} Given a $d$-dimensional range $R = [X_1 \ldots Y_1] \times [X_2 \ldots Y_2] \ldots [X_d \ldots Y_d]$, report the number of points in $D \cap R$.
    \item \textbf{$SUM(R)$:} Given a $d$-dimensional range $R = [X_1 \ldots Y_1] \times [X_2 \ldots Y_2] \ldots [X_d \ldots Y_d]$, report the sum of $v(p)$ taken over the points in $D \cap R$. The sum of an empty set is defined to be 0.
    \item \textbf{$REPORT(R)$:} Given a $d$-dimensional range $R = [X_1 \ldots Y_1] \times [X_2 \ldots Y_2] \ldots [X_d \ldots Y_d]$, return all the points in $D \cap R$.
\end{enumerate}
Queries 1 - 4 can be supported in $O(\log^d(n))$ and range report queries can be supported in $O(\log^{d}(n) + k)$ with $k$ being the amount of reported points and $n$ being the number of points in $D$. The data structure consumes $O(\log^{d-1}(n))$ space.
\end{fact}

\section{Close Suffix Selection and Close Suffix Rank}\label{s:prev}
Our algorithm improves the results of ~\cite{isaac_ab:20}. For self-containment, this section provides an overview of the terminology, techniques, and results of ~\cite{isaac_ab:20} that are relevant to our work. This section also identifies the sub-tasks that need to be efficiently implemented in order to improve the previous work.
\subsection{Close suffixes, far suffixes, and the $k$-words tree}

Let $k$ be an integer parameter. We start by defining a simple yet useful data structure.
\begin{definition}
For a string $S[1\ldots n]$, the $k$-words tree of $S$ is a balanced search tree with the following components:
\begin{enumerate}
    \item For every distinct $k$-length subword $w$ of $S$, the $k$-words tree contains a node $v=N(w)$ corresponding to $w$.
    \item The nodes of the $k$-words tree are sorted by lexicographical order of their corresponding words.
    \item $N(w)$ stores a balanced search tree containing the starting indexes of every occurrence of $w$ in $S$ in increasing order. $N(w)$ also stores the number of words in the subtree rooted in $N(w)$.
\end{enumerate}
\end{definition}

Let $v$ be a node in the $k$-words tree. We call the word $w$ such that $N(w) = v$ the \textit{word of $v$}. We denote the word of $v$ as $W(v)$. We point out the distinction between nodes and words: every node in the $k$-words tree represents a set of occurrences of certain sub-words. The number of words in a subtree of the $k$-words tree is the accumulated number of occurrences represented by the nodes of the sub-tree. The following theorem is proven in~\cite{isaac_ab:20}.

\begin{theorem}[\cite{isaac_ab:20}, Lemma 3]
Given a string $S[1 \ldots n]$, a data structure for dynamically maintaining the $k$-words tree of $S$ can be constructed in $\cOtilde(n)$ time using $O(n)$ space. When applying an edit operation update to $S$, the $k$-words tree can be updated in $\cOtilde(k)$ time.
\end{theorem}

In \cite{isaac_ab:20}, Lemma 3 is proved with respect to substitution updates. It can be easily modified to support deletion and insertions as well.

We proceed to define {\em far} and {\em close} suffixes.
\begin{definition} \label{d:closFarSuff}
Let $S^i = S[i \ldots n]$ and $S^j = S[j \ldots n]$ be two suffixes of a text $S[1 \ldots n]$. $S[i\ldots n]$ and $S[j \ldots n]$ are called {\em close suffixes} (or {\em $k$-close suffixes}) if $LCP(i,j) \ge k$, and {\em far suffixes} (or {\em $k$-far suffixes}) otherwise. 
\end{definition}

When we discuss $k$-close and $k$-far suffixes throughout the paper, the value of $k$ is clear from context. We therefore use the notations `close' and `far' suffixes, omitting the specific value of $k$. 

The primary role of the $k$-words tree is to apply a reduction from a query about the suffix array to a query about close suffixes. Note that the starting indexes of close suffixes are stored within the same node of the $k$-words tree while the starting indexes of far suffixes are stored within different nodes.

\subsection{Dynamic Inverted Suffix Array}
Reporting $iSA_S[i]$ is equivalent to counting the number of suffixes that are lexicographically smaller than $S[i \ldots n]$.
Upon a query integer $i$ for $iSA[i]$, we can find the node $v$ in the $k$-words tree with $W(v) = S[i \ldots i + k -1]$ in $O(\log^2(n))$ time by applying a classic binary tree search. The lexicographical comparison required in every node can be executed in $O(\log(n))$ time by employing Lemma \ref{lem:dynLCE}.

By counting the words to the left of the path from the root to $v$, we obtain the amount of suffixes of $S$ that are lexicographically smaller than $S[i\ldots n]$ and are far from $S[i\ldots n]$. Given the path, this can be easily executed in logarithmic time. The remaining problem is to find the number of suffixes close to $S[i \ldots n]$ that are lexicographically smaller than $S[i\ldots n]$.

We formalize the problem of finding the close smaller suffixes as follows:

{\defproblema{\textsc{Close Suffixes Rank}}{Given an index $i$ representing a suffix $S[i \ldots n]$,  return lexicographic rank of $S[i \ldots n]$ among the suffixes close to $S[i\ldots n]$.}}

In this paper, we present a dynamic data structures for Close Suffix Rank queries. Specifically, we prove the following in Section \ref{s:dcsrq}.
\begin{theorem}\label{t:countClose}
Given a dynamic text $S[1 \ldots n]$, there is a data structure that supports Close Suffixes Rank queries in $O(\log^4(n))$ time for $k = \sqrt{n}$. The data structure can be maintained in $\cOtilde(\sqrt{n})$ when a substitution update is applied to $S$.
\end{theorem}

With Theorem \ref{t:countClose}, and the above reduction using the $k$-words tree, our main result for inverted suffix array is trivially achieved:
\begin{theorem}
Given a dynamic string $S[1\ldots n]$ there is a data structure that reports $iSA_S[i]$ in $O(\log^4(n))$ time. The data structure can be maintained in $\cOtilde(\sqrt{n})$ time when a substitution update is applied to $S$. The data structure uses $\cOtilde(n)$ space.
\end{theorem}

\subsection{Dynamic Suffix Array}
A query $i$ for $SA_S[i]$ can also be reduced to a query about close suffixes using the $k$-words tree by applying the following recursive algorithm:

\noindent {\bf$Findvr(Root,i)$: }\\
\fbox{\begin{minipage}{12cm}
{\sf Input:} An integer $i$ and a node $Root$ in the $k$-words tree. Initially the root.
\\\\
{\sf Output:} A node $v$ and an integer $r$.
\\\\
Let $R$ and $L$ be the right and the left children of $Root$, respectively. Let $|T(R)|$, $|T(L)|$ be the number of words in the trees rooted in $R$ and in $L$, respectively. Let $|Root|$ be the number of words contained in the root node. 
\begin{enumerate}
    \item If $|T(L)| \ge i$  return $Findvr(L,i)$.
    \item If $|T(L)| + |Root| \ge i$ return the tuple $(Root, i - |T(L)|)$.
    \item If $|T(L)| + |Root| < i$ return $Findvr(R,i - |T(L)| - |Root|)$
\end{enumerate}
\end{minipage}}

The following can be easily verified:
\begin{fact}
Applying Algorithm $Findvr(R,i)$ with $R$ the root of the $k$-words tree of $S$ yields the node $v$ containing $SA[i]$ and the lexicographic rank $r$ of $SA_S[i]$ within the suffixes close to $SA_S[i]$. The time complexity of Algorithm $Findvr$ is $O(\log(n))$.
\end{fact}

With that, reporting $SA[i]$ is reduced to a Close Suffix Selection Query defined below.

{\defproblema{\textsc{Close Suffixes Select}}{Given a subword of size $k$ represented as an index $i \in [1 \ldots n - k + 1]$ and a rank $r$, return the suffix $S[i \ldots n]$ with lexicographic rank $r$ among the suffixes starting with $S[i\ldots i + k - 1]$}}

In Section \ref{s:dcssq}, we prove the following: 
\begin{theorem}\label{t:selectClose}
Given a dynamic text $S[1\ldots n]$, there is a data structure that support close suffixes selection queries in $O(\log^5(n))$ time. The data structure can be maintained in $\cOtilde((\frac{n}{k})^2 + k)$ time when an edit operation is applied to $S$. The data structure uses $\cOtilde(n + (\frac{n}{k})^2)$ space.
\end{theorem}

By setting $k = n^{\frac{2}{3}}$ and applying the above reduction with a $k$-words tree, we obtain the following:
\begin{theorem}\label{t:dsa}
For a dynamic string $S[1\ldots n]$, there is a data structure that supports look up queries to $SA_S[i]$ in $O(\log^5(n))$ time. When an edit operation update is applied to $S$, the data structure can be maintained in $\cOtilde(n^{\frac{2}{3}})$ time. The data structure uses $\cOtilde(n)$ space.
\end{theorem}

\section{Dynamic Close Suffix Select Queries}\label{s:dcssq}
In this section, we present the main idea for proving Theorem \ref{t:selectClose}. 
Let $w$ be a word of size $t = \frac{k}{2}$. We denote as $POR_w$ the periodic occurrences representation of $w$ in $S[1\ldots n]$ and $Occ_w$ the set of occurrences of $w$. We denote as $A_w[1 \ldots |Occ_w|]$ the array consisting of the starting indexes of occurrences of $w$ sorted by lexicographic order of the suffixes starting in these indexes. The size of $POR_w$ is bounded by $O(\frac{n}{k})$ while the size of $A_w$ may be $\Omega(n)$. 

Let $C = (a,b,p) \in POR_w$ be a periodic cluster of occurrences of $w$ with period $per(w)=p$.

\begin{definition}
Let $w_t = S[s_t \ldots e_t]$ be an occurrence implied by $C$ of $w=S[s\ldots e]$ and let $w_s = w[e - p + 1 \ldots e]$ be the suffix of size $p$ of $w$. The \textit{period rank} of $w_t$, denoted as $r_p(w_t)$, is the maximal integer $x$ such that $S[e_t +1 \ldots e_t + p \cdot x] = (w_s)^x$. Equivalently, $r_p(w_t)$ is the maximal number of consecutive occurrences of $w_s$ following $w_t$.
\end{definition}
 One can easily observe that the $t$'th occurrence in $C$ has $r_p(w_t) = |C| - t - 1$. (Recall that the occurrences within a cluster are indexed with zero based indexes)

\begin{definition}[Cluster Head, Cluster Tail, Increasing and Decreasing Clusters]
Let $w_0 = S[ s_0 \ldots e_0]$ and $w_{|C|-1} = [s_{|C|-1} \ldots e_{|C|-1}]$ be the leftmost and the rightmost occurrence of $w$ represented by $C$, respectively (with respect to starting index). Let $w_s = w[e - p + 1 \ldots e]$ be the suffix of size $p$ of $w$. We call the prefix $S[1\ldots s_0 - 1]$ the \textit{head} of $C$ denoted as $Head(C)$ and the suffix $S[e_{|C|-1} + 1 \ldots n]$ the \textit{tail} of $C$ denoted as $Tail(C)$. We call $C$ an \textit{increasing} cluster if $Tail(C) >_L w_s$, or a \textit{decreasing} cluster if $Tail(C) <_L w_s$. 
\end{definition}

\begin{figure}
    \centering
      \includegraphics[width=\textwidth]{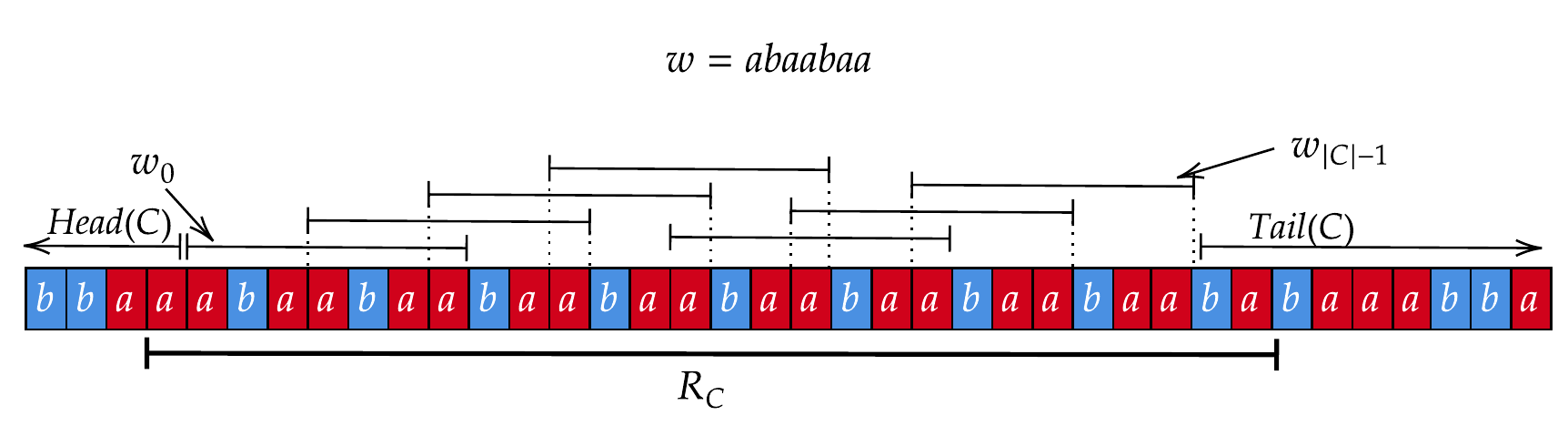}
    \caption{Periodic cluster of occurrences and its properties. The lines above the string segment denote the occurrences implied by the cluster. Note that in this example, $Tail(C) = bab \ldots > baa = w_s$. Therefore, the cluster is increasing. } \label{fig:clusterexample}
\end{figure}

See Figure \ref{fig:clusterexample} for an example of a cluster and its properties. In Section \ref{s:ersq}, we prove that $A_w$ has the following structure (derived from Lemma \ref{l:decthaninc} and Lemma \ref{l:decRanks}):

$A_w$ can be described as a concatenation $A_w = D[1\ldots |D|] , I[1\ldots |I|]$ with $D$ being the sub array containing the set of occurrences implied by decreasing clusters, and $I$ being the set of occurrences implied by increasing clusters. 

$D$ can be written as the concatenation $D = D_0, D_1 \ldots D_{m-1},D_{m}$. $m$ is the maximal period rank of an occurrence in a decreasing cluster. For every $r\in [0\ldots m]$ , $D_{r}$ contains the occurrences $w'$ from decreasing clusters with $r_p(w') = r$ sorted by lexicographic order of the tails of their clusters.

Similarly, $I = I_t ,I_{t-1} \ldots I_1,I_0$. $t$ is the maximal period rank of an occurrence in an increasing cluster. For every $r\in [0\ldots t]$ , $I_{r}$ contains the occurrences $s$ from increasing clusters with $r_p(s) = r$ sorted by lexicographic order of the tails of their clusters (See Figure \ref{fig:awdemons}).

Let $w = S[s\ldots e]$ be a subword of $S$. Let $l$ and $r$ be two non-negative integers. 
\begin{definition}
An occurrence $w_1 = S[s_1\ldots e_1]$ of $w$ is an $(l,r)$-extendable occurrence of $w$ if $lcs(s-1,s_1-1) \ge l$ and $lcp(e+1,e_1+1) \ge r$. We define the array $A^{(l,r)}_w$ as the lexicographically sorted array of suffixes of $S$ starting with an $(l,r)$-extendable occurrence of $w$.
\end{definition}

It is easy to see that given a word $w^c = S[s - l \ldots e + r]$ containing $w$, an occurrence of $w^c$ in index $i$ is equivalent to an $(l,r)$-extendable occurrence of $w$ in $i + l$. Therefore, $A_{w^c}[i] = A^{(l,r)}_w[i] - l$ for every $i \in [1 \ldots |A_{w^c}|]$. We consider the following problem:

{\defproblema{\textsc{Extension Restricting Selection}}
{
\textbf{Preprocess:} A word $w=S[s\ldots e]$ and $POR_w$
\\
\textbf{Query:} upon input $l,r,i$. Report $A^{(l,r)}_w[i]$ }
}

In Section \ref{s:ersq}, we prove the following:
\begin{theorem}\label{t:extSelect}
If $LCE$ queries can be evaluated in $O(\log(n))$ time, the periodic occurrences representation $POR_w$ of a word $w=S[s\ldots e]$ can be preprocessed in time $\cOtilde(|POR_w|)$ to answer Extension Restricting Selection queries in $O(\log^5(n))$ time. The data structure uses $\cOtilde(|POR_w|)$ space.
\end{theorem}

Constructing the data structure described in Theorem \ref{t:extSelect} is the most technically involved part in the construction of our dynamic suffix array data structure. It is achieved by exploiting the characterization of the structure of $A_w$. Our insights on the structure of $A_w$ allow us to reduce an Extension Restricting Selection query to a small set of multidimensional range queries.

We are now ready to present the algorithm for dynamic Close Suffix selection queries.
We initialize a data structure for supporting dynamic $LCE$ queries on $S$ by employing Lemma \ref{lem:dynLCE}. We also initialize a dynamic data structure for obtaining periodic occurrences representations by employing Fact \ref{l:fastCluster}.
Let $t=\lfloor \frac{k}{2} \rfloor$. We partition $S$ into words of size $t$ starting in indexes $i = 0 \bmod t$. For every word $w_x=S[x\cdot t \ldots (x+1)\cdot t - 1]$ for $x \in [0 \ldots \lfloor \frac{n}{t} \rfloor]$, we obtain $POR_{w_x}$ in $\cOtilde(\frac{n}{t})$ time. We preprocess $POR_{w_x}$ for Extension Restricting Selection queries using Theorem \ref{t:extSelect}. We keep an array $ERSQ[1 \ldots  \lfloor \frac{n}{t} \rfloor]$ with $ERSQ[x]$ containing a pointer to the Extension Restricting Selection queries data structure of $w_x$.

Upon an edit operation update, we update the data structures for $LCE$ queries and for obtaining the periodic occurrence representation. We then partition the updated $S$ in the same manner as in the initialization step. Note that the words in the partition may change, as the size of $S$ may change due to an insertion or a deletion update, and $k$ may be a function of $|S|=n$. For every word $w_x$ in the updated partition, we extract $POR_{w_x}$ and evaluate the extension restricting selection queries data structures for every word $w_x$ from scratch. We also construct the array $ERSQ$ from scratch as in the initialization step.

Upon a Close Suffix Selection query $(i,r)$, the word $w_{\lceil \frac{i}{t} \rceil}$ is completely contained within $w = S[i \ldots i + k -1]$. Let $L$ be the distance between the starting indexes of  $w_{\lceil \frac{i}{t} \rceil}$ and $w$ and let $R$ be the distance between the ending indexes. We execute an Extension Restricting Selection query $(L,R,r)$ on the Extension Restricting Selection queries data structure of $w_{\lceil \frac{i}{t} \rceil}$. Let the output of the query be $z$. We return $z  -L$ .

\textbf{Correctness:} A Close Suffix Selection query $(i,r)$ is equivalent to reporting $A_w[r]$. Denote the indexes of $w_{\lceil \frac{i}{t} \rceil}=S[s \ldots e]$. The word $w = S[i \ldots i + k - 1]$ can be written as $w = S[s - L , e + R]$. Therefore, $A_w[r] =  A^{(L,R)}_{w_{\lceil \frac{i}{t} \rceil}}[r] - L$. 

\textbf{Preprocessing time:} The preprocessing times of the $LCE$ data structure and the periodic occurrences data structure are both bounded by $\cOtilde(n)$. The remaining operations are, in fact, an initial update. The complexity analysis for these is provided below.

\textbf{Update time:} Evaluating the partition can be executed in $O(\frac{n}{k})$ time in a straightforward manner. There are $\frac{n}{t} \in O(\frac{n}{k})$ words in our partition. For every word $w$, we find $POR_w$ in $\cOtilde(\frac{n}{k})$ time. Since $|w| = \frac{k}{2}$, we have $|POR_w| \in O(\frac{n}{k})$. So constructing the Extension Restricting Selection data structure from $POR_w$ takes $\cOtilde(\frac{n}{k})$. The update time for the $LCE$ data structure is polylogarithmic and the update time for the periodic occurrences representation data structure is $\cOtilde(k)$. The overall update complexity is dominated by $\cOtilde((\frac{n}{k})^2 + k)$.

\textbf{Space:} The space consumed by the Extension Restricting Selection data structures is bounded by $\cOtilde((\frac{n}{k})^2)$. The space consumed by the array $ERSQ$ is bounded by $O(\frac{n}{k})$. The space consumed by the data structure for periodic occurrences representation is bounded by $O(n)$. The space consumed by the $LCE$ data structure is bounded by $\cOtilde(n)$. The overall space complexity is bounded by $\cOtilde(n + (\frac{n}{k})^2)$.

\textbf{Query time:} The query consists of a constant number of basic arithmetic operations, a single array lookup and a single Extension Restricting Selection query. The query time is dominated by $O(\log^5(n))$. \QED

\section{Dynamic Close Suffixes Rank Queries}\label{s:dcsrq}
In this section, we set $k = \sqrt{n}$ to be the threshold parameter for distinguishing between close and far suffixes.

We denote the rank of a suffix $S[i \ldots n]$ among the suffixes close to $S[i \ldots n]$ as $r(i)$. The following lemma is required to initialize our data structure. 
\begin{lemma}\label{l:staticrankqueries}
$r(i)$ can be evaluated for all the suffixes of $S[1\ldots n]$ in $O(n)$ time    
\end{lemma}
\textbf{Proof:}
Construct the suffix array $SA[1\ldots n]$ and $LCP$-Array $LA[1\ldots n]$, where $LA(i)=LCP(SA[i-1],SA[i])$, in $O(n)$ time ~\cite{4976463,fischer2011inducing}. We set $r(1) = 0$ and proceed to set $r(i)$ for $i \in [2 \ldots n]$ in increasing order of $i$. 

If $LA[i] < k$ then $SA[i]$ is the lexicographically minimum among the suffixes close to $SA[i]$. Therefore, we set $r(SA[i]) = 0$. If $LA[i] \ge k$ then $SA[i]$ is close to $SA[i-1]$ and is the lexicographical successor of $SA[i-1]$. Therefore, $SA[i]$ is larger than exactly $r(SA[i]) = r(SA[i-1]) +1$ close suffixes. Every index is treated in constant time and the overall time is linear. \QED

Our dynamic data structure for close suffixes rank queries is an `array' of integers $R[1\ldots n]$, initialized as the initial close suffix ranks calculated by Lemma \ref{l:staticrankqueries} ($R[i] = r(i)$). When a substitution update is applied to $S$  the ranks $r(i)$ of the suffixes change. We wish to change $R[i]$ in a corresponding manner.

We provide an analysis of the updates that are need to be applied to $R$ when a symbol substitution is applied to $S$. Unsurprisingly, these changes are too wild to be efficiently applied to an ordinary array (e.g. - $ \Omega (\sqrt{n}) $ indexes may have to be updated). However, the updates to the ranks can be described as a small set of \textit{interval increment} and \textit{stairs} updates on $R$.

\begin{definition}
An {\em interval increment update} is denoted by $I = (i,j,x)$. Applying $I$ to $R$ increases the value of $R[a]$ by $x$ for $a \in [i \ldots j]$.  
\end{definition}

\begin{figure} 
    \centering
      \includegraphics[width=\textwidth]{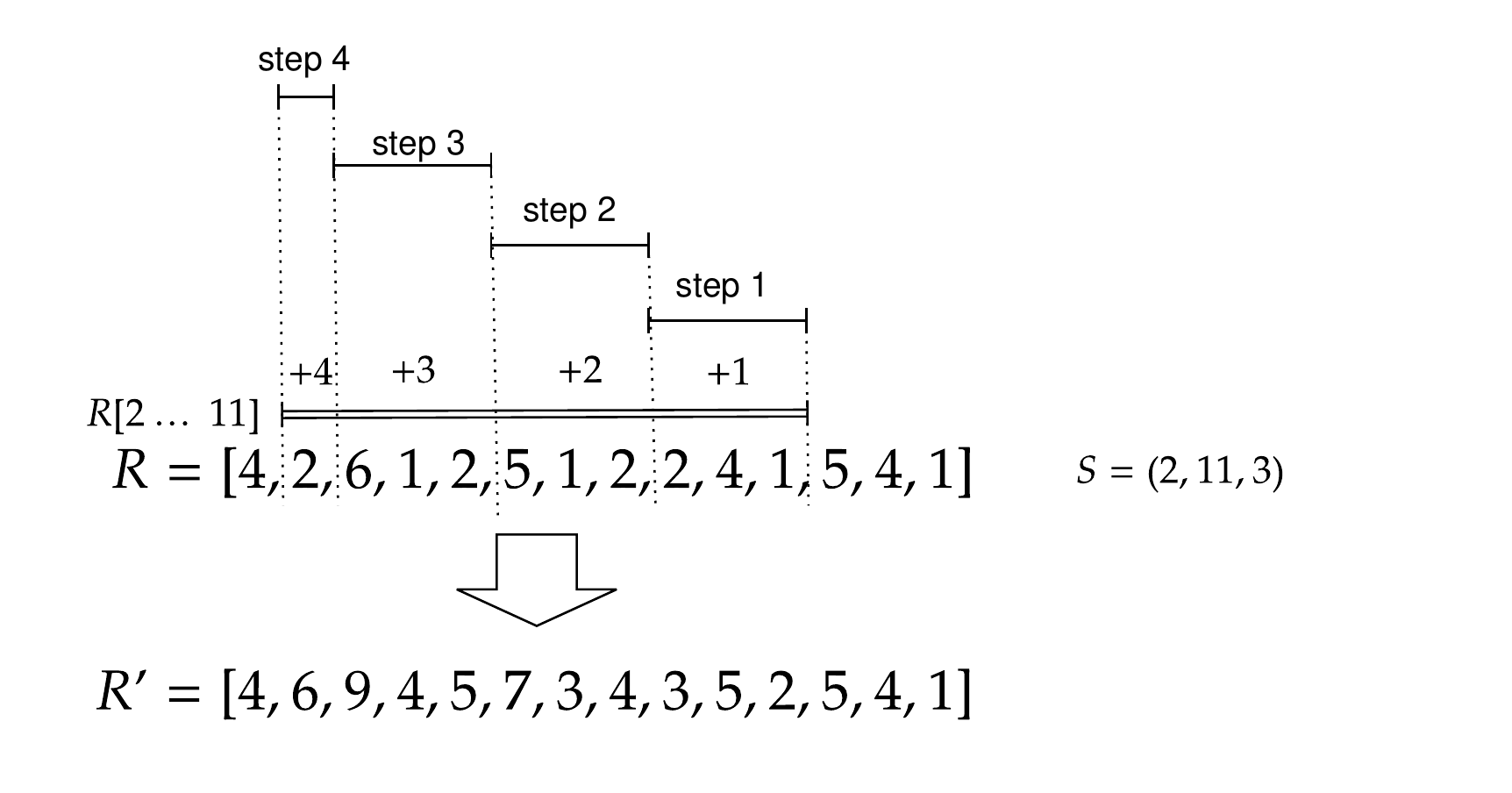}
    \caption{A visualization of an application of a decreasing stairs update and its properties. The update $S$ is applied to the array $R$ to obtain $R'$. } \label{fig:stairsexp2} 
\end{figure}

\begin{definition}
A {\em decreasing stairs update} (See Figure \ref{fig:stairsexp2}) is denoted by $S = (i,j,p)$. Applying $S$ to $R$ increases the value of the indexes in $R[i\ldots j]$ as follows: the rightmost $p$ indexes $R[j - p + 1 \ldots j]$ are increased by $1$, the $p$ indexes to their left are increased by $2$, and so on until we finally exceed $R[i\ldots j]$. 

Formally, for every $t \in [1 \ldots \lceil \frac{j - i + 1}{p} \rceil]$, the counters $R[a]$ with $a \in [max(j - t\cdot p + 1,i) \ldots j - (t-1) \cdot p]$ are increased by $t$. 

We call $p$ the {\em step width} of the stairs update $S$. We call the interval increased by exactly $t$ the {\em $t$'th step} of the update.
\end{definition}

An increasing stairs update is defined symmetrically, with the `stairs' having increasing (rather than decreasing) values from left to right. A stairs update can also be a negative update, decreasing the values of the $t$'th step by $t$ rather than increasing it by $t$. A formal definition of these variants can be found in Section \ref{s:updates}.

\begin{example}
Let $R=(0,0,0,0,0,0,0,0,0,0)$. Applying a decreasing stairs update $(3,8,2)$ on $R$ will result in  $R=(0,0,4,3,3,2,2,1,1,0,0)$. Applying an increasing stairs update $(2,8,3)$ instead will result in $R'=(0,1,1,1,2,2,2,3,3,0,0)$. Applying a negative decreasing stairs update $(4,10,2)$ on $R'$ will result in $R'=(0,1,1,-3,-2,-1,-1,1,1,-1,-1)$.
\end{example}

In section \ref{s:updates}, we prove the following:

\stairsp*

For the sake of self containment, we provide a proof for the following (simple, possibly folklore) claim:
\begin{restatable}{claim}{intupdates}\label{c:intupdates}
There is a data structure for maintaining an indexed set of integers $R$ to support both applying an interval increment update to $R$ and look-up queries $R[i]$ in time $O(\log(n))$.
\end{restatable}

Our algorithm may require stairs updates with varying values of $p$.  We will eventually describe the details for how the restricted data structure of Theorem \ref{t:stairsp} is utilized. Until then, we refer to stairs update application and query as an $O(\log^3(n))$ time procedure. We want stairs updates to be executed in $O(\log^3(n))$ rather than $O(\log^3(U))$. This can be achieved in amortized time by rebuilding our data structure from scratch after every $O(n)$ updates. The running time can be deamortized using standard techniques.

From now on, we fix $x \in [1\ldots n]$ to be the index within $S[1\ldots n]$ such that $S[x]$ was substituted. By $S$ we refer to the text after the update was applied, and by $T$ we refer to the text prior to the substitution. For an index $i\in [1\ldots n ]$, we denote the word $w^{new}_i = S[i\ldots i + k - 1]$ and $w^{old}_i= T[i\ldots i + k - 1]$. For $w^{new}_i$ and $w^{old}_i$ to be defined for every $i\in [1\ldots n]$, we process $S$ as if it has the string $\$^k $ appended to its end for some symbol $\$ \notin \Sigma$ that is lexicographically smaller than every symbol $\sigma \in \Sigma$. Note that this does not affect the lexicographical order between the suffixes of $S$.

Consider the following clustering of the suffixes of $S$: two suffixes $T^i$ and $T^j$ are in the same cluster if and only if they are closed. Equivalently, every cluster $C$ corresponds to a word $w_C$ and a suffix $T^i$ is in $C$ if and only if $w^{old}_i = w_C$.  Even though we do not explicitly maintain such a clustering in our data structure, it is helpful to think of the suffixes of $S$ as if they are stored in this manner.

Observe that $R[i]$ is the lexicographic rank of $T^i$ among the suffixes in $C_{w^{old}_i}$. After an update is applied to $T$, $R[i]$ needs to be modified to be the rank of $S^i$ among the suffixes in a possibly different cluster $C_{w^{new}_i}$.  We distinguish between two types of suffixes:
\begin{enumerate}
    \item \textbf{Dynamic suffixes: } $S^i$ with $x \in [i \ldots  i +k - 1]$: For these suffixes, $w^{old}_i \neq w^{new}_i$. 
    \item \textbf{Static suffixes:} $S^i$ with $x \notin [i \ldots i + k - 1]$. For these suffixes, $w^{old}_i= w^{new}_i$
\end{enumerate}

\begin{figure} 
    \centering
      \includegraphics[width=\textwidth]{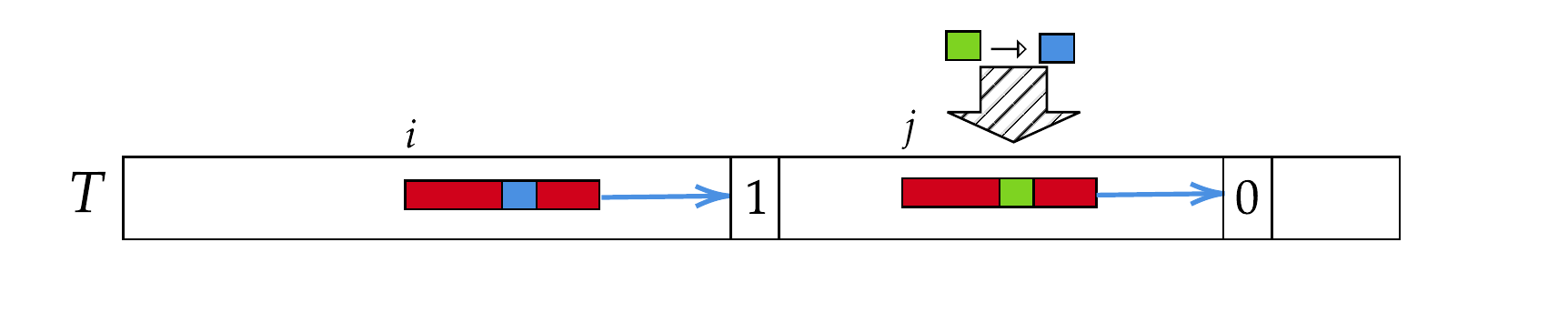}
    \caption{A visualization of a shift update. The (mostly) red rectangles resemble $w^{old}_i$ and $w^{old}_j$. In $T$, the words $w^{old}_i$ and $w^{old}_j$ are identical, except for the marked non-red rectangle, which represents a `blue' symbol in $w^{old}_i$ and a `green' symbol in the corresponding location in $w^{old}_j$. The blue arrows to the right of $w^{old}_i$ and $w^{old}_j$ resemble the $LCP$ between $i+k-1$ and $j+k-1$. In this example, the green symbol in $w^{old}_j$ is modified by an update to a blue symbol, so we have $w^{new}_i = w^{new}_j$. The symbol following the $LCP$ after $i$ is $1$, and the symbol following the $LCP$ after $j$ is $0$, so we have $S^j <_L S^i$. The suffix $S^j$ became close to $S^i$ as a result of the last substitution, so $r(i)$ is increased by $1$. The reason is that we have a new suffix close to $S^i$ that is lexicographically smaller than $S^i$. } \label{fig:ShifExample}
\end{figure}

\begin{figure} 
    \centering
      \includegraphics[width=\textwidth]{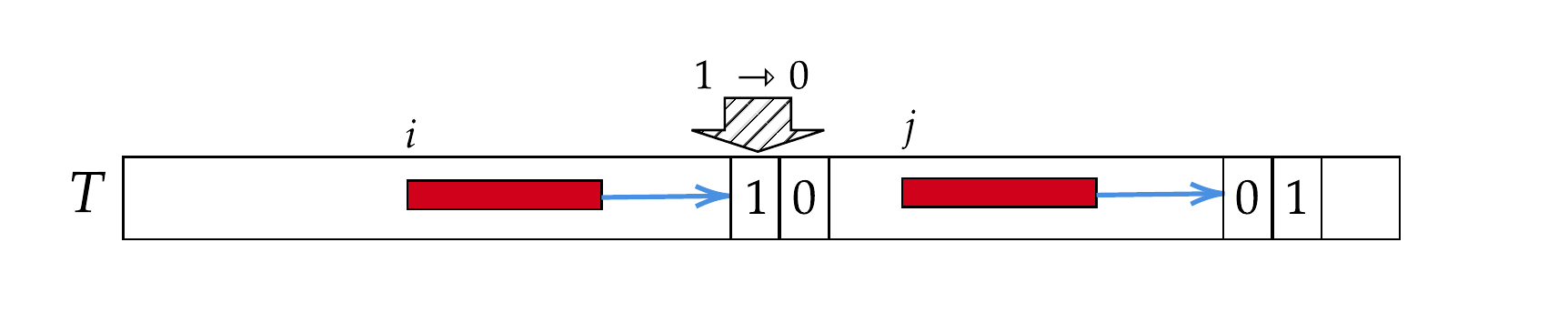}
    \caption{A visualization of an overtake update. The red rectangles resemble $w^{old}_i$ and $w^{old}_j$, and we have $w^{old}_i = w^{old}_j$, so $T^i$ and $T^j$ are close suffixes. The substitution occurs in an index that is not contained either in $w^{old}_i$ or in $w^{old}_j$, therefore they are both static suffixes. The $lcp$ between $i+k-1$ and $j+k-1$ is denoted by the blue arrows. Prior to the update, the $lcp$ terminates in a $1$ following $S^i$ and in a $0$ following $j$. Therefore, $T^i >_L T_j$. After the update is applied, the $lcp$ is extended by $1$ symbol and we have $S^i <_L S_j$, so $S^j$ overtakes $S^i$. The rank of $S^i$ should be decreased by $1$ and the rank of $S^j$ should be increased by $1$ } \label{fig:OvertakeExample}
\end{figure}

Note that dynamic suffixes are suffixes that are moved to a different cluster as a result of the update, and static suffixes are the suffixes that stay in the same cluster after the update is applied. We partition the updates applied to $R$ into $3$ types:
\begin{enumerate}
    \item \textbf{Evaluation Updates} - Evaluating $R[i]$ for every dynamic suffix $S^i$.
    \item \textbf{Shift Updates} - Increasing (resp. decreasing) by $1$ the values of $R[i]$ for a static suffixes $S^i$ as a result of a dynamic suffix $S^j$ that has $w^{new}_j = w^{new}_i$ (resp. $w^{old}_j = w^{old}_i$) and $S^j <_L S^i$ (resp. $T^j < T^i$) (See Figure \ref{fig:ShifExample}).
    \item \textbf{Overtakes} - For a pair of close static suffixes $S^i$,$S^j$ such that $S^i >_L S^j$ and $T^i <_L T^j$, increasing the value of $R[i]$ by 1 and decreasing the value of $R[j]$ by 1 (See Figure \ref{fig:OvertakeExample}).
\end{enumerate}

An evaluation update corresponds to evaluating the rank of a dynamic suffix within the new cluster for which it was recently inserted. A shift update corresponds to an increase (resp. decrease) in the rank of a static suffix $S^i$ as a result of a dynamic suffix that is lexicographically smaller than $S^i$ being inserted (resp. deleted) from the cluster of $S_i$. An overtake corresponds to a change in lexicographical order between two static suffixes within the same cluster.

It should be clear that if we apply all the evaluation updates, all the shift updates, and all the overtake updates to $R$, then $R$ is properly set and $R[i] = r(i)$ for every $i\in [1\ldots n]$ with respect to the close suffix ranks $r(i)$ in the updated text $S$. In the following sections we show how to efficiently apply these types of updates to $R$. 

\subsection{Evaluation updates}\label{ss:eval}
In this section, we present an $O(\sqrt{n})$ time algorithm for evaluating $r(i)$ for $i \in [x - k + 1 \ldots x -\frac{k}{2} + 1]$. These are only half of the dynamic suffixes. We conclude by explaining how the same techniques can be used to construct a similar algorithm to evaluate $r(i)$ for $i\in [x - \frac{k}{2} \ldots x]$.

The task of evaluating $r(i)$ can be equivalently described as a counting problem. Specifically, $r(i)$ is the number of suffixes $S^j$ starting with $w^{new}_i$ such that $S^j <_L S^i$. With that characterization in mind, a naive approach would be the following: Iterate the values $i\in [x -k + 1\ldots x]$ and for every value, evaluate the set of close suffixes and count how many of these are lexicographically smaller than $S^i$. An explicit straightforward execution of this idea would result in an algorithm with running time $\cOtilde( n^{1.5})$, since the number of dynamic suffixes is $O(k) = O(\sqrt{n})$ and the number of suffixes close to a certain dynamic suffix can not be bounded by anything smaller than $O(n)$. Of course, this is not a satisfactory running time. We present an implementation of the naive algorithm that exploits the structure of the problem to achieve a time complexity of $\cOtilde(\sqrt{n})$. 

Our enhanced implementation of the naive approach relies on exploiting the proximity of the starting indexes of the words $w_i^{new}$ corresponding to dynamic suffixes $S^i$. Instead of evaluating the set of occurrences of $w_i^{new}$ for every $i \in [x - k + 1 \ldots x -\frac{k}{2} + 1]$, we consider the occurrences of the word $w^c_l =S[x-\frac{k}{2} + 1 \ldots x]$. Note that $w^c_l$ is a subword of all the words $w_i^{new}$ with $i \in [x - k + 1 \ldots x -\frac{k}{2} + 1]$. We use an occurrence of $w^c_l$ as an anchor that may be extendable to an occurrence of $w_i^{new}$ for some values of $i$. Since $w^c_l$ has length $\Theta(k)$, the occurrences of $w^c_l$ in $S$ can be represented as a set of $O(\frac{n}{k}) =O( \sqrt{n})$ clusters (Fact \ref{f:perrep}). We provide a method to efficiently process a cluster $C$ of occurrences of $w^c_l$ to deduce the modifications to $r(i)$ that are required as a result of occurrences represented by $C$.

To simplify notations, we define the function $D(\delta) = x - \frac{k}{2} + 1 - \delta$. Upon input integer $\delta$, the function $D(\delta)$ outputs the index in $S$ that is $\delta$ indexes to the left from the starting index of $w^c_l$. The following fact captures the relation between the occurrences of $w^c_l$ and the suffixes that should be counted by our algorithm.

(The indexes of $w^c_l$ are renamed to $S[s \ldots e]$ for clearer presentation)
\begin{fact}\label{f:wcl}
Let $S^i$ be a dynamic suffix with $i= D(\delta )$ for some $\delta \in [0 \ldots \frac{k}{2}]$. For an index $j\in [1 \ldots n]$, the following are equivalent:
\begin{enumerate}
    \item $S^j$ is close to $S^i$ (starts with an occurrence of $w^{new}_{i}$) and $S^j <_L S^i$.
    \item There is an occurrence $w_1=S[s_1\ldots e_1]$ of $w^c_l=S[s \ldots e]$ in index $s_1 = j+ \delta$ such that $S^{s_1} <_L S^s$, $lcs(s_1 -1, s-1) \ge \delta$ and $lcp(e_1+1,e+1) \ge \frac{k}{2} - \delta$. 
\end{enumerate}
\end{fact}

\begin{figure}[htpb!] 
    \centering
    \includegraphics[width=\textwidth]{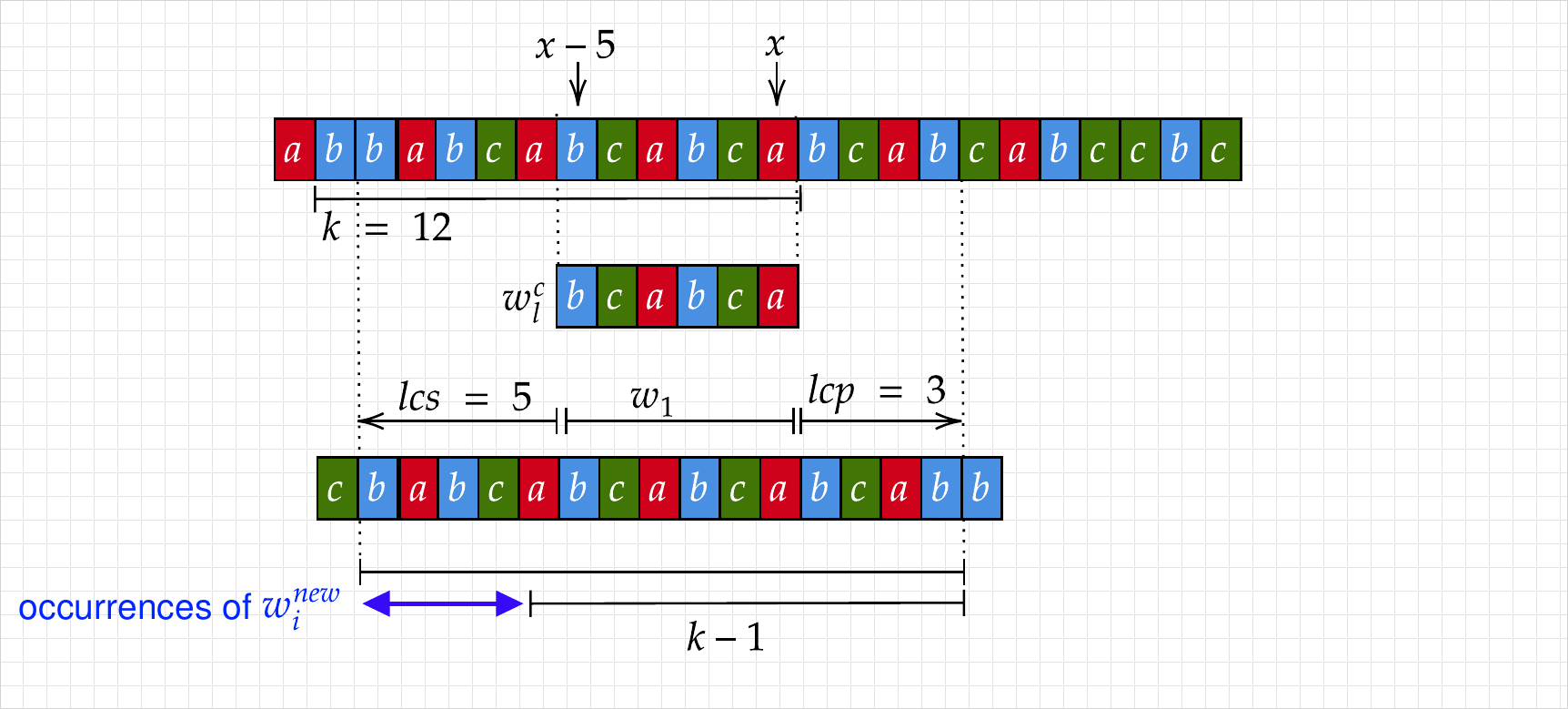}
    \caption{A visualization of Fact \ref{f:wcl} with $k=12$. The upper segment is a substring of $S$ surrounding the changed index $x$. The word $w^c_l$ is ``$bcabca$" ending in index $x$. The lower segment is a separated substring of $S$ containing an occurrence of $w^c_l$ (aligned with the occurrence in the upper segment). Every index within the blue interval marked in the lower string is the starting index of an occurrence of its corresponding aligned $w^{new}_i$} \label{fig:wclexample2}
\end{figure}

Fact \ref{f:wcl} holds due to the fact that $w^c_l$ is a substring of $w^{new}_i$ for these values of $i$. An occurrence of $w^{new}_i$ is equivalent to an occurrence of $w^c_l$ that has a sufficiently long common extension with $w^c_l$. See Figure \ref{fig:wclexample2} for a visualization.

For $i = D(\delta)$ with $\delta \in [0 \ldots \frac{k}{2}]$, we say that an occurrence of $w^{new}_i$ in index $j$ is \textit{induced} by the occurrence of $w^c_l$ in index $j + \delta$.

We use Fact \ref{f:wcl} to evaluate $r(i)$ for every  $i \in [x - k + 1 \ldots x -\frac{k}{2} + 1]$
in a parallel manner. Instead of processing occurrences of $w^{new}_i$ for every dynamic suffix $S^i$ with $i \in [D(\frac{k}{2}) \ldots D(0)]$, we process the occurrences of $w^c_l=S[s \ldots e]$. For every occurrence of $w_1 = S[s_1 \ldots e_1]$, we identify values of $i$ for which $w^{new}_i$ is induced by $w_1$ according to Fact \ref{f:wcl}. If $S^{s_1} <_L S^s$, we count the induced suffixes as smaller close suffixes of their corresponding dynamic suffixes.

A technical implementation of the above approach is given as Algorithm \ref{alg:1}

\begin{algorithm}\label{alg:1}

\fbox{\begin{minipage}{12cm}
{\sf Initialize an array of counters $Re[0\ldots \frac{k}{2}]$ with zeroes. }. 

{\sf For every} occurrence $w=S[s_1 \ldots e_1]$ of $w^c_l$:
\begin{enumerate}
    \item Check if $S^{s_1} <_L S^s$. If it is not - stop processing this occurrence.
    \item Compute $l =lcs(s_1 - 1,s-1)$ and $r = lcp(e_1 + 1, e+1)$.
    \item Increase the counters of $Re$ with indexes in $[max(0, \frac{k}{2} - r) ,  \min(l, \frac{k}{2})]$ by $1$.
\end{enumerate}
{\sf For every} $\delta \in [0 \ldots \frac{k}{2}]$, set $R[D(\delta)] \longleftarrow Re[\delta ]$.
\end{minipage}}
\end{algorithm}

\begin{claim}\label{c:alg1cor}
By the end of the inner loop of Algorithm \ref{alg:1}, $Re[\delta] = r(D(\delta))$
\end{claim}
\textbf{Proof:} The counter $Re[\delta]$ is increased by $1$ as many times as an occurrence $w= S[s_1 \ldots e_1]$ of $w^c_l$ that satisfies the following conditions is processed:
\begin{enumerate}
    \item $lcs(s_1-1,s-1) = l \ge \delta$ 
    \item $lcp(e_1 + 1, e + 1)= r \ge \frac{k}{2} - \delta$
    \item $S^{s_1} <_L S^s$
\end{enumerate}

According to Fact \ref{f:wcl}, there are exactly as many occurrences of $w^c_l$ satisfying these conditions as suffixes close to $S^{D(\delta)}$ that are lexicographically smaller than $S^{D(\delta)}$. \QED

Assuming that we have the occurrences of $w^c_l$ in hand, Algorithm \ref{alg:1} can be straightforwardly implemented in $\cOtilde(|occ(w^c_l)|)$ time, with $occ(w^c_l)$ being the set of occurrences of $w^c_l$ in $S$. Since we may have $|w^c_l| \in \Theta(n)$, this is not sufficiently efficient. To further reduce the time complexity, we process the periodic occurrences representation of $w^c_l$ rather than the explicit set of occurrences.

We use Lemma \ref{l:fastCluster} with $x=\frac{k}{2} = |w^c_l|$ to obtain the periodic occurrences representation of $w^c_l$ in $\cOtilde(\frac{2n}{k}) = \cOtilde(\sqrt{n})$ time. The following is proven in section \ref{s:comsrq}:

\begin{restatable}{lemma}{clusterlceprog}\label{l:clusterlceprog}

Let $C = (a,b,p)$ be a cluster of occurrences of a word $w=S[s\ldots e]$ with $per(w) = p$ in $S[1\ldots n]$. Let the occurrences represented by $C$ be $w_0, w_1 \ldots w_{|C|-1}$ with $w_t = S[s_t \ldots e_t]$ and $s_t = a + t\cdot p$. Let $l_t = lcs(s_t - 1,s - 1)$ and $r_t = lcp(e_t + 1, e + 1)$. Let $Ex_l$ and $Ex_r$ be the lengths of the extension of the run with period $p$ containing $w$ to the left of $s-1$ and to the right of $e+1$, respectively. Let $Ex^C_l$ and $Ex^C_r$ be the lengths of the extension of the run with period $p$ containing $w_0$ to the left of $s_0-1$ and to the right of $e_0+1$, respectively.
\begin{enumerate}
    \item If $Ex_l \neq Ex^C_l + p\cdot t$, we have $l_t = \min(Ex_l, Ex^C_l + p\cdot t)$.
    \item If $Ex_r \neq Ex^C_r - p\cdot t$, we have $r_t = \min(Ex_r, Ex^C_r - p\cdot t)$
\end{enumerate}
The values of $Ex_l$, $Ex_r$, $Ex^C_l$ and $Ex^C_r$ can be evaluated in $O(\log(n))$ time given $C$, the indexes of $w$, and a data structure for $O(\log (n))$-time LCE queries. 
\end{restatable}

For most of the occurrences $w_t$ represented by a cluster $C$, Lemma \ref{l:clusterlceprog} provides a compact, efficiently computable  representation of the common extensions of $w_t$ and $w$. We note that every cluster may contain a constant number of occurrences for which Lemma \ref{l:clusterlceprog} does not provide a representation of $l_t$ and $r_t$. We formally define these occurrences as follows.

\begin{definition}
Let $C = (a,b,p)$ be a cluster of occurrences of a word $w=S[s\ldots e]$ with $per(w) = p$ in $S[1\ldots n]$. Let $Ex_l$, $Ex_r$, $Ex^C_l$ and $Ex^C_r$ be the lengths of the extension of the runs as in Lemma \ref{l:clusterlceprog}. $w_t$ is an {\em aligned occurrence} if $Ex_l = Ex^C_l + p\cdot t$ or  $Ex_r = Ex^C_r - p\cdot t$.
\end{definition}

The aligned occurrences within a cluster $C$ can be easily found in constant time given $C$, $w$ and the runs extension lengths. The two constrains defining aligned occurrences are linear equations with respect to $t$. Therefore, there is at most one value $t$ satisfying each equation and it can be found in constant time.

For evaluation updates, we are only interested in the $LCE$ values that correspond to occurrences $w_t$ with $S^{s_t} <_L S^s$. To extract these values, we employ the following fact proven in Section \ref{s:comsrq}.

\begin{restatable}{claim}{perfastlex}
\label{f:perfastlex}
Let $s$ be an index and let $C= (a,b,p)$ be a periodic cluster with $s_t = a + t\cdot p$. There are consecutive intervals $I_<,I_> \subseteq [0 \ldots |C-1|]$ such that $S^{s_t} <_L S^s$ iff $t \in I_<$ and $S^{s_t} >_L S^s$ iff $t \in I_>$. The intervals $I_<$ and $I_>$ can be evaluated in $O(\log(n))$ time given $C$, $s$ and a data structure for $O(\log(n))$ LCE queries. 

\end{restatable}

Let $l_t = lcs(s-1, s_t-1)$ and $r_t=lcp(e+1,e_t+1)$. We evaluate a compact representation of the list $L^e = l_0,l_1,l_2 \ldots l_{|C|-1}$ by applying Lemma \ref{l:clusterlceprog}. We can obtain $I_<=[i \ldots j]$ by applying Claim \ref{f:perfastlex}. Recall that we are interested in occurrences $w_t$ such that $S^{s_t} <_L S^s$. So the relevant $lcs$ values are $l_t$ with $t \in I_<$. We proceed to work on the reduced list $L_< = l_i, l_{i+1} \ldots l_j$. For similar arguments, we proceed to work on the reduced list $R_< = r_i, r_{i+1} \ldots r_j$ corresponding to the relevant values of $r_t$.

So far, we characterized the structure of $lcp$ and $lcs$ values between a word and a set of its occurrences represented by a cluster. In what follows, we provide a method to efficiently apply the updates that are derived from these values to $R$.

\begin{definition}[$p$-Normal Sequence]
Let $B=b_0,b_1 \ldots b_t$ be a sequence of integers. $B$ is a $p$-normal sequence for some integer $p$ if:
\begin{enumerate}
    \item $B$ is a \textbf{fixed} sequence: $b_i = b_0$ for every $i\in[1\ldots t]$, or
    \item $B$ is an \textbf{arithmetic progression} with difference $p$: $b_i= b_0 + p \cdot t$ for every $i\in[1\ldots t]$, or
    \item $B$ is the \textbf{maximum} of two $p$-normal sequences: $b_i=(max(a^1_i,a^2_i))$ for two $p$-normal sequences $\{a^1_i\}$ and $\{a^2_i\}$, or
    \item $B$ is the \textbf{minimum} of two $p$-normal sequences: $b_i=(\min(a^1_i,a^2_i))$ for two $p$-normal sequences $\{a^1_i\}$ and $\{a^2_i\}$ 
\end{enumerate}

The degree of a $p$-normal sequence is denoted as $deg(B)$. If $B$ is either fixed or an arithmetic progression then $deg(B)=1$. If $B$ is a maximum sequence with $b_i=(max(a^1_i,a^2_i))$, or a minimum sequence with $b_i=(\min(a^1_i,a^2_i))$ then  $Deg(B)= Deg(\{a^1_i\}) + Deg(\{a^2_i\})$.

Finally, the construction tree of a $p$-normal sequence $B$ is a full binary tree denoted as $CT(B)$. It aims to represent the structure of $B$. If $B$ has $deg(B)=1$ then $CT(B)$ consist of a single node. If $B$ is a maximum sequence with $b_i=(max(a^1_i,a^2_i))$, or a minimum sequence with $b_i=(\min(a^1_i,a^2_i))$ then $CS(B)$ consist of a root with the left child being $CT(\{a^1_i\})$ and the right child being $CT(\{a^2_i\}) $
\end{definition}

\begin{example}
The sequence $A=4,7,10,13,16$ is a $3$-normal sequence with degree $1$, as it is an arithmetic progression with difference $p=3$. $CT(A)$ is a single root node $\{A\}$. The sequence $B=b_0,b_1 \ldots b_t$ with $b_i = max(13, 4 + 7\cdot t)$ is a $7$-normal sequence. The left input in the $max$ expression is a fixed sequence with respect to $t$, which is a $p$-normal sequence with degree $1$. The right sequence is an arithmetic progression with difference $7$ with respect to $t$, which is a $7$-normal sequence with degree $1$. therefore, $deg(B) = 1+1 = 2$. $CS(B)$ is a root node with two leaf children corresponding to the construction trees of $[7,7,7\ldots ]$ and of $[4,11,18 \ldots ]$.
\end{example}

The following lemma is proven in Section \ref{s:comsrq}
\begin{restatable}{lemma}{applynorfast}\label{l:applypnormfast} 
Let $U=u_1,u_2 \ldots u_{|U|-1}$ be a set of interval increment updates with $u_t = (i_t,j_t,1)$ and both $\{i_t\}$ and $\{j_t\}$ being $p$-normal sequences for some integer $p$ with degrees $Deg(\{i_t\}), Deg(\{j_t\}) \in O(1)$. Applying all the interval updates in $U$ is equivalent to applying a constant number of interval increment updates and stairs updates with stair width $|p|$ to $R$. The set of equivalent stairs and interval updates can be evaluated in constant time given the representation of $i_t$ and $j_t$ as $p$-normal sequence (The input can be given, for example, as the construction trees of $\{i_t\}$ and $\{j_t\}$).

\end{restatable}

Consider the occurrence $w_t$ of $w^c_l$ starting in index $s_t$ represented by $C$ for some $t \in I_<= [i_< \ldots j_<]$. When $w_t$ is iterated in Algorithm \ref{alg:1}, the interval increment $(max(0, r_t - \frac{k}{2}) , \min(l_t, \frac{k}{2}),1)$ is applied to $Re$. Denote the left border of the interval increment applied when $w_t$ is iterated as $i_t =max(0, r_t - \frac{k}{2})$ and the right border as $j_t = \min(l_t, \frac{k}{2})$. We make the following claim:

\begin{claim}\label{c:fastEvaluasios}
Given $C$, $w=S[s\ldots e]$ and $I_<$, all the updates $u_t=(i_t,j_t,1)$ for $t\in I_<$ can be applied to $Re$ in time $O(\log^3(n))$. 
\end{claim}
\textbf{Proof:} Let $Al$ be the set of aligned occurrences within $C$. It is known that $|Al| \le 2$. Denote the set of aligned occurrences as $Al= \{t_1,t_2\}$ with $t_1 < t_2$. If $|Al|\le 1$, we set $t_2 = \infinity{}$. If $|Al| = 0$, we set $t_1 = -\infinity{}$. For $t \notin Al$ we have $Ex_r \neq Ex^C_r - p\cdot t$ and  $Ex_l \neq Ex^C_l - p\cdot t$ . Therefore, $i_t = max(0, \min(Ex_r - \frac{k}{2},Ex^C_r - \frac{k}{2} - p\cdot t))$ and $j_t = \min(\min(Ex_l , Ex^C_l - p\cdot t), \frac{k}{2})$ according to Lemma \ref{l:clusterlceprog}. These are $(-p)$-normal sequence with degree $d_i = 3$. For the integers in $I_1 = [i_< \ldots t_1)$, both $\{i_t\}$ and $\{j_t\}$ are $p$-normal sequences. The same applies for the intervals $I_2 = (t_1 \ldots t_2)$ and $I_3 = (t_2 \ldots j_<]$. All the updates $u_t$ for $t\in I_1 \cup I_2 \cup I_3$ can be expressed as a constant number of stairs updates by applying Lemma \ref{l:applypnormfast} for every interval. Interval updates corresponding to aligned occurrences can be applied in $O(\log(n))$ time each by employing Claim \ref{c:intupdates}. Applying a constant number of stairs and interval updates can be executed in $O(\log^3(n))$ time by  Theorem \ref{t:stairsp}. The representations of $r_t$ and $l_t$ can be obtained in $O(\log(n))$ time by employing Lemma \ref{l:clusterlceprog}. \QED

We are now ready to present an efficient implementation of Algorithm \ref{alg:1}.
\begin{algorithm}\label{alg:2}

\fbox{\begin{minipage}{12cm}
{\sf Initialize an array of counters $Re[0\ldots \frac{k}{2}]$ with zeroes. } 

{\sf Obtain} the periodic occurrences representation $por$ of $w^c_l$.\\
{\sf For every} singular occurrence $w=S[s_1 \ldots e_1]$ in $por$:
\begin{enumerate}
    \item Check if $S^{s_1} <_L S^s$. If it is not, then stop processing this occurrence.
    \item Compute $l =lcs(s_1 - 1,s-1)$ and $r = lcp(e_1 + 1, e+1)$.
    \item Increase the counters of $Re$ with indexes in $[max(0, r - \frac{k}{2}) ,  \min(l, \frac{k}{2})]$ by $1$.
\end{enumerate}

{\sf For every} cluster $C=(a,b,c)$ of occurrences in $por$:
\begin{enumerate}
    \item Obtain $I_<$ using Claim \ref{f:perfastlex}.
    \item Apply all the updates corresponding to occurrences $w_t$ with $t \in I_<$ implied by $C$ by employing Claim \ref{c:fastEvaluasios}.
\end{enumerate}

{\sf For every} $\delta \in [0 \ldots \frac{k}{2}]$, set $R[D(\delta)] \longleftarrow Re[\delta ]$.
\end{minipage}}

\end{algorithm}

It is straightforward that the updates applied to $Re$ in Algorithm \ref{alg:2} are equivalent to the updates applied to $Re$ in Algorithm \ref{alg:1}. Therefore, the close suffix ranks evaluated in $Re$ are correct directly from Claim \ref{c:alg1cor}.

\begin{claim}\label{c:alg2comp}
Algorithm \ref{alg:2} runs in $\cOtilde(\sqrt{n})$ time.
\end{claim}
\textbf{Proof:} Since $|w^c_l| = \frac{k}{2}$, the size of the periodic occurrences representation is $\frac{2n}{k} \in O(\sqrt{n})$ and it can be obtained in time $\cOtilde(\sqrt{n})$ by applying Lemma \ref{l:fastCluster}.

For a singular occurrence in the representation, we process in logarithmic time just as in Algorithm \ref{alg:1}. For a cluster, we use Claim \ref{f:perfastlex} and Claim \ref{c:fastEvaluasios} to apply all the updates corresponding to occurrences implied by $C$. The bottleneck of handling a cluster is the application of Claim \ref{c:fastEvaluasios} in $O(\log^3(n))$ time.  The overall time to process an element, either a singular occurrence or a cluster, is polylogarithmic. Since there are $O(\sqrt{n})$ elements in the representation. Algorithm \ref{alg:2} runs in $\cOtilde(\sqrt{n})$ time. \QED

Note that Algorithm \ref{alg:2} initializes the data structure $Re$ for stairs updates with a constant stair size $p$, uses it to evaluate the ranks and inserts its values to the matching indexes in $R$. After the execution of Algorithm \ref{alg:2},  $Re$ can be discarded, since it is not used in future updates. The restricted result of Theorem \ref{t:stairsp} is sufficient for this case.

This concludes the evaluation of the ranks for dynamic suffixes starting in $[x-k + 1 \ldots x - \frac{k}{2}]$. We still need to handle the suffixes starting in $[x - \frac{k}{2} +1\ldots x]$. This is done in the same manner, but instead of $w^c_l = S[ x - \frac{k}{2} + 1 \ldots x]$ we consider the occurrences of $w^c_r= S[x \ldots x + \frac{k}{2}-1]$. One can easily verify that $w^c_r$ is a subword of every $w^{new}_i$ with $i \in [x - \frac{k}{2} +1\ldots x]$.

\subsection{Shift updates} 

Shift updates are handled by applying the same observations that we used for evaluation updates. We partition shift updates into two subtypes:
\begin{enumerate}
    \item \textbf{Right shifts:} A static suffix $S^i$ whose rank is increased by $1$ since $w^{new}_j = w^{new}_i$ for some dynamic suffix $S^j <_L S^i$.
    \item \textbf{Left shifts:} A static suffix $S^i$ whose rank is decreased by $1$ since $w^{old}_j = w^{old}_i$ for some dynamic suffix $S^j$ with $T^j<_L T^i$.
\end{enumerate}

We start by considering right shifts. For every static suffix $S^i$, we wish to count the dynamic suffixes $S^j$ with $w^{new}_j = w^{new}_i$ and $S^j <_L S^i$, and increase $R[i]$ by this number. Similarly to Section \ref{ss:eval}, we show how to increase the value of $R[i]$ for static suffixes $S^i$ by the amount of dynamic suffixes $S^j$ with $w^{new}_i = w^{new}_j$ such that $S^j < S^i$ and $j\in [x - k +1 \ldots x - \frac{k}{2}]$. The task of increasing $R[i]$ by the amount of dynamic suffixes $S^j$ with $w^{new}_i = w^{new}_j$ and $S^j <S^i$ for some $j\in [x - \frac{k}{2} + 1 \ldots x]$ can be handled in a similar way by replacing $w^c_l= S[x - \frac{k}{2} + 1 \ldots x]$ with $w^r_l= S[x \ldots x + \frac{k}{2} - 1]$.

We consider the following variation of Fact \ref{f:wcl}:
\begin{fact}\label{f:wcl2}
Let $S^i$ be a dynamic suffix with $i= D(\delta )$ for some $\delta \in [0 \ldots \frac{k}{2}]$. For an index $j\in [1 \ldots n]$, the following are equivalent:
\begin{enumerate}
    \item $S^j$ is close to $S^i$ (starts with an occurrence of $w^{new}_{i}$) and $S^j >_L S^i$.
    \item There is an occurrence $w_1=S[s_1\ldots e_1]$ of $w^c_l=S[s \ldots e]$ in index $s_1 = j+ \delta$ such that $S^{s_1} >_L S^s$, $lcs(s_1 -1, s-1) \ge \delta$ and $lcp(e_1+1,e+1) \ge \frac{k}{2} - \delta$. 
\end{enumerate}
\end{fact}

Similarly to Algorithm \ref{alg:1}, we can construct an algorithm for increasing the ranks of suffixes smaller than $S^i$ for every $i\in [1 \ldots n]$ by processing the occurrences of $w^c_l$ according to Fact \ref{f:wcl2}.

\begin{algorithm}\label{alg:3}
\fbox{\begin{minipage}{12cm}
{\sf For every} occurrence $w=S[s_1 \ldots e_1]$ of $w^c_l$:
\begin{enumerate}
    \item Check if $S^{s_1} >_L S^s$. If it is not - stop processing this occurrence.
    \item Compute $l =lcs(s_1 - 1,s-1)$ and $r = lcp(e_1 + 1, e+1)$.
    \item Increase the counters $R[i]$ with $i \in [max(s_1 - l,s_1 - \frac{k}{2}) , \min(e_1 + r - k+1,s_1)]$ by $1$.
\end{enumerate}
\end{minipage}}
\end{algorithm}

Algorithm \ref{alg:3} is very similar to Algorithm \ref{alg:1}. The main difference between these two algorithms is that Algorithm \ref{alg:3} applies updates directly to $R$, unlike Algorithm \ref{alg:1} that applies updates to a temporary array $Re$ and inserts the final values to the matching indexes in $R$. This change is required because shift updates may affect all the suffixes in $S$ while evaluation updates only affects a small set of suffixes. We prove the following claim:

\begin{claim}\label{c:alg3cor}
For every $i\in [1\ldots n]$, the value of $R[i]$ is increased in the execution of Algorithm \ref{alg:3} by the amount of dynamic suffixes $S^j$ with $w^{new}_i=w^{new}_j$ and $j\in [x - k +1 \ldots x - \frac{k}{2}]$ such that $S^j <_L S^i$.
\end{claim}
\textbf{Proof:} Let $i \in [1\ldots n]$ be an index in $S$. Algorithm \ref{alg:3} increases the value of $R[i]$ by $1$ every time it processes an occurrence $w_1=S[s_1 \ldots e_1]$ of $w^c_l$ with the following properties:
\begin{enumerate}
    \item \label{item:al31} $S^{s_1} >_L S^s$
    \item \label{item:al32} $max(s_1 - l, s_1 - \frac{k}{2}) \le i$
    \item \label{item:al33} $\min(e_1 +r - k + 1,s_1) \ge i$
\end{enumerate}

Recall that Fact \ref{f:wcl2} suggests that for every dynamic suffix $S^j$ with $j= D(\delta)$ for some $\delta \in [0 \ldots \frac{k}{2}]$ such that $w^{new}_i = w^{new}_j$ and $S^j <_L S^i$, we have a unique occurrence of $w^c_l$ starting in index $i + \delta$. We show that $i$ is increased by Algorithm \ref{alg:3} when iterating an occurrence $w_1$ iff $w_1$ is an occurrence corresponding to one of these dynamic suffixes.

From the second component in the $max$ expression in \ref{item:al32} and \ref{item:al33} we get $i \in [s_1 -\frac{k}{2}, s_1]$. Therefore, we have $s_1 = i + \delta$ for some $\delta \in [0 \ldots \frac{k}{2}]$. The first expression in \ref{item:al32} yields $l \ge s_1 - i$ and therefore $\delta \le lcs(s_1 - 1,s - 1)$. The first expression in \ref{item:al33} yields $r \ge i + k - e_1 - 1$. Since $e_1 = s_1 + \frac{k}{2}-1$, we have $lcp(e_1+1,e+1) = r \ge \frac{k}{2} - \delta$. We have shown that $i$ is increased by Algorithm \ref{alg:3} when processing $w_1$ iff an occurrence $w_1$ satisfies Proposition 2 of Fact \ref{f:wcl2} with respect to $i$. This concludes the proof.  \QED

Similarly to Algorithm \ref{alg:1}, Algorithm \ref{alg:3} can be easily implemented in time $O(|occ(w^c_l)|)$. Again, this is not sufficiently efficient. The solution is the same - we process the periodic occurrences representation of $w^c_l$ rather than the explicit occurrences. 

Let $C=(a,b,p)$ be a cluster of occurrences $s_t = a + t\cdot p$ of $w^c_l$. Let $w_t =S[s_t \ldots e_t]$ be the $t$'th occurrence in the cluster $C$. Let $l_t = lcs(s_t - 1,s - 1)$ and $r_t = lcp(e_t + 1, e)$. Let $I_>$ be the interval of $t$ values that satisfies $S^{s_t} >_L S^s$. Let $i_t = max(s_t - l_t, s_t - \frac{k}{2})$ and $j_t = \min(e_t +r_t - k + 1,s_t)$ be the ends of the interval update applied to $r$ when the occurrence $w_t$ is processed in Algorithm \ref{alg:3}. We make the following claim:

\begin{claim}\label{c:fastshifts}
Given $C$ and $w=S[s\ldots t]$, all the updates $u_t=(i_t,j_t,1)$ for $t\in I_>$ can be applied to $R$ in time $O(\log^3(n))$. 
\end{claim}

\textbf{Proof:} Similar to the proof of Claim \ref{c:fastEvaluasios}. We start by employing Lemma \ref{l:clusterlceprog} to obtain $l_t = \min(Ex_l, Ex^C_l + p\cdot t)$ and $r_t = \min(Ex_r, Ex^C_r - p\cdot t)$ for non-aligned occurrences. \QED 

With that, we can prove the following:
\begin{claim}
For non-aligned occurrences, $\{i_t\}$ and $\{j_t\}$ are $p$-normal sequences with degree $3\in O(1)$.
\end{claim}
\textbf{Proof:} By substituting Lemma \ref{l:clusterlceprog} and $s_t = s_0 + t\cdot p$, we get $i_t = max(s_0 + p\cdot t - \min(Ex_l, Ex^C_l + p\cdot t), s_0 + p\cdot t - \frac{k}{2})$. Since $s_0 - \frac{k}{2}$ is a constant, the second argument of this max expression is an arithmetic progression with difference $p$. The first argument can be rewritten as $max(s_0 + t \cdot p - Ex_l, s_0 - Ex^C_l)$, which is a $p$-normal sequence with degree $2$. With that, we proved that $\{i_t\}$ is a $p$-normal sequence with degree $3$. Similar arguments can be made to show that $\{j_t\}$ is a $p$-normal sequence with degree $3$ as well.\QED

It follows from the above claim that applying all the updates corresponding to non-aligned occurrences can be reduced to applying a constant number of stairs with stair width $p$ updates and interval increment updates (Lemma \ref{l:applypnormfast}). The (at most two) aligned occurrences can be explicitly handled by applying two interval increment updates in $O(\log(n))$ time. The time for applying a constant number of stairs and interval increment updates corresponding to the non-aligned occurrences is bounded by $O(\log^3(n))$ \QED

With this, an efficient implementation of Algorithm \ref{alg:3} can be constructed in an identical manner to the construction of Algorithm \ref{alg:2}.

\begin{algorithm}

\label{alg:4}
\fbox{\begin{minipage}{12cm}

{\sf Obtain} the periodic occurrences representation $por$ of $w^c_l$.
\\\\
{\sf For every} singular occurrence $w=S[s_1 \ldots e_1]$ in $por$:
\begin{enumerate}
    \item Check if $S^{s_1} >_L S^s$. If it is not, then stop processing this occurrence.
    \item Compute $l =lcs(s_1 - 1,s-1)$ and $r = lcp(e_1 + 1, e+1)$.
    \item Increase the counters of $R$ with indexes in  $[max(s_1 - l,s_1 - \frac{k}{2}) , \min(e_1 + r - k+1,s_1)]$ by $1$
\end{enumerate}

{\sf For every} cluster $C=(a,b,c)$ of occurrences in $por$:
\begin{enumerate}
    \item Obtain $I_>$ using Claim \ref{f:perfastlex}.
    \item Apply to $R$ all the interval updates corresponding to $w_t$ with $t\in I_>$ by employing Claim \ref{c:fastshifts}.
\end{enumerate}
\end{minipage}}
\end{algorithm}
It is easy to see that Algorithm \ref{alg:4} applies exactly the same updates to $R$ as Algorithm \ref{alg:3} does. The same arguments as in the proof of Claim \ref{c:alg2comp} can be made to prove that Algorithm \ref{alg:4} runs in time $\cOtilde(\sqrt{n})$. Recall that Algorithm \ref{alg:4} only counts dynamic suffixes $S^j$ with $j \in [x-k + 1\ldots x-\frac{k}{2}]$, and a symmetric procedure needs to be executed for the remaining dynamic suffixes. This concludes the evaluation of right shifts.

Finally, we need to handle \textit{left shifts}. For every static suffix $S^i$, we need to count the amount of dynamic suffixes $S^j$ for which $w^{old}_j=w^{old}_i$ and $T^j <_L T_i$, and subtract that number from $R[i]$.

This can be executed with an algorithm almost identical to Algorithm \ref{alg:4}, with the following two minor modifications:

\begin{enumerate}
    \item Instead of considering the occurrences of $w^c_l=S[x - \frac{k}{2} +1 \ldots x]$, consider occurrences of $u^c_l = T[x - \frac{k}{2} +1 \ldots x]$
    \item Replace every interval and stairs update with its \textbf{negative} variant.
\end{enumerate}

Since we consider the same indexes of $w^c_l$ in $T$ rather than in $S$, an occurrence of $u^c_l$ is extendable to occurrences of $w^{old}_i$ in the same manner as an occurrence of $w^c_l$ is extendable to occurrences of $w^{new}_i$. Similarly to right shifts, Algorithm \ref{alg:4} with the above modification would only count dynamic suffixes $S^j$ with $j \in [x-k + 1\ldots x-\frac{k}{2}]$, and a symmetric procedure needs to be executed for the remaining dynamic suffixes.

We conclude this section with the following.
\begin{lemma}\label{l:effshift}
Upon a substitution update on text $S$, all the shift updates can be applied to $R$ in time $\cOtilde(\sqrt{n})$.
\end{lemma}

\subsection{Overtake updates}\label{ss:overtakes}
In order to apply overtake updates to $R$, we need to increase the value of $R[i]$ by the amount of static suffixes $S^j$ for which $T^i <_L T^j$ and $S^i >_L S^j$ ($S^i$ overtakes $S^j$) and decrease $R[i]$ by the amount of static suffixes $S^j$ for which  $T^i >_L T^j$ and $S^i <_L S^j$ ($S^j$ overtakes $S^i$). These updates are required for every $i\in [1\ldots n]$ (excluding dynamic suffixes). For this purpose, we consider the word $w^{swap} = S[x - k \ldots x-1]$.

We introduce the notations $lcs_X(i,j)$ and $lcp_X(i,j)$ denoting the $LCS$ (resp. $LCP$) of the prefixes (resp. suffixes) of string $X$ ending (resp. starting) in indexes $i$ and $j$ (Until now, we used $lcs$ and $lcp$ to refer to $lcp_S$ and $lcs_S$). We present the following lemma:

(We rename the indexes of $w^{swap}$ to $w^{swap}=S[s\ldots e]$ for clearer presentation)
\begin{lemma}\label{l:wswap}
Let $S^i$ be a static suffix. A static suffix $S^j$ overtakes $S^i$, iff one of the following conditions is satisfied.

\textbf{Condition 1:}
All of the following sub-conditions are satisfied for $s_1 = j + s -i$:
\begin{enumerate}
    \item \label{it:sc1} $s - i \ge 0$
    \item \label{it:sc2} There is an occurrence $w_1 = S[s_1 \ldots e_1]$ of $w^{swap}$ 
    \item \label{it:sc4} $S^{s_1}$ is a static suffix
    \item \label{it:sc5} $S^{s_1}$ overtakes $S^s$
    \item \label{it:sc6} $m(s , s_1) \ge s - i$.
\end{enumerate}
\textbf{Condition 2:}
All of the following sub-conditions are satisfied for $s_1 = i + s -j$:
\begin{enumerate}
    \item $s-j \ge 0$
    \item There is an occurrence $w_1 = S[s_1 \ldots e_1]$ of $w^{swap}$ 
    \item $S^{s_1}$ is a static suffix
    \item $S^s$ overtakes $S^{s_1}$
    \item $m(s , s_1) \ge s - j$.
\end{enumerate}

In both conditions, $m(s,s_1) = \min(lcs_S(s-1 , s_1 - 1),lcs_T( s-1,s_1-1))$
\end{lemma}

\begin{figure} 
    \centering
      \includegraphics[width=\textwidth]{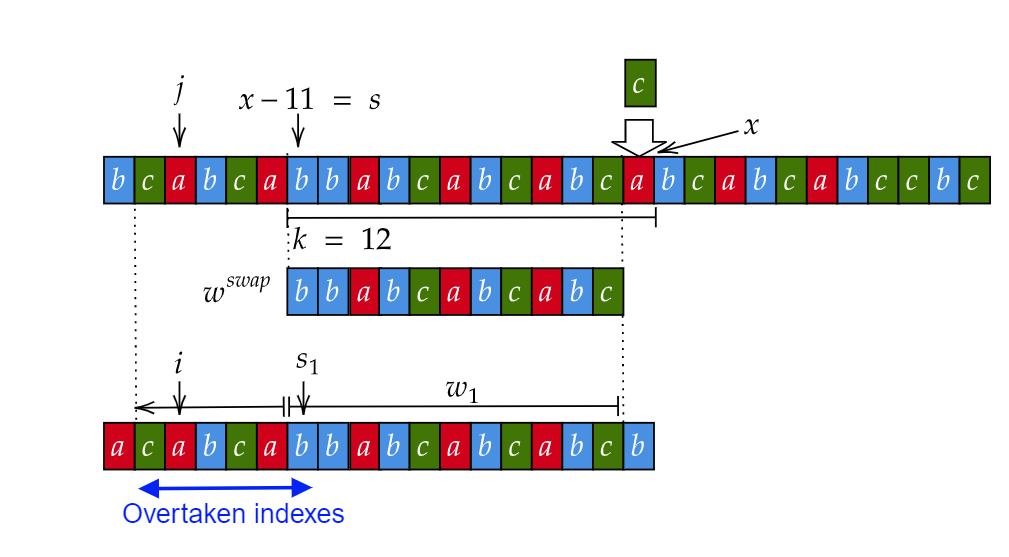}
    \caption{A visualization of Lemma \ref{l:wswap}. The upper segment is a segment of the text around the substitution index $x$. The lower segment is a segment of the text around an occurrence $w_1$ of $w^{swap}$. The arrow to the left of $w_1$ denotes the $lcs$ between $w_1$ and $w^{swap}$. The indexes $i$ and $j$ are selected as an arbitrary example for $i$ and $j$ in the statement of the lemma. Every two aligned indexes within the blue interval would be a valid demonstration of Lemma \ref{l:wswap}. } \label{fig:overtakeLemma}
\end{figure}

Lemma \ref{l:wswap} may seem technical and involved, but it aims to describe a fairly straightforward intuition. In a high level, Lemma \ref{l:wswap} states that the common extension to the right of two suffixes $S^i$ and $S^j$ such that one overtake the other must reach the substitution index $x$. The common extension must therefore contain an occurrence $w_1$ of $w^{swap}$, and there must be an overtake between $w^{swap}$ and $w_1$. For an example, see Figure \ref{fig:overtakeLemma}.

\textbf{Proof:} We start by proving the first direction. Namely, we show that if a static suffix $S^j$  overtakes $S^i$, we have an occurrence $w_1$ satisfying one of the conditions presented in the Lemma.

Let $r=lcp_S(i,j)$ and $r' = lcp_T(i,j)$. We start by proving the following claim:
\begin{claim}\label{c:swaoconx}
Either $x = i + \min(r',r)$ or $x = j +\min(r,r')$.
\end{claim}
\textbf{Proof:}
We assume that $r \le r'$, the case in which $r > r'$ can be treated symmetrically. From the definition of $r$ we have $S[i \ldots i + r - 1] = S[j \ldots j + r - 1]$. Since $S^j$ overtakes $S^i$, we have $S[i + r] < S[j + r]$. We consider two cases:
\begin{enumerate}
    \item $r = r'$. In this case, from the definition of $r'$ we have $T[i \ldots i + r - 1] = T[j \ldots j + r -1]$. And from the fact that $S^j$ overtakes $S^i$, $T[i+r] > T[j +r]$.
    \item $r < r'$. In this case, we  have $T[i \ldots i + r' - 1] = T[j \ldots j + r' -1]$. Since $r<r'$, we must have $T[i+r] = T[j+r]$.
\end{enumerate}
In both cases, the lexicographic order between $T[i+r]$ and $T[j+r]$ is different from the lexicographic order between $S[i+r]$ and $S[j +r]$. Therefore, either $T[i+r]$ or $T[j+r]$ is the symbol modified by the substitution update.
\QED

We partition our proof into two cases, depending on which of the options for the value of $x$ presented in Claim \ref{c:swaoconx} applies.
\begin{enumerate}
    \item $x = i + \min(r,r')$. In this case, we show that condition 1 is satisfied. Since $S^i$ is static and $i \le x$, we must have $i \le x - k = s$ (sub-condition \ref{it:sc1}). Therefore, $S[i \ldots i + \min(r,r') -1]$ and $T[i \ldots i + \min(r,r') - 1]$ both contain the substring $S[x - k + 1 \ldots x -1] = w^{swap}$.
    
    Let $d_s $ and $d_e $ be two integers such that $i + d_s = s$ and $i+ d_e = e$. Note that  $d_s = \min(r,r')-k$ and $d_e = \min(r,r') - 1$. Since $r$ is the longest common prefix of $S^i$ and $S^j$, we have $S[i \ldots i + \min(r,r') -1] = S[j \ldots j + \min(r,r') - 1]$. And in particular, $S[j + d_s \ldots j + d_e] = S[i + d_s \ldots i + d_e] = w^{swap}$. Since $\min(r,r') \le r'$, we have equality of the substrings with the same indexes in $T$. So we have an occurrence of $w^{swap}$ in index $j + d_s$ both in $S$ and in $T$. Denote $s_1 = j + d_s$ and $e_1 = j + d_e$ and $w_1 = S[s_1 \ldots e_1]$. Note that $s_1 = j + s - i$, $s_1$ is a starting index of an occurrence of $w^{swap}$ in $S$ (sub-condition \ref{it:sc2}), and since $s_1$ is a starting index of the same word $w^{swap}$ with length $k$ both in $S$ and in $T$, $S^{s_1}$ is a static suffix (sub-condition \ref{it:sc4}).
    
    From $S[i \ldots i + r - 1]= S[j \ldots j + r -1]$ we get $S[i + d_s \ldots i+r-1] = S[j +d_s \ldots j+r-1]$. Recall that $S[i + r] <_L S[j +r]$. Therefore, $S^s <_L S^{s_1}$. From $T[i \ldots i + r' - 1] = T[j \ldots j + r' -1]$ we get $T[i + d_s \ldots i + r' - 1] = T[j + d_s \ldots j + r' -1]$ and therefore $T[s \ldots i + r' -1] = T[s_1 \ldots j + r'-1]$. Since $S^j$ overtakes $S^i$, we have $T[i + r'] >_L T[j+r']$ and therefore $T^s >_L T^{s_1}$. We showed that $T^s >_L T^{s_1}$ and $S^s <_L S^{s_1}$. Therefore, $S^{s_1}$ overtakes $S^s$ (sub-condition \ref{it:sc5}). Since $S[i \ldots s] = S[j \ldots s_1]$ and $T[i \ldots s] = T[j \ldots s_1]$ we have $m(s,s_1) = \min(lcs_S(s-1 , s_1 - 1),lcs_T( s-1,s_1-1)) \ge s-i$.
    
    \item $x = j + \min(r,r')$. Symmetrical arguments can be made to show that in this case, condition 2 is satisfied.
\end{enumerate}

We proceed to prove the second direction of the lemma. Namely, if there is an occurrence $w_1 = S[s_1 \ldots e_1]$ of $w^{swap}=S[s\ldots e]$ satisfying one of the conditions in the statement of the Lemma, $S^j$ overtakes $S^i$. We provide a proof for the first condition. A proof for the second condition can be constructed in a similar manner.

Assume that condition 1 holds, so we have $s-i \ge 0$ , and an occurrence $w_1 = S[s_1 \ldots e_1]$ of $w^{swap}$ such that $s_1 = j + s -i$. We also know that $S^{s_1}$ is a static suffix overtaking $S^s$ and $m(s,s_1) \ge s - i $.

From the definition of $m(s,s_1)$ we have:
\begin{enumerate}
    \item $S[s_1 - m(s,s_1) \ldots s_1-1] = S[s -m(s,s_1) \ldots s - 1]$
    \item $T[s_1 - m(s,s_1) \ldots s_1-1] = T[s -m(s,s_1) \ldots s - 1]$
\end{enumerate}
Since $m(s,s_1) \ge s-i \ge 0$, the above particularly implies:
\begin{enumerate}
    \item $S[j \ldots s_1-1] = S[i \ldots s-1]$
    \item $T[j \ldots s_1 - 1]=T[i \ldots s-1]$
\end{enumerate}

Let $r = lcp_S(s,s_1)$ and $r' = lcp_T(s,s_1)$. From the definitions of $r$ and $r'$ we have $S[s \ldots s + r - 1] = S[s' \ldots s' + r - 1]$ and $T[s \ldots s +r' -1] = T[s_1 \ldots s_1 + r' -1]$. Putting it together with the above, we get:

\begin{enumerate}
    \item $S[j \ldots s_1 + r - 1] = S[i \ldots s + r -1]$
    \item $T[j \ldots s_1 + r'-1] = T[i \ldots s + r'-1]$
\end{enumerate}
Since $S^{s_1}$ is a static suffix overtaking $S^s$, we have $S[s_1 + r] > S[s + r]$ and $T[s_1 + r'] < T[s + r']$. Therefore, $S^j$ overtakes $S^i$. 
\QED

Given two static suffixes $S^i$ and $S^j$ such that $S^j$ overtakes $S^i$, we say that an occurrence of $w^{swap}$ satisfying one of the conditions of Lemma \ref{l:wswap} with respect to $S^i$ and $S^j$ is \textit{implying} the overtake between $S^j$ and $S^i$. The  following Lemma is required to ensure that we count every overtake exactly once.

\begin{lemma}\label{l:uniqueimpswap}
Given two static suffixes $S^i$ and $S^j$ such that $S^j$ overtakes $S^i$, there is exactly one occurrence $w_1$ of $w^{swap}$ implying the overtake between $S^j$ and $S^i$.
\end{lemma}
\textbf{Proof:} The existence of at least one implying occurrence is immediate from Lemma \ref{l:wswap}. Given $i$, $j$ and $s$, there are only two possible starting indexes for the starting index of $w_1$: $s_1 =j + s - i$ and $s_2=i + s - j$. Assume to the contrary that both $s_1$ and $s_2$ are starting indexes of an occurrence of $w^{swap}$ satisfying the first and the second conditions in Lemma \ref{l:wswap} respectively. 

Let $d = i -j$. Note that $s_1 = s -d$ and $s_2 = s+d$. Therefore, one of the indexes $s_1$, $s_2$ is to the right of $s$. We assume that $s_1 > s$, the case in which $s_2 > s$ can be treated in a similar way. Since $s_1$ is to the right of $s= x-k$, and is the starting index of a static suffix, it must be the case that $s_1 > x$.

Since both of the conditions of Lemma \ref{l:wswap} hold, we have $s-i \ge 0$ and $j-s\ge 0$. Therefore, both $i$ and $j$ are not to the right of $s$. Since $m(s,s_1)\ge s-i$ we have the following:
\begin{enumerate}
    \item \label{it:meq1} $S[j \ldots s_1-1] = S[i \ldots s-1]$
    \item \label{it:meq2} $T[j \ldots s_1 - 1] = T[i\ldots s-1]$
\end{enumerate}
Recall that $s_1 >x$ and $j < s < x$. Therefore, $x \in [j \ldots s_1 - 1]$. Since $x$ is the modified index, $T[x] \neq S[x]$. In particular, $T[j \ldots s_1 - 1]\neq  S[j \ldots s_1-1]$ and it can not be the case that both (\ref{it:meq1}) and (\ref{it:meq2}) hold. Therefore, we have reached a contradiction. \QED

Lemmas \ref{l:wswap} and \ref{l:uniqueimpswap} enable the construction of an algorithm with the structure of Algorithms \ref{alg:1} - \ref{alg:4}. It allows us to identify overtake updates using the occurrences of $w^{swap}$ in a similar manner to how we use Fact \ref{f:wcl2} to identify shift updates using occurrences of $w^c_l$. 

As before, we start by showing an inefficient algorithm for the sake of presenting the modifications that are required to be applied to $R$ in order to represent the overtake updates corresponding to an individual occurrence of $w^{swap}$. We present Algorithm \ref{alg:5}.

\begin{algorithm} \label{alg:5}
\fbox{\begin{minipage}{12cm}

{\sf For every} occurrence $w=S[s_1 \ldots e_1]$ of $w^{swap}$:
\begin{enumerate}
    \item {\sf If} $S^{s_1}$ is a dynamic suffix - stop processing $S^{s_1}$
    \item {\sf If} $S^{s_1} <_L S^s$ and $T^{s_1} >_L T^s$ ($S^s$ overtakes $S^{s_1}$)
    \begin{enumerate}
        \item Compute $l_S =lcs_S(s_1 - 1,s-1)$ and $l_T=lcs_T(s_1 - 1,s-1)$.
        \item \label{it:alg6st11} Increase the counters of $R$ with indexes in $[s -\min(l_S,l_T) , s]$ by $1$.
        \item \label{it:alg6st12} Decrease the counters of $R$ with indexes in $[s_1 - \min(l_S,l_T), s_1]$ by $1$
    \end{enumerate}
    \item {\sf If} $S^{s_1} >_L S^s$ and $T^{s_1} <_L T^s$ ($S^{s_1}$ overtakes $S^s$)
    \begin{enumerate}
        \item Compute $l_S =lcs_S(s_1 - 1,s-1)$ and $l_T=lcs_T(s_1 - 1,s-1)$.
        \item \label{it:alg6st21} Decrease the counters of $R$ with indexes in $[s -\min(l_S,l_T) , s]$ by $1$.
        \item \label{it:alg6st22} Increase the counters of $R$ with indexes in $[s_1 - \min(l_S,l_T), s_1]$ by $1$
    \end{enumerate}
\end{enumerate}
\end{minipage}}

\end{algorithm}

We start by proving that Algorithm \ref{alg:5} applies overtake updates correctly.
\begin{lemma}\label{l:alg5cor}
For every $i \in [1 \ldots n]$, let $\uparrow^i$  be the number of times that $R[i]$ is increased during the execution of Algorithm \ref{alg:5}. Let $\downarrow^i$ be the number of times that $R[i]$ is decreased during the execution of Algorithm \ref{alg:5}. $\uparrow^i$ equals to the number of suffixes $S^j$ that $S^i$ is overtaking. $\downarrow^i$ equals to the amount of suffixes $S^j$ that are overtaking $S^i$.
\end{lemma}
\textbf{Proof:} Let $j \in [1\ldots n]$ be an index (we use $j$ instead of $i$ in order to be consistent with the statement of Lemma \ref{l:wswap}). $R[j]$ is increased in step \ref{it:alg6st11} (second 'if') every time an iterated occurrence $w_1=S[s_1\ldots e_1]$ of $w^{swap}$ satisfies the following conditions:
\begin{enumerate}
    \item $S^{s_1}$ is static.
    \item $S^s$ overtakes $S^{s_1}$
    \item $j \le s$
    \item $j \ge s-\min(lcs_S(s_1-1,s-1),lcs_T(s_1-1,s-1))$
\end{enumerate}
Let $i = s_1 - s + j$. It can be easily verified that all the sub-conditions of Condition 2 of Lemma \ref{l:wswap} are satisfied for $s_1$, making $w_1$ a unique occurrence implying that $S^j$ overtakes $S^i$.
\\\\
An index $j$ is increased in step \ref{it:alg6st22} (third 'if') every time an iterated occurrence $w_1 =S[s_1\ldots e_1]$ satisfies the following conditions:
\begin{enumerate}
    \item $S^{s_1}$ is static.
    \item $S^{s_1}$ overtakes $S^s$.
    \item $j \le s_1$
    \item $j \ge s_1 -\min(lcs_S(s_1-1,s-1),lcs_T(s_1-1,s-1))$
\end{enumerate}

Let $i = s - s_1 + j$. It can be easily verified that all the sub-conditions of condition 1 of Lemma \ref{l:wswap} are satisfied for $s_1$, making $w_1$ a unique occurrence implying that $S^j$ overtakes $S^i$.

Lemma \ref{l:uniqueimpswap} guarantees that every suffix $S^i$ such that $S^j$ overtakes $S^i$ corresponds to a unique occurrence of $w^{swap}$ satisfying one of the above conditions. It directly follows that $\uparrow^j$ is exactly the amount of suffixes overtaken by $S^j$.

Symmetrical arguments can be made to show the $\downarrow^i$ is equal to the amount of suffixes that overtake $S^i$. \QED

As in evaluation and shift updates, a straightforward implementation of Algorithm \ref{alg:5} is not efficient. In order to implement Algorithm \ref{alg:5} efficiently, we need the following claims:
\begin{claim}\label{c:fastotint}
Let $C=(a,b,p)$ be a cluster of occurrences of a word $w=S[s\ldots e]$ with the $t$'th occurrence represented by $C$ denoted as $w_t = S[s_t \ldots e_t]$. There are consecutive intervals $I^{\uparrow}$ and $I^{\downarrow}$ such that $S^{s_t}$ overtakes $S^s$ iff $t \in I^{\uparrow}$ and $S^s$ overtakes $S^{s_t}$ iff $t \in I^{\downarrow}$. The intervals $I^{\uparrow}$ and $I^{\downarrow}$ can be found in $O(\log(n))$ time.
\end{claim}
\textbf{Proof:} We can apply Claim \ref{f:perfastlex} to find the intervals $I^S_>=[i_1 \ldots j_1]$ and $I^T_< = [i_2 \ldots j_2]$ such that $S^{s_t} >_L S^s$ iff $t \in I^S_>$ and $T^{s_t} <_L T^s$ iff $t \in I^T_<$. By definition, $I^{\uparrow} = I^S_> \cap I^T_< = [max(i_1,i_2) \ldots \min(j_1,j_2)]$. We can apply Claim \ref{f:perfastlex} to find the intervals $I^S_<=[i_3 \ldots j_3]$ and $I^T_>=[i_4 \ldots j_4]$ such that $S^{s_t} <_L S^s$ iff $t \in I^S_<$ and $T^{s_t} >_L T^s$ iff $t \in I^T_>$. By definition, $I^{\downarrow} = I^S_< \cap I^T_> = [max(i_3,i_4) \ldots \min(j_3,j_4)]$. \QED

We need to further 'filter' the occurrences represented by a cluster to exclude dynamic suffixes. We employ the following simple claim.
\begin{claim}\label{c:findintstatic}
Let  $I=[i\ldots j]$ be a subinterval of $[0 \ldots |C|-1]$ representing a subset of the indexes of the occurrences implied by a cluster $C$. $I$ can be processed in $O(1)$ time to evaluate at most two subintervals $I_1$ and $I_2$ such that $t\in I_1 \cup I_2$ iff $S^{s_t}$ is a static suffix and $t\in I$. 
\end{claim}
\textbf{Proof:} We need to find the maximal value $t_1$ in $I$ such that $s_{t_1} \le x-k$ and the minimal value $t_2$ such that $s_{t_2} \ge x+1$. Straightforwardly, $I_1 = [i \ldots t_1]$ and $I_2 = [t_2 \ldots j]$ satisfy the desired condition. Since $s_t$ is given as an arithmetic progression, $t_1$ and $t_2$ can be found in constant time.\QED

We adopt the notations of $I^{\uparrow}$ and $I^{\downarrow}$ from Claim \ref{c:fastotint} for the following claim. we also present the notation $I^1_{st} = [i^1_{st} \ldots j^1_{st}]$ and $I^2_{st} = [i^2_{st} \ldots j^2_{st}]$ denoting the intervals of indexes within $[0 \ldots |C|-1]$ corresponding to static suffixes.

Let $C=(a,b,p)$ be a cluster of occurrences of $w^{swap}=S[s \ldots e]$ with the $t$'th occurrence represented by $C$ denoted as $w_t = S[s_t \ldots e_t]$. We employ the notations $l^S_t = lcs_S(s-1, s_t - 1)$ and $l^T_t = lcp_T(e+1,e_t+1)$. For $t\in I^{\downarrow} \cap (I^1_{st} \cup I^2_{st})$, we define the following sequences of interval increment updates: 
\begin{itemize}
    \item $u^1_t =(i^1_t \ldots j^1_t,1) = (s - \min(l^S_t,l^P_t), s,1)$ the interval increment update applied to $R$ when $w_t$ is processed in step \ref{it:alg6st11} of Algorithm \ref{alg:5}.
    \item $u^2_t=(i^2_t,j^2_t,-1) = (s_t -\min(l^S_t,l^P_t),s_t,-1)$ the (negative) interval increment applied to $R$ when $w_t$ is processed in step \ref{it:alg6st12} of Algorithm \ref{alg:5} 
\end{itemize}

For $t\in I^{\uparrow} \cap (I^1_{st} \cup I^2_{st})$, we define the following sequences of interval increment updates:
\begin{itemize}
    \item $u^3_t=(i^3_t \ldots j^3_t,1)=(s-\min(l^S_t,l^P_t),s,1)$ the interval increment update applied to $R$ when $w_t$ is processed in step \ref{it:alg6st21} of Algorithm \ref{alg:5}.
        \item $u^4_t=(i^4_t \ldots j^4_t,-1)=(s_t-\min(l^S_t,l^P_t),s_t,-1)$ the (negative) interval increment update applied to $R$ when $w_t$ is processed in step \ref{it:alg6st22} of Algorithm \ref{alg:5}.
\end{itemize}
 
 We make the following claim.
\begin{claim}\label{c:fastovertakes}
Given $C$ and $w=S[s\ldots t]$, all the updates $u^y_t$ for $t\in I^{\downarrow} \cap (I^1_{st} \cup I^2_{st})$ and $y \in \{1,2\}$ can be applied to $R$ in time $O(\log^3(n))$. All the updates $u^y_t$ for $t\in I^{\uparrow} \cap (I^1_{st} \cup I^2_{st})$ and $y \in \{3,4\}$ can be applied to $R$ in time $O(\log^3(n))$ 
\end{claim}

\textbf{Proof:} As in the proof of Claim \ref{c:fastEvaluasios},  we start by applying Lemma \ref{l:clusterlceprog} to obtain $l_t = \min(Ex_l, Ex^C_l + p\cdot t)$ for non-aligned occurrences $w_t$. We proceed to prove the following claim:
\begin{claim}
$i^2_t$ is a $p$-normal sequence with degree $4$ for non-aligned occurrences.
\end{claim}
\textbf{Proof:} By Lemma \ref{l:clusterlceprog}, both $l^S_t$ and $l^T_t$ can be written as $l^S_t = \min(f_S,a_S + t\cdot p)$ and $l^T_t = \min(f_T, a_T + t\cdot p)$ for some constants $f_S$, $f_T$, $a_S$ and $a_T$ for non-aligned occurrences. Therefore, $i^2_t = s_0 + t\cdot p -  \min(\min(f_T, a_T + t\cdot p),\min(f_S,a_S + t\cdot p)) $ . It follows that $i^2_t = max(max(s_0 + t \cdot p - f_T,s_0 - a_T), max(s_0 + t\cdot p - f_S, s_0 - a_S))$. This is a $p$-normal sequence with degree $4$. \QED

It can be easily proven that $i^z_t$ and $j^z_t$, excluding aligned occurrences, are $p$-normal sequences with constant degrees for every $z\in [1\ldots 4]$ by applying the same reasoning.

We apply Lemma \ref{l:applypnormfast} in the same manner as in the proof of Claim \ref{c:fastEvaluasios} to obtain a constant number of interval and stairs updates that are equivalent to applying all the updates corresponding to non-aligned occurrences. Note that we need to treat  $I^{\downarrow} \cap I^1_{st}$ and $I^{\downarrow} \cap  I^2_{st}$ independently, as \ref{l:applypnormfast} deals with a consecutive $p$-normal sequence. Applying the constant set of stairs updates and interval increment updates can be executed in $O(\log^3(n))$ time. The (at most two) updates corresponding to aligned occurrences can be applied to $R$ in $O(\log(n))$ time by employing Claim \ref{c:intupdates}. \QED

With Claim \ref{c:fastotint}, Claim \ref{c:findintstatic} and Claim \ref{c:fastovertakes}, we are ready to present the efficient variant of Algorithm \ref{alg:5}.

\begin{algorithm}\label{alg:6}
\fbox{\begin{minipage}{12cm}

{\sf Obtain} the periodic occurrences representation $por$ of $w^{swap}$:
\\\\
{\sf For every} singular occurrence $w=S[s_1 \ldots e_1]$ in $por$: Process $w_1$ as in Algorithm \ref{alg:5}.

{\sf For every} cluster of occurrence $C=(a,b,p)$ in $por$:
\\\\
{\sf Obtain} $I^{\uparrow}_{st} =I^{\uparrow} \cap (I^1_{st} \cup I^2_{st})$ and $I^{\downarrow}_{st} =I^{\downarrow} \cap (I^1_{st} \cup I^2_{st}) $ using Claims \ref{c:fastotint} and \ref{c:findintstatic}.
\\\\
{\sf If} $I^{\uparrow}_{st}$ is not empty: Efficiently apply the interval updates corresponding to $w^t$ with $t\in I^{\uparrow}$ using Claim \ref{c:fastovertakes}.
\\\\
{\sf If} $I^{\downarrow}_{st}$ is not empty: Efficiently apply the interval updates corresponding to $w^t$ with $t\in I^{\downarrow}$ using Claim \ref{c:fastovertakes}.
\end{minipage}}
\end{algorithm}

It is straightforward that Algorithm \ref{alg:6} applies the same updates to $R$ as Algorithm \ref{alg:5} does - as it only processes occurrences corresponding to static suffixes that either overtake $S^s$ or are overtaken by $S^s$. We conclude this section with the following.

\begin{lemma}\label{l:fastovertakes}
Upon a substitution update in index $x$ of $S[1\ldots n]$. All the overtake update can be applied to $R$ in $\cOtilde(\sqrt{n})$ time.
\end{lemma}
\textbf{Proof:} Applying Algorithm \ref{alg:6} results in the same changes to $R$ as applying Algorithm \ref{alg:5}. This is equivalent to applying the overtake updates according to Lemma \ref{l:alg5cor}. Algorithm \ref{alg:6} processes every singular occurrence or cluster in the periodic occurrences representation of $w^{swap}$ in polylogarithmic time. Therefore, it runs in time $\cOtilde(\frac{n}{k}) = \cOtilde(\sqrt{n})$ \QED. 

Recall that for shift and overtake updates we manipulate the existing ranks in $R$ by applying interval increment and stairs updates to $R$, but for evaluation we set the updated ranks explicitly. Therefore, shifts and overtake modification needs to be counted and applied to $R$ \textbf{prior} to the evaluation of the ranks of dynamic suffixes. This is required to avoid the case in which a rank is set for a dynamic suffix, and then modified by a stairs or an interval update.  

With that, we have shown how to efficiently apply every type of updates to $R$. Upon a substitution update to $S$ in index $x$, we apply algorithms \ref{alg:2}, \ref{alg:4} and \ref{alg:6} to update $R[i]$. After doing so, $R[i]$ contains the updated close suffix rank of $S^i$, for every $i\in [1\ldots n]$.
\subsection{Handling Stairs Updates}

In this section, we show how to apply the restricted data structure provided by Theorem \ref{t:stairsp} to our settings. We recall that the data structure of Theorem \ref{t:stairsp} enables the application of a stairs update to a set of counters and an evaluation of a counter in polylogarithmic time, but only if every stairs update $(i,j,p)$ has the same stair width $p$.

Our dynamic algorithm require stairs update with various values of $p$, so the data structure of Theorem \ref{t:stairsp} is not sufficient. We start by showing that for every stairs update $(i,j,p)$ that is applied by our algorithm, $S[i\ldots j]$ is periodic. We then proceed to show how this property, alongside with the restricted stairs updates data structure, can be used to efficiently support stairs updates in our settings.

We start with the following observations concerning the structure of the interval updates applied in Algorithm \ref{alg:3} and in Algorithm \ref{alg:5}.

\begin{observation}\label{o:lcpwithinrunalg4}
Let $C=(a,b,p)$ be a cluster of occurrences of a word $w^c_l = S[s \ldots e]$ with period $p$ that is processed in Algorithm \ref{alg:3}. Let $R_C = S[i_C \ldots j_C]$ be the run with period $p$ containing the occurrences implied by $C$. For an occurrence $w_t = S[s_t \ldots e_t]$ implied by $C$, let $u_t = (i_t,j_t,1)$ be the interval increment update applied to $R$ when $w_t$ is processed by Algorithm \ref{alg:3}. Every occurrence $w_t$ implied by $C$ that is not an aligned occurrence has $[i_t \ldots j_t] \subseteq [i_C \ldots j_C]$.
\end{observation}
\textbf{Proof:} In algorithm \ref{alg:3}, the left border of the interval update applied to an occurrence $w_t$ of $w^c_l$ is $i_t = max(s_t - l_t,s_t - \frac{k}{2})$. The right border of the update is $j_t = \min(e_t + r_t - k+1,s_t)$. According to Lemma \ref{l:clusterlceprog}, for an occurrence that is not aligned we have $i_t = max(s_t - \min(Ex_l, Ex^C_l + p\cdot t) ,s_t - \frac{k}{2})$ which is at least $s_t - \min(Ex_l, Ex^C_l + p\cdot t)\ge s_t - Ex^C_l - p\cdot t = s_0 - Ex^C_l = i_C$.

According to Lemma \ref{l:clusterlceprog}, we have $j_t = \min(e_t + \min(Ex_r, Ex^C_r - p\cdot t) - k+1,s_t) \le s_t$. Since all the occurrences implied by $C$ are contained within $R_C$, we have $s_t \le j_C$ and therefore $j_t \le j_C$. We showed that $i_t \ge i_C$ and $j_t \le j_C$. Therefore, $[i_t \ldots j_t] \subseteq [i_C \ldots j_C]$. \QED
\\\\
We proceed with a similar observation concerning Algorithm \ref{alg:5}.

\begin{observation}\label{o:lcpwithinrunalg6}
Let $C=(a,b,p)$ be a cluster of occurrences of a word $w^{swap} = S[s \ldots e]$ with period $p$ that is processed in Algorithm \ref{alg:5}. Let $Run_C = S[i_C \ldots j_C]$ and $Run_w=S[i_w \ldots j_w]$ be the runs with period $p$ containing the occurrences implied by $C$ and the word $w^{swap}$, respectively. For an occurrence $w_t = S[s_t \ldots e_t]$ implied by $C$, let $u_1$, $u_2$, $u_3$ and $u_4$ be the interval increment update applied to $R$ when $w_t$ is processed by Algorithm \ref{alg:5} in step \ref{it:alg6st11}, step \ref{it:alg6st12}, step \ref{it:alg6st21} and step \ref{it:alg6st22}, respectively. Let $u^z_t = (i^z_t,j^z_t,b_z)$ for $z \in [1\ldots 4]$ and $b_z \in \{1,-1\}$. The following holds:
\begin{enumerate}
    \item for $z=1$ and $z=3$, every non-aligned occurrence $u_t$ has $[i^z_t,j^z_t] \subseteq [i_w \ldots j_w]$.
    \item for $z=2$ and $z=4$, every non-aligned occurrence $u_t$ has $[i^z_t,j^z_t] \subseteq [i_C \ldots j_C]$.
\end{enumerate}
\end{observation}
\textbf{Proof:}
For $z=2$ and $z=4$, the left border of the interval update applied to an occurrence $w_t$ of $w^{swap}$ in step \ref{it:alg6st11} in both parts is $i^2_t = i^4_t = s_t - \min(l^S_t,l^T_t) \ge s_t - l_t$ and the right border is $j^2_t=j^4_t = s_t$. We showed that $s_t - l_t \ge i_C$ and $s_t \le j_C$ for a non-aligned occurrence in the proof of Observation \ref{o:lcpwithinrunalg4}. So in this case, $[i^z_t \ldots j^z_t] \subseteq [i_C \ldots j_C]$.

For $z=1$ and $z=3$, the left border of the interval update applied to an occurrence $w_t$ of $w^{swap}$ is in step \ref{it:alg6st11} in both parts is $i^1_t = i^3_t = s - \min(l^S_t,l^T_t) \ge s - l_t$ and the right border $j^1_t = j^3_t = s$. Since the run $Run_w$ contains $w^{swap}$, we have $j_w \ge e > s$. According to Lemma \ref{l:clusterlceprog}, if $w_t$ is not an aligned occurrence we have $i_t \ge s - \min(Ex_l, Ex^C_l + p\cdot t) \ge s- Ex_l = i_w$. Therefore, we have $[i^z_t \ldots j^z_t] \subseteq [i_w \ldots j_w]$ \QED.

The following is directly derived from Observations \ref{o:lcpwithinrunalg4} and \ref{o:lcpwithinrunalg6}

\begin{lemma}\label{l:stairsinruns}
For every stairs update $(i,j,p)$ applied by Algorithm \ref{alg:4} or by Algorithm \ref{alg:6}, $S[i\ldots j]$ is contained within a run with period $p$. Either $Run_w$ or $Run_C$.
\end{lemma}
\textbf{Proof:} Recall that we only use stairs updates to apply interval increment updates corresponding to non-aligned occurrences. 

Let $C$ be a cluster of occurrences of a word $w$ with period $p$. Let $U$ be the set of intervals updates $U$ that are applied to $R$ when the occurrences implied by $C$ are processed either in algorithm \ref{alg:3}, or in one of the steps in Algorithm \ref{alg:5} in which an interval increment update is applied. In the proofs of Claims \ref{c:fastshifts} and \ref{c:fastovertakes}, we show that for non-aligned occurrences, $U$ is a $p$-normal sequence with a constant degree, and therefore can be reduced to a constant sized set $St$ of stairs updates with stair width $|p|=p$.

According to Observations \ref{o:lcpwithinrunalg4} and \ref{o:lcpwithinrunalg6}, either interval update $u=(i,j,1) \in U$ has $[i \ldots j] \subseteq [i_C \ldots j_C]$ or every $u\in U$ has $[i \ldots j] \subseteq [i_w \ldots j_w]$. Naturally, every stairs update $(i_1,j_1,p)\in St$ has $[i_1 \ldots j_1] \subseteq [i_C \ldots j_C]$ or $[i_1 \ldots j_1] \subseteq [i_w \ldots j_w]$. That is due to the fact that applying these stairs updates is equivalent to applying $U$ and therefore they only affects the indexes that are affected by the updates in $U$. 

To be more precise, it is theoretically possible that $St$ contains stairs updates that affect an index $i$ that was not modified by the updates in $U$, and the updates in $St$ cancel each other in index $i$. An examination of Lemma \ref{l:intToStairs} in Section \ref{s:updates} (in which the method for reducing a sequence of interval increment updates to a small set of stairs updates and interval updates is presented) easily shows that this is never the case. \QED 

Recall that in every stairs update $(i,j,p)$ applied by our algorithm, $p$ is the period of a word with length at most $k$. The following directly follows.

\begin{observation}
Every stairs update $(i,j,p)$ applied by our algorithm has $p \in [1\ldots \frac{k}{2}]$
\end{observation}

We proceed to show how to use the restricted data structure to maintain stairs updates with various stair widths in our settings.

\textbf{The idea:} We only apply interval increment updates directly to $R$. Additionally to $R$, we maintain a restricted data structure $St_p$ for applying and querying stairs updates with fixed stair width $p$ for $p \in [1 \ldots \frac{k}{2}]$. We also keep a data structure $ActInt$ for 2-dimensional range searching queries. When we wish to apply a stairs update $(i,j,p)$ to $R$, we apply it to $St_p$ instead. Additionally, we add a point $(i,j)$ with value $v(i,j) = p$ to $ActInt$. This point represent the interval $[i\ldots j]$ that was affected by the stairs update. Upon a query for the close suffixes rank of $S^i$, we extract from $ActInt$ the intervals touching the index $i$.  The values of the extracted points form the set $P$ of values of $p$ such that $St_p$ received an update affecting index $i$. We evaluate $\Sigma_P = \Sigma_{p \in P}St_p[i]$ and report $\Sigma_p + R[i]$. The periodicity of the intervals that are affected by $St_p$ ensures that $|P| \in O(\log(n))$, so the process of reporting the rank of $S^i$ remains polylogarithmic. However, further work needs to be dedicated to the maintenance of these intervals and their periodicity properties.

We present the notation of an extremely periodic string.
\begin{definition}
A string $S$ is \textbf{extremely periodic} if $per(S) \le \frac{|S|}{5}$
\end{definition}

We are interested in runs of $S$ that are also extremely periodic. The following property of extremely periodic runs of a string $S$ is proven in Section \ref{s:comsrq} 

\begin{restatable}{theorem}{hprcx}
\label{t:hprcx}
Given a string $S[1 \ldots n]$, let $x\in [1 \ldots n]$ be an index within $S$. There are $O(\log(n))$ extremely periodic runs containing $x$.
\end{restatable}

We start by slightly modifying algorithms \ref{alg:4} and \ref{alg:6}. Specifically, we replace some of the stairs updates applied in these algorithms with the set of interval update they represent. We only do that in a situation in which there are not many such interval updates, so applying them explicitly is not too costly. The details for this adjustment are as follows.

Let $w$ be a word we process in one of our algorithms for modifying the ranks ($w^c_l$, $w^{swap}$, etc). Let $Run_w = S[s_{rw} \ldots e_{rw}]$ the run with period $p$ containing $w$. If $Run_w$ is not extremely periodic, it must be the case that $\frac{|w|}{5} < per(w)$. If that is the case, there are at most $5\cdot \frac{n}{|w|} \in O(\frac{n}{k}) = O(\sqrt{n})$ occurrences of $w$ in $S$ (~\cite{amir_et_al:LIPIcs:2019:11126}, Fact 5). We can afford to process them explicitly and apply the interval increment updates corresponding to them rather than using stairs updates (as in algorithms \ref{alg:3} and \ref{alg:5}).

Otherwise, assume that we process the cluster $C$ of occurrences of $w$. Let $R_C = S[s_r \ldots e_r]$ be the run with period $p$ containing $C$. Assume that we need to apply a stairs update $(i,j,p)$ corresponding to a subset of the occurrences within $C$ such that $S[i \ldots j]$ is contained within the run $R_C$. We check if $R_C$ is extremely periodic. If it is not - there are at most $4$ occurrences of $w$ in $C$. We apply the interval increment updates derived from these occurrences to $R$ instead of processing them as stairs updates.

If $R_C$ is extremely periodic, we apply the stair update as follows. We add the point $p_{R_C} = (s_r , e_r)$ with value $v(p_{R_C}) = p$ to $ActInt$ (while avoiding duplicates), and apply the stairs update $(i,j,p)$ to $St_p$. If $S[i\ldots j]$ is contained within $RUN_w$, we add the point $p_{w} = (s_{rw} , e_{rw})$ with the same value $v(p_{w}) = p$ instead of $p_{R_c}$. 

\begin{observation}
With the above modifications, the running times of algorithms \ref{alg:4} and \ref{alg:6} remains $\cOtilde(\sqrt{n})$. For every $i\in [1\ldots n]$, the modifications applied to $R[i]$ by Algorithm \ref{alg:4} and Algorithm \ref{alg:6} prior to these modifications are equivalent to the updates applied to $R[i] + \Sigma_{p=1}^{\frac{k}{2}}St_p[i]$ with the modifications. 
\end{observation}

At this point, it may seem like we are done. We only insert extremely periodic runs to $ActInt$, so according to Theorem \ref{t:hprcx} every index $i\in [1 \ldots n]$ is only touched by $O(\log(n))$ runs in $ActInt$. Upon a query for the close suffix rank of $S^i$, we can find all the runs in $ActInt$ containing $i$, sum $\Sigma_p$ the values of $St_p[i]$ with the corresponding $p$ values, and return $\Sigma_p + R[i]$. However, there is another subtle point to handle. The intervals $S[s_r \ldots e_r]$ such that $(s_r,e_r)$ is inserted to $ActInt$ are extremely periodic runs \textbf{in the time of their insertion}. Future updates may change that by cutting an extremely periodic run or by extending it.

Therefore, we need to dedicate some extra work to maintain the following two invariants:

\begin{enumerate}
    \item Every interval (represented as a point) within $ActInt$ is an extremely periodic run
    \item If $x$ is not contained in interval (represented as point) $p_1 = [s_1 \ldots e_1]$ with value $p$, we have $St_p[x] = 0$.
\end{enumerate}

\begin{lemma}\label{l:preserveextreme}
After a substitution in index $x$, the data structures $ActInt$ and $\{St_p\}_{p\in [1\ldots \frac{k}{2}]}$ can be updated such that the above invariants are preserved, and the accumulated value of $R[i] + \Sigma_{p=1}^{\frac{k}{2}} St_p[i]$ is not affected for every $i\in [1\ldots n]$. The update can be executed in $\cOtilde(\sqrt{n})$ time.
\end{lemma}
\textbf{Proof:} Upon an update in index $x$, we assume that the invariants are satisfied prior to the application of the substitution update in $x$. we query $ActInt$ for all the intervals containing the index $x$ we get a set $P'$ of points $p_1 = (s_1 , e_1)$ such that $S[s_1 \ldots e_1]$ was an extremely periodic run with period $v(p_1)$ in $T$. Therefore, according to Theorem \ref{t:hprcx}, $|P'| \in O(\log(n))$. 

For every point $p_1 = (s_1 , e_1) \in P'$ with $v(p_1) = p$, we initialize a set $Exclude_{p_1} = \{x\} $ of candidates for exclusion. The update in $x$ cuts the run $T[s_1 \ldots e_1]$ into two maximal substrings with period $p$ - $R_l = S[s_1 \ldots x-1]$ and $R_r = S[x+1 \ldots e_1]$. We start by deleting $p_1$ from $ActInt$. If $|R_l| \ge 5\cdot p$ , we insert $p_l = [s_1 \ldots x-1]$ with $v(p_l) = p$ to $ActInt$. Otherwise, we insert the indexes $[s_1 \ldots x-1]$ to $Exclude_{p_1}$ (And do not add $p_l$ to $ActInt$). We treat $R_r$ the same. 

We proceed to treat the candidates for exclusion. For every index $i \in Exclude_{p_1}$, we check if there is an interval $p_2$ in $ActInt$ containing $i$ with $v(p_2) = p$. If there is a point with that property - we simply proceed to the next index. If there is not - we evaluate the current value $a = St_p[i]$, set $R[i] = R[i] + a$ and set $St_p[i] = 0$. By doing that, we ensure that the value of $St_p$ is no longer required for computing the rank of $S^i$ (second variant), while not changing the value of $R[i] + \Sigma_{p=1}^{\frac{k}{2}} St_p[i]$.

We also query $ActInt$ for intervals starting with $x+1$ or ending with $x-1$. The extremely periodic runs corresponding to these intervals may have been extended as a result of the update. For every point $p_1 = (s_1 ,x-1)$ with $v(p_1) = p$, we query $r = lcp(x-1, x-1-p)$. If $r > 0$ - the run $S[s_1 ,x-1]$ was extended. In this case, we remove $p_1$ from $ActInt$ and add $p_2 = (s_1, x-1 + r)$ instead (while avoiding duplicates). We do the symmetric procedure for every point $p_1 = (x + 1, e_1)$.

\textbf{Complexity:} The bottleneck of maintaining the runs within $ActInt$ is the exclusion procedure. Apart from the exclusion, we use a constant number of $lcp$, $lcs$, basic arithmetic operations and range queries to $ActInt$ for every point. Since there are $O(\log(n))$ points touching $x$, $x-1$ and $x+1$, this is polylogarithmic. We note that the data structure for maintaining and querying stairs updates can be easily modified to allow setting a certain index to $0$ without hurting its performance.

As for the exclusion, we only exclude an interval of indexes if its size is less than $5 \cdot p \le  \frac{5k}{2} \in O(k) = O(\sqrt{n})$. We exclude at most two such intervals for every contested point $p_1$, so $|Exclude_{p_1}| \in O(\sqrt{n})$. For every $i\in Exclude_{p_1}$, we query $ActInt$ at most twice and execute a constant number of queries on $St_p$ and $R$, all polylogarithmic operations. So the overall complexity of the exclusion procedure is $\cOtilde(\sqrt{n})$ per point, which is $\cOtilde(\sqrt{n})$ over all the contested points. \qedsymbol

With that, the proof of Theorem \ref{t:countClose} is completed. Applying the stairs updates in the proofs of Claims \ref{c:fastshifts} and \ref{c:fastovertakes} to $R$ is done implicitly by applying them to $St_p$ while maintaining the extremely periodic invariants after every update using Lemma \ref{l:preserveextreme}.

As for the query time, for the close suffixes rank of $S^i$ we query $ActInt$ for $Report([1 \ldots i] \times [i \ldots n])$ to obtain the set $P$ of the intervals containing $i$. The time complexity of this query is $O(\log^2(n) + |P|) = O(\log(n))$. We query every $St_p$ such that $per(p_1) = p$ for some point $p_1 \in P$ for $St_p[i]$ in $O(\log^3(n))$ time each and sum the results to obtain $\Sigma_p$. We query $R$ for $R[i]$ in $O(\log(n))$ time and return $R[i] + \Sigma_p$. The complexity is dominated by $O(\log^4(n))$.

\section{Complementary Proofs for Close Suffixes Rank queries} \label{s:comsrq}

\subsection{Proof of Lemma \ref{l:clusterlceprog}}
\clusterlceprog*

\textbf{Proof:} For $t\in [0 \ldots |C|-1]$, let $Ex^l_t$ and $Ex^r_t$ be the extension of the run with period $p$ containing $C$ to the left and to the right from $w_t$, respectively. Since the starting index of $w_t$ is $s_t = s_0 + t \cdot p$, we have $Ex^l_t = Ex^C_l + t \cdot p$ and $Ex^r_t = Ex^C_r - t \cdot p$. We proceed to prove the correctness of the formula for $lcs(s-1,s_t-1)$. The formula for $lcp(e+1 , e_t + 1)$ can be treated similarly.

The run with period $p$ extends at least $m= \min(Ex_l, Ex^l_t)$ symbols to the left of both $s$ and $s_t$. Let $\delta \in [1\ldots m]$ be an integer. Let $r_{\delta} \in [0\ldots p-1]$ and $q_{\delta}$ be the two unique integers such that $\delta = q_{\delta}\cdot p + r_{\delta}$. Due to the extensions of the run, we have $S[s - \delta] = S[s - r_{\delta} + p] = w[p - r_{\delta}]$. Since the extension of the period is of size at least $m$ to the left of $s_t$ as well, we have $S[s_t - \delta] = S[s_t  - r_{\delta} + p] = w_t[p - r_{\delta}] =w[p -r_{\delta}]$. From transitivity, we get $S[s -\delta] = S[s_t - \delta]$ for every $\delta \in [1 \ldots m]$ and therefore $lcs(s - 1, s_t - 1) \ge m$.
        
From our assumption that $Ex_l \neq Ex^C_l + t\cdot p$ we get $Ex_l \neq Ex^l_t$. If $Ex_l < Ex^l_t$, the extension of the run with period $p$ to the left of $s$ terminates in index $s - (m+1)$. Therefore, $S[s-(m+1)] \neq S[s-(m +1) + p]$. However, the extensions of the run with period $p$ extends at least $m+1$ indexes to the left of $s_t$. Therefore we have $S[s_t - (m+1)] = S[s_t - (m+1) + p]$. We know from $lcs(s-1,s_t-1) \ge m$ that $S[s_t +(m+1) - p] = S[s + (m+1) - p]$. From substitution and transitivity we get $S[s - (m+1)] \neq S[s_t - (m+1)] $ and therefore $lcp(s-1,s_t-1) = m$. Symmetric arguments can be made in the case in which $Ex_l > Ex^l_t$.

Evaluating $Ex_l$, $Ex_r$, $Ex^C_l$ and $Ex^C_r$ can be efficiently executed by querying $LCS(s-1,s+p-1)$, $LCP(e-1,e-p - 1)$, $LCS(s_0 - 1,s_0+p - 1)$ and $LCP(e_0 + 1,e_0 - p +1)$  respectively. \QED

\subsection{Proof of Claim \ref{f:perfastlex}}
\perfastlex*
\textbf{Proof:} We start by proving the following claim:
\begin{claim}\label{c:clustemonoton}
For every cluster $C=(a,b,p)$ of occurrences, let $Seq = S^{s_0},S^{s_1} \ldots S^{s_{|C|-1}}$ be the sequence of suffixes starting in the starting indexes of occurrences of $W$ implied by $C$. $Seq$ is either an increasing or a decreasing lexicographic sequence.
\end{claim}
\textbf{Proof:} Let $s_i = a + i\cdot p$ and $s_{i + 1} = s_i + p$ be two successive starting indexes of occurrences implied by $C$. Let $Ex_r$ be the extension of the run with period $p$ to the right of $s_0$. For every $j \in [s_0 + p \ldots s_0 + Ex_r]$ we have $S[j] = S[j - p]$. In particular, $S[s_{i+1}+x] = S[s_i + p + x] = S[s_i + x]$ for $x \in [0 \ldots Ex_r - (i+1)\cdot p]$. Since the run with period $p$ does not extend to the index $s_0 + Ex_r + 1$, we have $S[s_{i+1} +Ex_r - p\cdot (i+1) +1] = S[s_0 + Ex_r+1] \neq S[s_0 + Ex_r + 1 - p] =  S[s_i + Ex_r - (i+1)p + 1]$. Therefore, $lcp(s_i,s_{i+1}) = Ex_r - p\cdot (i+1)$ and the lexicographic order between $s_i$ and $s_{i+1}$ is only dependent on the lexicographic order between $S[s_0 + Ex_r + 1]$ and $S[s_0 + Ex_r + 1 - p]$. The two characters deciding the order between $s_i$ and $s_{i+1}$ are independent from the value of $i$. Therefore, it is either the case that $S^{s_i} <_L S^{s_{i+1}}$ for every $i \in [0 \ldots |C|-2]$ or $S^{s_i} >_L S^{s_{i+1}}$ for every $i \in [0 \ldots |C|-2]$. \QED

Claim \ref{c:clustemonoton} guarantees that the values $i\in [0 \ldots |C|-1]$ such that $S^s <_L S^{s_i}$ (resp. $S^s >_L S^{s_i}$) form a consecutive sub interval $I_<$ (resp. $I_>$). Specifically either a suffix or a prefix of $[0\ldots |C|-1]$. We proceed to show how to efficiently evaluate $I_>$. Similar technique can be used to evaluate $I_<$. 

Let $r_i = lcp(s_i,s)$. for $i\in [0 \ldots |C|-1]$. We start by treating the case in which $r_0<|w|$.

Since every $s_i$ is a starting index of an occurrence of $w$, we must have $r_i =r_0$ for every $i\in [0 \ldots |C|-1]$. Therefore, the symbols deciding the lexicographic order between $S^s$ and $S^{s_i}$ are $S[s+r_0]$ and $S[s_i + r_0]$. Since $r_0 < p$, $S[s_i + r_0] = w[r_0]$. It follows that the lexicographic order between $S^s$ and $S^{s_i}$ is independent of $i$, and therefore either $I_> = [1 \ldots |C|-1]$ or $I_> = \varnothing$. We can decide which of these two options apply by comparing $S[s + r_0]$ and $w[r_0]$.

We proceed to treat the case in which $r_0 \ge |w|$. If $l_0 \ge |w|$, we have that $s$ is a starting index of an occurrence of $w$. Therefore, Lemma \ref{l:clusterlceprog} can be applied to obtain $r_i = \min(Ex_r, Ex^C_r - p\cdot i) + |w|$ for $Ex_r \neq  Ex^C_r - p\cdot i$ (with the notations of $Ex_r$ and $Ex^C_r$ as in the statement of Lemma \ref{l:clusterlceprog}). Let $i' = \frac{Ex_r - Ex^C_r}{p}$. We treat three distinct sub intervals of $[1\ldots |C|-1]$. For every sub interval. we show that the lexicographical order between $S^s$ and $S^{s_i}$ is fixed within the sub interval.

\begin{claim}\label{c:inzidis}
The interval $[0 \ldots i')$ is either fully contained in $I_>$ or disjoint from $I_>$. 
\end{claim}
For $i\in [0 \ldots i')$ we have $Ex_r < Ex^C_r - p \cdot i$ and therefore $r_i = Ex_r +|w|$. Therefore, the lexicographical order between $S^s$ and $S^{s_i}$ is decided by the values of $S[s + |w|+ Ex_r]$ and $S[s_i + |w| + Ex_r]$. For $i < i'$, $s_i +|w| + Ex_r < s_0 + p\cdot i + |w| + Ex^C_r - p\cdot i = s_0 +|w|+ Ex^C_r$. Therefore, $S[s_i + |w| + Ex_r]$ is within the periodic run containing $w_0$. It follows that $S[s_i + |w| + Ex_r] = S[s_0 + p\cdot i + |w| + Ex_r]=S[s_0 + |w| + Ex_r]$. Both of the symbols that decide the lexicographic order between $S^s$ and $S^{s_i}$ are independent from $i$. Therefore, Either $[0 \ldots i') \subseteq I_>$ or $[0 \ldots i'] \cup I_> = \varnothing$.\QED

\begin{claim}\label{c:inzidis2}
The interval $(i' \ldots |C|-1]$ is either fully contained in $I_>$ or disjoint from $I_>$. 
\end{claim}
For $i\in (i' \ldots |C|-1]$ we have $Ex_r > Ex^C_r - p \cdot i$ and therefore $r_i = Ex^C_r - p \cdot i + |w|$. Therefore, the lexicographical order between $S^s$ and $S^{s_i}$ is decided by the values of $c^i_1 = S[s + |w|+ Ex^C_r - p\cdot i]$ and $c^i_2=S[s_i + |w| + Ex^C_r-p\cdot i]$. For $i > i'$, $s +|w| + Ex^C_r - p \cdot i < s + |w| + Ex^C_r$. Therefore, the index $s+ |w| + Ex^C_r -p\cdot i$ is contained within the run with period $p$ containing $w$. It follows that $c^i_1 = S[s+|w| + Ex^C_r - p\cdot i]$ is independent of $i$. Note that $c^i_2 = S[s_0 + p\cdot i + |w| + Ex^C_r - p \cdot i] = S[s_0 + |w| + Ex^C_r]$. Both of the symbols that decide the lexicographic order between $S^s$ and $S^{s_i}$ are independent from $i$. Therefore, Either $(i' \ldots |C|-1] \subseteq I_>$ or $(i' \ldots |C|-1] \cup I_> = \varnothing$. \QED

We use a single $LCP$ query to find $r_0$. If $r_0<|w|$, we can either report $I_< = [0 \ldots |C|-1]$ or $I_<= \varnothing$ according to the lexicographical order between $S^s$ and $S^{s_0}$.

If $r_0 \ge |w|$, we employ Lemma \ref{l:clusterlceprog} to find $Ex^C_r$ and $Ex_r$ in $O(\log(n))$ time. With these in hand, $i'$ can be evaluated in constant time. We lexicographically compare between $S^s$ and a single suffix from each of the intervals $[0 \ldots i')$, $(i' \ldots |C|-1]$. According to Claim \ref{c:inzidis} and Claim \ref{c:inzidis2}, the results of these comparisons can be used to decide if $[0 \ldots i')$ and $(i' \ldots |C|-1]$ form a part of $I_>$.

If $i'$ is an integer, the single suffix $S^{s_{i'}}$ needs to be compared to $S^s$ in order to decide if $i' \in I_>$.  We can evaluate $i'$ in constant time. We report $I_>$ to be the union of the intervals that were found to be contained in $I_>$ and possibly $i'$.

In the process of evaluating $I_>$ we executed a constant number of arithmetic operations and $lcp$ queries, as well as a single application of Lemma \ref{l:clusterlceprog}. The time complexity is bounded by $O(\log(n))$. \QED

\subsubsection{Proof of Theorem \ref{t:hprcx}}
\hprcx*
The following well known fact is helpful for proving Theorem \ref{t:hprcx}, and is derived directly from the periodicity lemma.
\begin{fact}\label{f:runoverlap}
Two runs $r_1$ and $r_2$ in a string $S$ can have an overlap with length at most $per(r_1) + per(r_2) - 1$
\end{fact}

To prove Theorem \ref{t:hprcx}, we prove the following lemma:
\begin{lemma}\label{l:runscount}
For every $k \in [1 \ldots \log_{\frac{3}{2}}(n)]$, there can be at most two distinct runs $R$ containing the index $x$ with $(\frac{3}{2})^{k-1} \le  Per(R) < (\frac{3}{2})^k$.
\end{lemma}
\textbf{Proof:} Assume to the contrary that there are three distinct runs $R_1 = S[s_1 \ldots e_1]$, $R_2 =[s_2 \ldots e_2]$ and $R_3 = S[s_3 \ldots e_3]$ containing the index $x$ with $per(R_i) = p_i$ and $(\frac{3}{2})^{k-1} \le  p_i < (\frac{3}{2})^k$ for $i\in \{1,2,3\}$. Assume w.l.o.g that $s_1 \le s_2 \le  s_3$. If one of the runs in our settings, $R_i$, fully contains one of the other runs, $R_j$. The overlap between them is $|R_j| \ge 5\cdot p_j > p_j + \frac{3}{2}\cdot p_j > p_j + p_i$, in contradiction to Fact \ref{f:runoverlap}. Therefore, no run is containing the other and it must be the case that $e_1 < e_2 < e_3$. It follows that $R_2$ is completely covered by $R_1$ and $R_3$ and therefore $|R_2| \le |R_1 \cap R_2| + |R_2 \cap R_3|$.  Due to Fact \ref{f:runoverlap}, $|R_2 \cap R_1| <p_2 + p_3$ and $|R_2 \cap R_3| < p_2 + p_3 $. Summing the two yields  $|R_1 \cap R_2| + |R_2 \cap R_3| < 2 \cdot p_2 + p_1 + p_3$ and from transitivity we get $|R_2| < 2 \cdot p_2 + p_1 + p_3$. Since $5 \cdot p_2 \le |R_2|$, we have $3 \cdot p_2 < p_1 + p_3$. Let $p_m$ be the maximal period among $p_1$ and $p_3$. We must have $3 \cdot p_2 < 2\cdot p_m$ which finally yields $\frac{3}{2} \cdot p_2 < p_m$, which is a contradiction to the assumption that $p_m,p_2 \in [(\frac{3}{2})^{k-1} \ldots (\frac{3}{2})^k - 1]$ \QED

With Lemma \ref{l:runscount}, the proof of Theorem \ref{t:hprcx} is straightforward. For every  $k \in [1 \ldots \log_{\frac{3}{2}}(n)]$, there are at most 2 runs with period $(\frac{3}{2})^{k-1} \le  Per(R) < (\frac{3}{2})^k$ containing $x$. \QED

\subsection{Proof of Lemma \ref{l:applypnormfast}}
\applynorfast*

\textbf{Proof:} We start by proving the following claim
\begin{claim}\label{c:partition1}
Let $U = u_0,u_1 \ldots u_{|U|-1}$ be a sequence of interval increment updates with $u_t = (i_t,j_t,1)$ and let $p$ be an integer. If $\{i_t\}$ and $\{j_t\}$ are both $p$-normal sequences with $deg(\{i_t\}) = d_i$ and $deg(\{j_t\}) = d_j$, Applying all the interval updates in $U$ to an array can be reduced to applying $O(2^{d_i + d_j})$ interval increment updates and stairs updates with stair width $|p|$. These stairs updates can be obtained in time $O(2^{d_i + d_j})$ given the representation of $\{i_t\}$ and $\{j_t\}$ as $p$-normal sequences.
\end{claim}

\textbf{Poof:} By strong induction on $d_i + d_j$. Our base case is when $d_i=d_j=1$. If it is the case, one of the following must occur:
\begin{enumerate}
\item \label{it:pnorm1} Both $i_t$ and $j_t$ are arithmetic progressions with difference $p$.
\item \label{it:pnorm2} One of $i_t$,$j_t$ is an arithmetic progression with difference $p$ and the other is fixed.
\item \label{it:pnorm3} Both $i_t$ and $j_t$ are fixed.
\end{enumerate}

Lemma \ref{l:intToStairs} in Section \ref{s:updates} suggests that applying all the interval updates $(i_t,j_t,1)$ for $t\in [0 \ldots |U|-1]$ can be reduced to applying a constant number of stairs updates. This concludes the induction base.

Assume $d_i + d_j = D > 2$. In particular, assume $d_i \ge 2$. The case in which $d_j \ge 2$ is treated symmetrically. We start with the following claim:

\begin{claim}
If $d_i \ge 2$, $CT(\{i_t\})$ contains a $p$-normal sequence $B'$ with $deg(B')=2$.
\end{claim}
\textbf{Proof:} It is easy to observe that a node corresponding to a $p$-normal sequence $B'$ has $deg(B')=2$ iff its construction tree is a root with two leaf children. Every finite full binary tree of size at least $3$ must contain a node with two leaf children \QED

Let $B'$ be a $p$-normal sequence in $CT(\{i_t\})$ with $deg(B')=2$. Assume that $B'$ is a maximum sequence $b'_t = max(a^1_t,a^2_t)$. The case in which $B'$ is a minimum sequence can be handled Similarly. $a^1_t$ and $a^2_t$ are both $p$-normal sequences with degree $1$. So one of the following must occur:

\begin{enumerate}
    \item \label{it:norrexstep1} Both $a^1_t$ and $a^2_t$ are fixed
    \item \label{it:norrexstep2} Both $a^1_t$ and $a^2_t$ are arithmetic progression with difference $p$.
    \item \label{it:norrexstep3} One of $a^1_t$, $a^2_t$ is an arithmetic progression with difference $p$, and the other is fixed.
\end{enumerate}

If (\ref{it:norrexstep1}) occurs, $B'$ can be replaced with the fixed value $max(a^1_0,a^2_0)$. If (\ref{it:norrexstep2}) occurs, $B'$ can be replaced with the arithmetic progression $a^y_0 + p\cdot t$ with the maximal value of $a^y_0$ for $y\in \{0,1 \}$.
In both (\ref{it:norrexstep1}) and (\ref{it:norrexstep2}), we replace every occurrence of $B'$ in $\{ i_t\}$ with a $p$-normal sequence with degree $1$. By doing so we obtain an equivalent sequence $\{i^*_t \}$ with degree $d^*_i < d_i$. From the induction hypothesis, the updates sequence $\{(i^*_t,j_t,1)\}$ can be reduced to $O(2^{d^*_i + d_j}) = O(2^{d_i + d_j})$ stairs updates. Since $\{i^*_t\}$ is equivalent to $\{i_t\}$, applying these stairs updates is equivalent to applying the updates of $U$.

If (\ref{it:norrexstep3}) occurs, assume w.l.o.g that $a^1_t$ is a fixed value sequence with value $a$ and $a^2_t = a^2_0 + p \cdot t$. Assume that $p \ge 0$ and let $t' = \lfloor \frac{a - a^2_0}{p} \rfloor$. It can be easily verified that $max(a^1_t,a^2_t) = a^1_t = a$ for $t\in [0 \ldots t']$ and that $max(a^1_t,a^2_t) = a^2_t = a^2_0 + p \cdot t$ for $t \in [t' + 1 \ldots |U|-1]$.

Therefore, in $t\in [0 \ldots t']$ we can replace every occurrence of $B'$ with a fixed sequence $a$. In $t\in [t' + 1 \ldots |U|-1]$ we can replace every occurrence of $B'$ with an arithmetic progression $a^2_0 + p$. In both cases, we replace an expression with degree $2$ with an expression with degree $1$ and therefore obtain a $p$-normal sequence with a strictly lower degree. By the recursion hypothesis, both the updates $u_0,u_1 \ldots u_{t'}$ and the updates $u_{t'+1} \ldots u_{|U|-1}$ can be represented by a set of $O(2^{d_i + d_j - 1})$ stairs updates. Applying both of these sets is equivalent to applying all the updates in $U$. The collective size of these sets is $O(2\cdot 2^{d_i + d_j -1 }) = O(2^{d_i + d_j})$.

An algorithm for finding the set of stairs updates in $O(2^{d_i + d_j})$ time (ignoring polynomial factors) can be directly derived from the proof \QED

With Claim \ref{c:partition1}, we are finally ready to prove Lemma \ref{l:applypnormfast}.

Let $d_i = Deg(\{i_t\})$ and $d_j = Deg(\{i_t\})$. According to Lemma \ref{c:partition1}, the updates of $U$ can be reduced to $O(2^{d_i + d_j})$ stairs or interval updates in $O(2^{d_i + d_j})$ time. Since $d_i,d_j \in O(1)$, we have $O(2^{d_i + d_j}) = O(1)$. \QED

\section{Extension Restricting Selection Queries} \label{s:ersq}

This section is dedicated to proving Theorem \ref{t:selectClose}.

\subsection{Occurrence Selection Queries}

We start by providing data structure for the following problem:

{\defproblema{\textsc{Occurrence Selection Query}}
{
\textbf{Input:} $POR_w$ the periodic occurrences representation of a word $w$.
\\
\textbf{Query:} upon input $i$. Return $A_w[i]$ }
}

The goal is to efficiently process $POR_w$ to enable an occurrence selection query in polylogarithmic time, without explicitly constructing $A_w$. Specifically, we wish to construct a data structure for OS queries in time $\cOtilde(|POR_w|)$.

In this section, we make an extensive use of the terms period rank, tail of a cluster, and head of a cluster as defined in Section \ref{s:dcssq}. We provide the following example to better familiarize the reader with these terms. Figure \ref{fig:clusterexample} is another example of these terms.
\begin{example}
Let $S = BAABBABBABBABBABBABAB$. The cluster $C = (4, 13 , 3)$ represents a periodic set of occurrence of $w = BBABBAB$ with period $3$. Consider the occurrence in $w_0 = S[4 \ldots 10]$. Its period rank is $r_p(w_0) = 3$, since there are 3 consecutive occurrences of $w[5 \ldots 7] = BAB$ following index $10$. The period rank of the second occurrence in the cluster, $w_1=S[7 \ldots 13]$, is $r_p(w_1) = 2$. One can easily observe that the $t$'th occurrence has $r_p(w_t) = |C| - t - 1$. The tail of $C$ is $Tail(C) = S[20 \ldots 21] = AB$ and the head is $Head(C) = S[1 \ldots 3] = BAA$.
\end{example}

The tail of a cluster can be equivalently described as the suffix following the rightmost consecutive occurrence of $w_s$ following $w_0$ (or equivalently, any other $w_t$). Since $w_{|C|-1}$ has the rightmost occurrence of $w_s$ as a suffix, the tail starts right after $w_{|C|-1}$.
Notice that we must have $Tail(C)[1 \ldots p] \neq w_s$. Otherwise, there would be another occurrence of $w$ exactly $p$ indexes to the right of $w_{|C|-1}$.


\begin{lemma} \label{l:decthaninc}
Let $C^i = (a^i,b^i,p), C^d = (a^d,b^d,p) \in POR_w$ be an increasing cluster and a decreasing cluster respectively (as defined in Section \ref{s:dcssq}). Let $w_1 = S[s_1 \ldots e_1]$ be an occurrence of $w$ implied by $C^i$ and $w_2 = [s_2 \ldots e_2]$ be an occurrence of $w$ implied by $C^d$. It must be the case that $S[s_1 \ldots n] >_L S[s_2 \ldots n]$. 
\end{lemma}
\textbf{Proof:} In what follows, we slightly abuse the notation of $<_L$. We use the notation $w_1<_L w_2$ between two equal words to denote that the suffixes starting in their corresponding starting indexes have $S[s_1 \ldots n] <_L S[s_2 \ldots n]$. 

\begin{figure} 
    \centering
      \includegraphics[width=\textwidth]{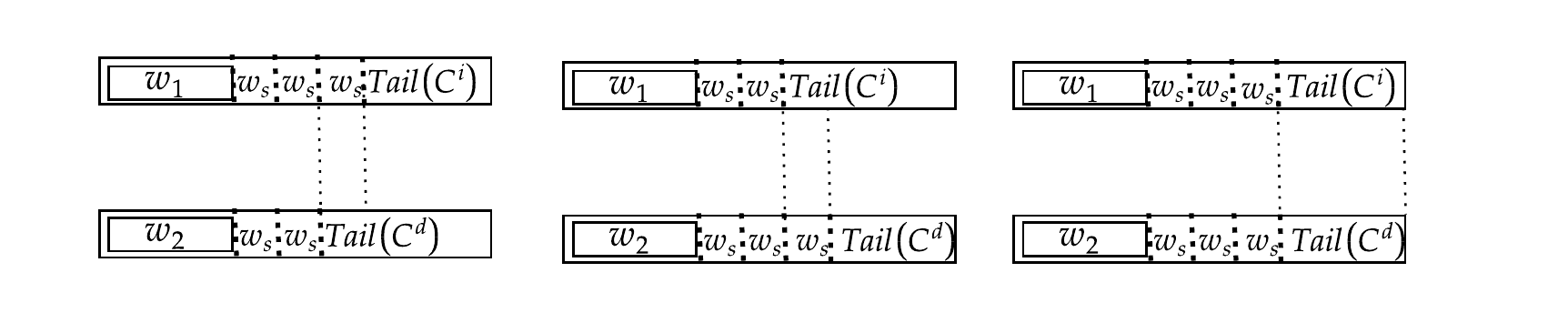}
    \caption{A visualization of the proof of Lemma \ref{l:decthaninc}. [Left] A demonstration of the first case. In the picture, $r_p(w_1) = 3$ and $r_p(w_2) = 2$. Following the first two instances of $w_s$, the comparison between $w_1$ and $w_2$ comes down to comparing $w_s$ and $Tail(C^d)$. [Center] A demonstration of the second case with $r_p(w_1) = 2$ and $r_p(w_2) = 3$. Symmetrically to the previous case, the comparison between $w_1$ and $w_2$ comes down to comparing $Tail(C^i)$ and $w_s$. [Right] A demonstration of the third case with $r_p(w_1) = r_p(w_2) = 3$. The comparison between $w_1$ and $w_2$ comes down to the comparison between the tails of the clusters. } \label{fig:diexampl1}
\end{figure}

We distinguish between three cases (See Figure \ref{fig:diexampl1} for a visualization): 
\begin{enumerate}
    \item $r_p(w_1) > r_p(w_2)$: In this case, the first $|w| + p \cdot r_p(w_2)$ symbols of both suffixes match. The $p$ symbols following $S[e_1 + p\cdot r_p(w_1)]$ are an occurrence of $w_s$. The $p$ symbols following $S[e_2 + p \cdot r_p(w_2)]$ are $Tail(C^d)$. Since $C^d$ is decreasing, $w_s > Tail(C^d)$ and we have $w_1 >_L w_2$. 
    \item $r_p(w_1) < r_p(w_2)$: In this case, the first $|w| + p \cdot r_p(w_1)$ symbols of both suffixes match. The $p$ symbols following $S[e_1 + p\cdot r_p(w_1)]$ are $Tail(C^i)$ and the $p$ symbols following $S[e_2 + p \cdot r_p(w_2)]$ are an occurrence of $w_s$. Since $C^i$ is increasing, $Tail(C^i) > w_s$ and we have $w_1 >_L w_2$.
    \item $r_p(w_1) = r_p(w_2)$ : In this case, the lexicographic order is determined by the lexicographic order between $Tail(C^i)$ and $Tail(C^d)$. We have $Tail(C^i) >_L w_s >_L Tail(C^d)$.
\end{enumerate}
\QED

Lemma \ref{l:decthaninc} suggests that $A_w$ can be expressed as the concatenation of two arrays $A= D[1\ldots |D|] I[1\ldots |I|]$ with $D$ the array of lexicographically sorted suffixes implied by decreasing clusters and $I$ the lexicographically sorted suffixes implied by increasing clusters. We can use this fact to reduce our task to either querying $D$ or $I$ for the element in a specified index. Upon a query for $A_w[i]$ with $i \le |D|$ , we query $D[i]$. For $i >|D|$ we query $I[i - |D| - 1]$. We can calculate the size of $D$ in $O(\frac{n}{k})$ time by accumulating the sizes of the decreasing clusters. From now on, we show how to construct a data structure for finding $I[i]$ for a given index $i$. A data structure for finding $D[i]$ can be constructed in a similar manner.

Using the same arguments as in the proof of Lemma \ref{l:decthaninc}, the following can be proven:

\begin{lemma} \label{l:decRanks}
$C_1 = (a_1,b_1,p),C_2=(a_2,b_2,p) \in Occ$ be two increasing clusters (not necessarily distinct). Let $w_1 = S[s_1 \ldots e_1]$ be an occurrence of $w$ implied by $C_1$ and $w_2 = [s_2 \ldots e_2]$ be an occurrence of $w$ implied by $C_2$.
\begin{enumerate}
    \item If $r_p(w_1) > r_p(w_2)$, $w_1 <_L w_2$.
    \item If $r_p(w_1) = r_p(w_2)$ and $Tail(C_1) <_L Tail(C_2)$, $w_1 <_L w_2$.
\end{enumerate}
\end{lemma}

\begin{figure} 
    \centering
      \includegraphics[width=\textwidth]{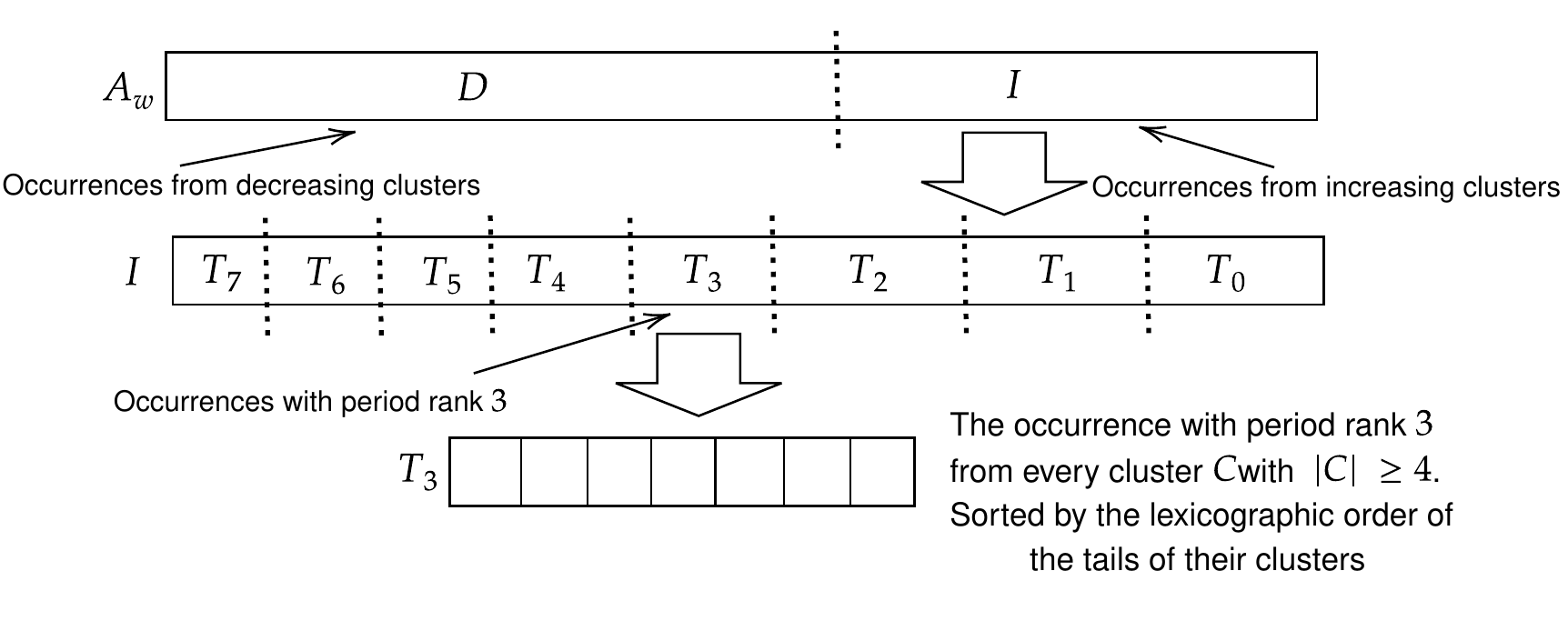}
    \caption{A demonstration of the structure of $A_w$. } \label{fig:awdemons}
\end{figure}

Lemma \ref{l:decRanks} suggest that sorting the suffixes of $I$ by lexicographic order is equivalent to sorting them primarily by decreasing order of their rank, and secondary by increasing lexicographic order of the tails of their clusters. Namely, $I$ can be written as the concatenation of $m$ arrays, with $m$ being the maximal rank of an occurrence within an increasing cluster. $I = T_{m},T_{m-1} \ldots T_1,T_0$. For every $r\in [0 \ldots m]$, $T_r$ is the set of occurrences with rank $r$ sorted by increasing lexicographic order of the tails of their clusters. The size of $T_r$ is the amount of clusters with size at least $r + 1$. Given an index $i$, we are interested in finding $r$ and $i'$ such that $I[i] \in T_r$ and in particular, $I[i] = T_r[i']$. See Figure \ref{fig:awdemons}

\begin{figure} 
    \centering
      \includegraphics[width=\textwidth]{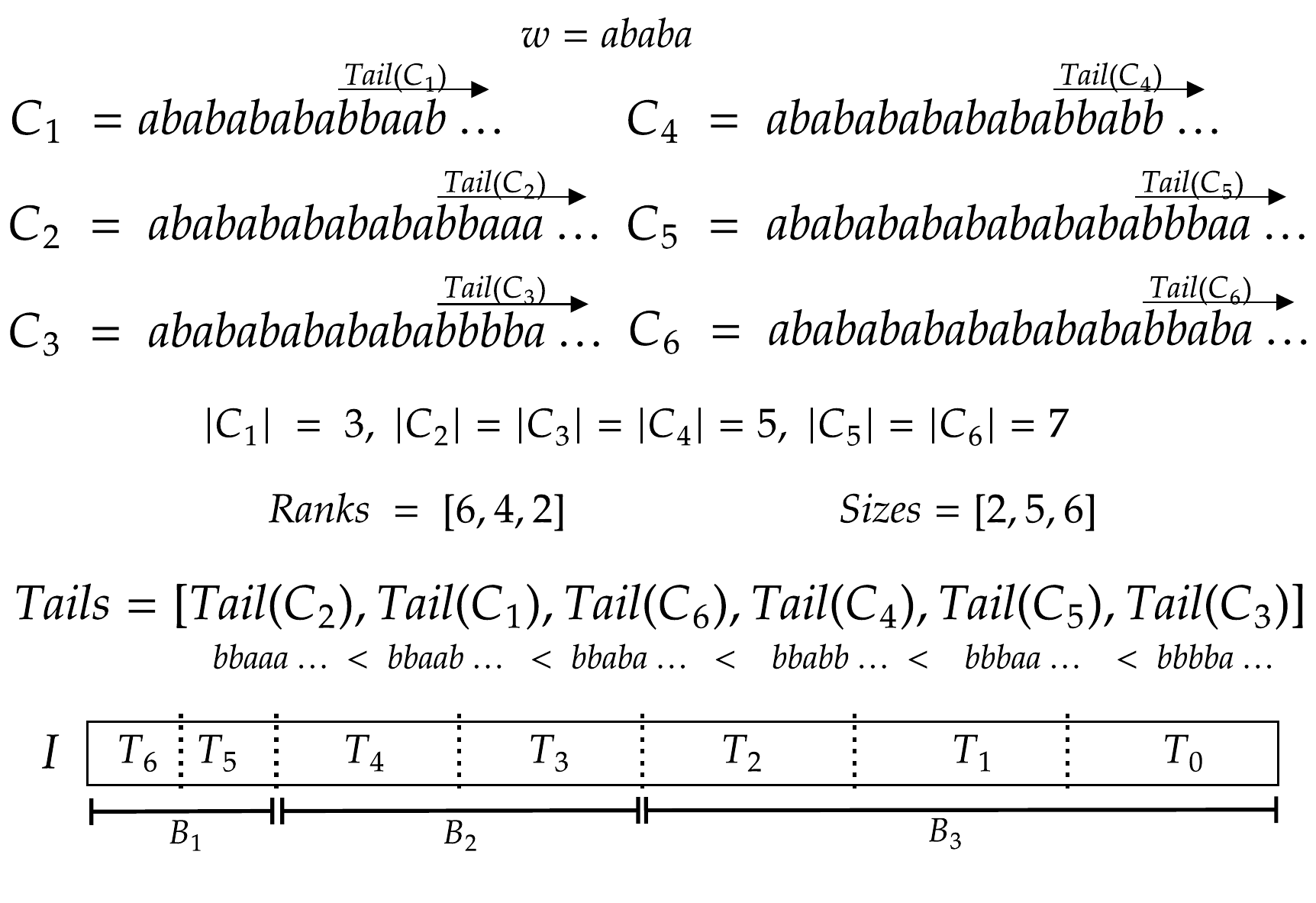}
    \caption{A demonstration of the arrays constructed from $POR_I$ and the structure of the buckets. Every cluster is presented as a suffix starting in the first occurrence represented by the cluster.}\label{fig:BucketsExmp}
\end{figure}

We process the set of increasing clusters in $POR_w$, denoted as $POR_I$, to obtain the following.
\begin{enumerate}
    \item $Ranks =r_1,r_2,r_3 \ldots r_d$ - the list of distinct values of the maximal ranks within an increasing cluster in $POR_I$ in decreasing order. Namely, if there is a cluster $C\in POR_I$ with size $x$, the number $x-1$ is in $Ranks$.
    \item $Sizes = s_1 , s_2, s_3 \ldots s_d$ - a list with $s_q$ being the amount of clusters with a maximal rank of at least $r_q$. Note that $Sizes$ is monotonically increasing. 
    \item $Tails = t_1,t_2,t_3 \ldots t_{|POR_I|}$ - a lexicographically sorted list of the tails of the clusters in $POR_I$. $Tail(C)$ is represented in $Tails$ as a pointer to $C$. 
\end{enumerate}
Note that $r_1 = m$ and $s_d = |POR_I|$. We define a set of $d$ subarrays (denoted as 'buckets') of $I$ such that $I = B_1,B_2 \ldots B_d$. The bucket $B_q$ for $q\in [1 \ldots d - 1]$ contains the concatenation of $T_{z}$ for $z \in [r_{q + 1} + 1 \ldots r_{q}]$ in decreasing order of $z$ (as they appear in $I$). $B_d$ is the concatenation of $T_z$ for $z \in [0\ldots r_d ]$ in decreasing order of $z$. See Figure \ref{fig:BucketsExmp} for a demonstration of $Ranks$, $Sizes$, $Tails$, and the buckets.

We make the following simple observation.
\begin{observation}
The size of every $T_z$ in the bucket $B_q$ is $s_q$.
\end{observation}

The following directly follows.
\begin{observation}
The size of $B_q$ is $s_q \cdot (r_{q} - r_{q+1})$ (with $r_{d+1} = -1$)
\end{observation}

Using the above, we calculate the sizes of the buckets, and use them to evaluate the starting index $i_q$ of every bucket $B_q$ within $I$. We insert the starting indexes to a data structure for predecessor queries $Bstarts$. For every cluster $C$, we denote the rank of $Tail(C)$ in $Tails$ as $tr(C)$. We initialize a data structure $Tlex$ for 2-d range counting queries. For every cluster in $POR_I$, we add the point $(tr(C), |C|)$ to $Tlex$. This concludes the construction of our data structure.

Upon query input $i$, we query $Bstarts$ for the predecessor of $i$ and obtain $i_z$. We deduce that $i$ is within the bucket $B_z$.
Let $\ell_z = r_z - r_{z+1}$ be the number of $T$ sub-arrays contained in $B_z$. For clearer presentation, we rename the $T$ sub-arrays contained in $B_z$ and denote $B_z = T^z_{0} \cdot T^z_{1} \ldots T^z_{\ell_z - 2} \cdot T^z_{\ell_z - 1}$. It can be easily verified that $T^z_q = T_{r_{z} - q}$ for every $q \in [0 \ldots \ell_z - 1]$. Let $T^z_{r'}$ be the sub array in the concatenation of $B_z$ containing $i$. Since every $T^z_q$ has the same size $s_z$, we can find the index $r'$ by evaluating $r' = \lfloor \frac{i - i_z}{s_z} \rfloor$. Via the aforementioned relation, we use $r'$ to obtain an index $r$ such that $T_r$ contains $i$. We also calculate the index $i' = i - i_z - s_z \cdot r'$ that satisfies $T_r[i'] = I[i]$.

Our next task is to find the $i'$th lexicographic suffix in $T_r$. Recall that $T_r$ consists of the suffixes with rank $r$ from the clusters of size at least $r + 1$ sorted by increasing lexicographic order of the tails of their corresponding clusters. 

\begin{figure} 
    \centering
      \includegraphics[width=\textwidth]{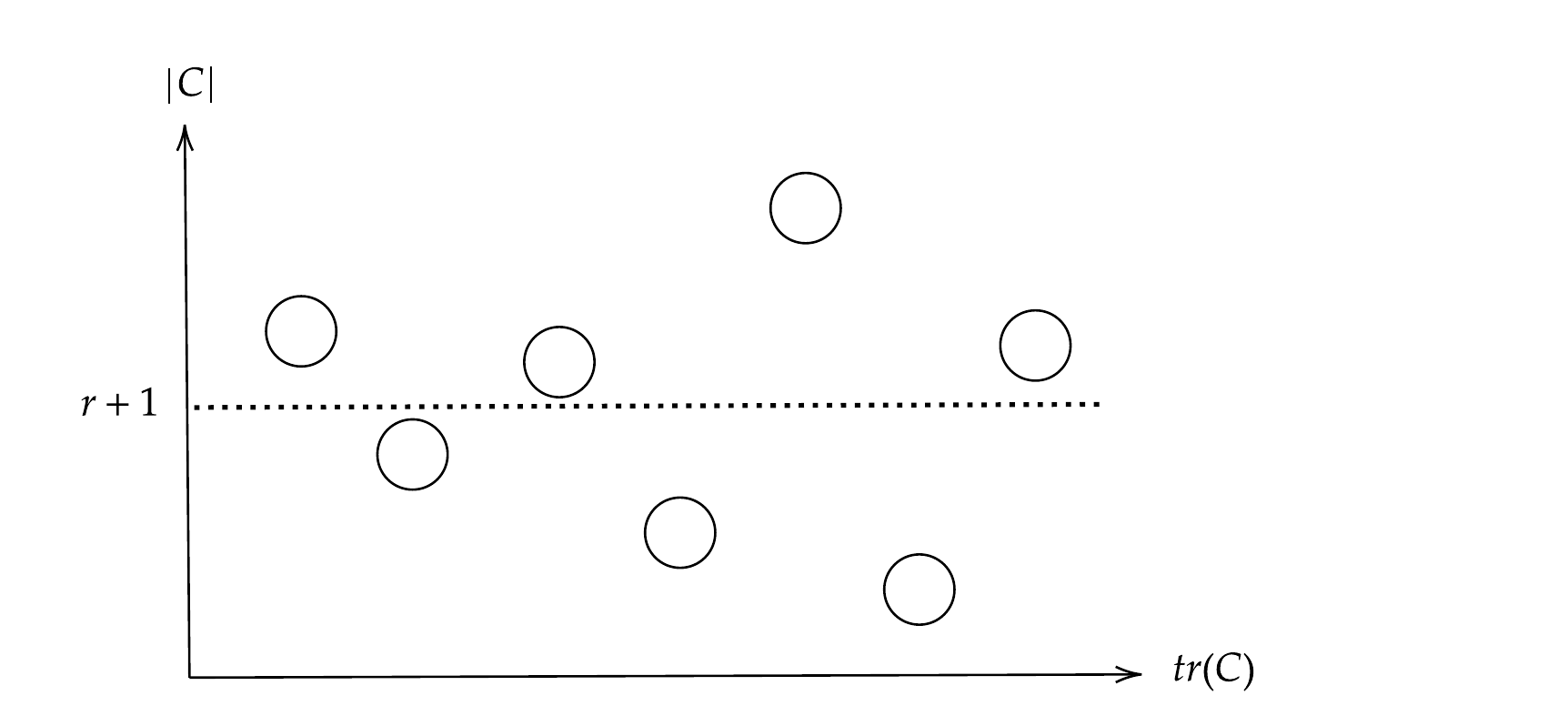}
    \caption{A demonstration of the binary search executed using $Tlex$. The clusters corresponding to circles above the $r+1$ horizontal line have a represented occurrence that participates in $T_r$. Since the occurrences appear in $T_r$ in lexicographic order of their clusters, we are interested in finding the $i'$ point from the left above this line.}\label{fig:tlex2d}
\end{figure}

In order to find $T_r[i']$, we perform a 'restricted' binary search on $Tails$ using $Tlex$ (See Figure \ref{fig:tlex2d} for a demonstration). We start by querying $Tlex$ for $L = Count([1 \ldots \frac{|POR_I|}{2}] \times [r + 1 \ldots \infinity{}])$. The condition on the second coordinate ensures that only clusters with size at least $r+1$ are counted. These are the clusters that have an implied occurrence in $T_r$. The second coordinate condition is fixes for all the queries of the binary search. If $L \ge i'$ we deduce that $T_t[i']$ is in the left half of $Tails$ and otherwise we deduce that it is in the right half. We shrink the searched interval accordingly and proceed recursively to finally obtain a tail $Tail(C)$ that is the $i'$th lexicographic tail among the tails of clusters participating in $T_r$. We deduce that $T_r[i']$ is from the cluster of $t_c$, denoted as $C = (a,b,p)$. Now we know the rank and the source cluster of $A[i]$, so we can finally report $I[i] = a + (|C| - r - 1) \cdot p$

\textbf{Preprocess complexity:} $Ranks$, $Sizes$ and the sizes and starting indexes of the buckets can be evaluated in $\cOtilde(|POR_I|)$ time by sorting the sizes of the clusters. Inserting the $O(|POR_I|)$ starting indexes of buckets to $BStarts$ takes $\cOtilde(|POR_I|)$ using, for example, a balanced search tree. $Tails$ can be obtained in $\cOtilde(|POR_I|)$ time by sorting the tails of the clusters with a standard $O(n\log n)$-comparisons sorting algorithm using $LCP$ queries to compare between tails. Insertions to $Tlex$ are also polylogarithmic. The space complexity is dominated by the space dedicated to the $2$ dimensional data structure $Tlex$. It is bounded by $O(|POR_I|\log(|POR_I|) = \cOtilde(|POR_I|)$. Recall that for queries on $A_w$, we need to do a symmetric preprocessing for the set of decreasing clusters too. The overall time and space complexity are both $\cOtilde(|POR_w|)$. 

\textbf{Query time:} In the process of finding  $r$ and $i'$, we use a predecessor query and constant number of array lookups and basic arithmetic operations to find $i_z$, $r'$ and $|B_z|$. The complexity for this part of the query is dominated by $(O(\log(|POR_w|)) \subseteq O(\log (n))$.

To find $T_r[i']$, we execute a restricted binary search on $Tails$. We use a range query on $Tlex$ in every iteration of the binary search. A 2-dimensional counting query is executed in $O(\log^2(|POR_w|))$, so the overall time complexity of this step is $O(\log^3(|POR_w|)) \subseteq O(\log ^3(n))$.

We conclude this subsection with the following:
\begin{theorem}\label{t:sortclusters}
We can process a periodic clusters of occurrences $POR_w$ of a word $w$ in $\cOtilde(|POR_w|)$ to support look-up queries for $A_w[i]$, in $O(\log^3(n))$ time. As a side effect, our query can also output the indexes $i'$ and $r$ such that $A_w[i] = T_r[i']$ as well as the cluster $C$ implying the occurrence starting in $A_w[i]$.
\end{theorem}

\subsection{Extensions Restricting Range Counting}

Let $POR_w$ be the periodic occurrences representation of a word $w= S[s \ldots e]$. In this subsection we introduce the final ingredient for Extensions Restricting Selection data structure:

{\defproblema{\textsc{Extension Restricting Range Counting}}
{
\textbf{Input:} $POR_w$
\\
\textbf{Query:} for input integers $l,r,i,j$, return the amount of $(l,r)$-extendable occurrences in $A_w[i \ldots j]$ }
}

Eventually, we will show that Extension Restricting Selection query can be reduced to $\log(n)$ Extension Restricting Range Counting queries via a binary search.

First, we recall that $A_w= D[1\ldots |D|] I[1\ldots |I|]$ with $D$ and $I$ being the sets of occurrences of $w$ represented by $POR_w$ implied by a decreasing and by an increasing cluster, respectively. $A_w[i\ldots j]$ is either a sub array of $I$, a sub array of $D$, or a concatenation of a suffix of $D$ and a prefix of $I$. Therefore, it is sufficient to count $(l,r)$-extendable occurrences in an interval within $D$ and an interval within $I$.

We provide a data structure for counting $(l,r)$-extendable occurrences in an interval within $I$. A similar algorithm can be constructed to count the $(l,r)$-extendable occurrences within an interval in $D$.

\subsubsection{Case 1: Periodic Extended Word}

Let $p$ be the period of $w=S[s \ldots e]$. In this subsection, we deal with the case in which $w' = S[s - l \ldots e + r]$ also has a period $p$. In these settings, we can use the following rule to decide if an occurrence of $w$ is $(l,r)$-extendable:
\begin{fact}\label{f:ruleextend}
Let $C \in POR_w$ be a periodic cluster of occurrences of $w$. Let $R_C=S[s_C \ldots e_C]$ and let $w_t = S[s_t\ldots e_t]$ be an occurrence represented by $C$. Let $Ex^t_l = s_1 - s_r$ be the length of the part of $R_C$ to the left of $w_t$ and $Ex^t_r = e_r - e_1$ be length of the part of $R_C$ to the right of $w_t$. $w_t$ is $(l,r)$-extendable iff $Ex^t_l \ge l$ and $Ex^t_r \ge r$. 
\end{fact}

\begin{figure}[htpb!]
    \centering
    \scalebox{0.9}{\tikzset{every picture/.style={line width=0.75pt}} 

\begin{tikzpicture}[x=0.75pt,y=0.75pt,yscale=-1,xscale=1]

\draw   (100,65) -- (535.5,65) -- (535.5,104) -- (100,104) -- cycle ;
\draw   (230.25,70.5) -- (405.25,70.5) -- (405.25,98.5) -- (230.25,98.5) -- cycle ;
\draw   (145.5,49.5) .. controls (148.68,53.34) and (151.72,57) .. (155.25,57) .. controls (158.78,57) and (161.82,53.34) .. (165,49.5) .. controls (168.18,45.66) and (171.22,42) .. (174.75,42) .. controls (178.28,42) and (181.32,45.66) .. (184.5,49.5) .. controls (187.68,53.34) and (190.72,57) .. (194.25,57) .. controls (197.78,57) and (200.82,53.34) .. (204,49.5) .. controls (207.18,45.66) and (210.22,42) .. (213.75,42) .. controls (217.28,42) and (220.32,45.66) .. (223.5,49.5) .. controls (226.68,53.34) and (229.72,57) .. (233.25,57) .. controls (236.78,57) and (239.82,53.34) .. (243,49.5) .. controls (246.18,45.66) and (249.22,42) .. (252.75,42) .. controls (256.28,42) and (259.32,45.66) .. (262.5,49.5) .. controls (265.68,53.34) and (268.72,57) .. (272.25,57) .. controls (275.78,57) and (278.82,53.34) .. (282,49.5) .. controls (285.18,45.66) and (288.22,42) .. (291.75,42) .. controls (295.28,42) and (298.32,45.66) .. (301.5,49.5) .. controls (304.68,53.34) and (307.72,57) .. (311.25,57) .. controls (314.78,57) and (317.82,53.34) .. (321,49.5) .. controls (324.18,45.66) and (327.22,42) .. (330.75,42) .. controls (334.28,42) and (337.32,45.66) .. (340.5,49.5) .. controls (343.68,53.34) and (346.72,57) .. (350.25,57) .. controls (353.78,57) and (356.82,53.34) .. (360,49.5) .. controls (363.18,45.66) and (366.22,42) .. (369.75,42) .. controls (373.28,42) and (376.32,45.66) .. (379.5,49.5) .. controls (382.68,53.34) and (385.72,57) .. (389.25,57) .. controls (392.78,57) and (395.82,53.34) .. (399,49.5) .. controls (402.18,45.66) and (405.22,42) .. (408.75,42) .. controls (412.28,42) and (415.32,45.66) .. (418.5,49.5) .. controls (421.68,53.34) and (424.72,57) .. (428.25,57) .. controls (431.78,57) and (434.82,53.34) .. (438,49.5) .. controls (441.18,45.66) and (444.22,42) .. (447.75,42) .. controls (451.28,42) and (454.32,45.66) .. (457.5,49.5) .. controls (460.68,53.34) and (463.72,57) .. (467.25,57) .. controls (467.33,57) and (467.42,57) .. (467.5,56.99) ;
\draw    (231,88) -- (190.5,88.95) ;
\draw [shift={(188.5,89)}, rotate = 358.65] [color={rgb, 255:red, 0; green, 0; blue, 0 }  ][line width=0.75]    (10.93,-3.29) .. controls (6.95,-1.4) and (3.31,-0.3) .. (0,0) .. controls (3.31,0.3) and (6.95,1.4) .. (10.93,3.29)   ;
\draw    (406,88) -- (442.5,88) ;
\draw [shift={(444.5,88)}, rotate = 180] [color={rgb, 255:red, 0; green, 0; blue, 0 }  ][line width=0.75]    (10.93,-3.29) .. controls (6.95,-1.4) and (3.31,-0.3) .. (0,0) .. controls (3.31,0.3) and (6.95,1.4) .. (10.93,3.29)   ;
\draw  [dash pattern={on 0.84pt off 2.51pt}]  (143.5,43) -- (143.5,104) ;
\draw  [dash pattern={on 0.84pt off 2.51pt}]  (466.5,42) -- (466.5,103) ;
\draw   (101,198) -- (536.5,198) -- (536.5,237) -- (101,237) -- cycle ;
\draw   (231.25,203.5) -- (406.25,203.5) -- (406.25,231.5) -- (231.25,231.5) -- cycle ;
\draw   (155.5,182) .. controls (158.68,186.1) and (161.72,190) .. (165.25,190) .. controls (168.78,190) and (171.82,186.1) .. (175,182) .. controls (178.18,177.9) and (181.22,174) .. (184.75,174) .. controls (188.28,174) and (191.32,177.9) .. (194.5,182) .. controls (197.68,186.1) and (200.72,190) .. (204.25,190) .. controls (207.78,190) and (210.82,186.1) .. (214,182) .. controls (217.18,177.9) and (220.22,174) .. (223.75,174) .. controls (227.28,174) and (230.32,177.9) .. (233.5,182) .. controls (236.68,186.1) and (239.72,190) .. (243.25,190) .. controls (246.78,190) and (249.82,186.1) .. (253,182) .. controls (256.18,177.9) and (259.22,174) .. (262.75,174) .. controls (266.28,174) and (269.32,177.9) .. (272.5,182) .. controls (275.68,186.1) and (278.72,190) .. (282.25,190) .. controls (285.78,190) and (288.82,186.1) .. (292,182) .. controls (295.18,177.9) and (298.22,174) .. (301.75,174) .. controls (305.28,174) and (308.32,177.9) .. (311.5,182) .. controls (314.68,186.1) and (317.72,190) .. (321.25,190) .. controls (324.78,190) and (327.82,186.1) .. (331,182) .. controls (334.18,177.9) and (337.22,174) .. (340.75,174) .. controls (344.28,174) and (347.32,177.9) .. (350.5,182) .. controls (353.68,186.1) and (356.72,190) .. (360.25,190) .. controls (363.78,190) and (366.82,186.1) .. (370,182) .. controls (373.18,177.9) and (376.22,174) .. (379.75,174) .. controls (383.28,174) and (386.32,177.9) .. (389.5,182) .. controls (392.68,186.1) and (395.72,190) .. (399.25,190) .. controls (402.78,190) and (405.82,186.1) .. (409,182) .. controls (412.18,177.9) and (415.22,174) .. (418.75,174) .. controls (422.28,174) and (425.32,177.9) .. (428.5,182) .. controls (431.68,186.1) and (434.72,190) .. (438.25,190) .. controls (441.78,190) and (444.82,186.1) .. (448,182) .. controls (451.18,177.9) and (454.22,174) .. (457.75,174) .. controls (461.28,174) and (464.32,177.9) .. (467.5,182) .. controls (470.68,186.1) and (473.72,190) .. (477.25,190) .. controls (480.78,190) and (483.82,186.1) .. (487,182) .. controls (489.16,179.21) and (491.27,176.51) .. (493.5,175.07) ;
\draw    (232,221) -- (191.5,221.95) ;
\draw [shift={(189.5,222)}, rotate = 358.65] [color={rgb, 255:red, 0; green, 0; blue, 0 }  ][line width=0.75]    (10.93,-3.29) .. controls (6.95,-1.4) and (3.31,-0.3) .. (0,0) .. controls (3.31,0.3) and (6.95,1.4) .. (10.93,3.29)   ;
\draw    (407,221) -- (443.5,221) ;
\draw [shift={(445.5,221)}, rotate = 180] [color={rgb, 255:red, 0; green, 0; blue, 0 }  ][line width=0.75]    (10.93,-3.29) .. controls (6.95,-1.4) and (3.31,-0.3) .. (0,0) .. controls (3.31,0.3) and (6.95,1.4) .. (10.93,3.29)   ;
\draw  [dash pattern={on 0.84pt off 2.51pt}]  (155.5,180) -- (155.5,237) ;
\draw  [dash pattern={on 0.84pt off 2.51pt}]  (493.5,175) -- (493.5,236) ;
\draw    (155.5,245) -- (229.5,245) ;
\draw [shift={(229.5,245)}, rotate = 180] [color={rgb, 255:red, 0; green, 0; blue, 0 }  ][line width=0.75]    (0,5.59) -- (0,-5.59)   ;
\draw [shift={(155.5,245)}, rotate = 180] [color={rgb, 255:red, 0; green, 0; blue, 0 }  ][line width=0.75]    (0,5.59) -- (0,-5.59)   ;
\draw    (406.5,246) -- (493.5,246) ;
\draw [shift={(493.5,246)}, rotate = 180] [color={rgb, 255:red, 0; green, 0; blue, 0 }  ][line width=0.75]    (0,5.59) -- (0,-5.59)   ;
\draw [shift={(406.5,246)}, rotate = 180] [color={rgb, 255:red, 0; green, 0; blue, 0 }  ][line width=0.75]    (0,5.59) -- (0,-5.59)   ;

\draw (307,73) node [anchor=north west][inner sep=0.75pt]   [align=left] {$\displaystyle w$};
\draw (208,69) node [anchor=north west][inner sep=0.75pt]   [align=left] {$\displaystyle l$};
\draw (417,69) node [anchor=north west][inner sep=0.75pt]   [align=left] {$\displaystyle r$};
\draw (308,206) node [anchor=north west][inner sep=0.75pt]   [align=left] {$\displaystyle w_{t}$};
\draw (147,154) node [anchor=north west][inner sep=0.75pt]   [align=left] {$\displaystyle s_{C}$};
\draw (485,151) node [anchor=north west][inner sep=0.75pt]   [align=left] {$\displaystyle e_{C}$};
\draw (209,202) node [anchor=north west][inner sep=0.75pt]   [align=left] {$\displaystyle l$};
\draw (418,202) node [anchor=north west][inner sep=0.75pt]   [align=left] {$\displaystyle r$};
\draw (248,15) node [anchor=north west][inner sep=0.75pt]   [align=left] {Run with period $\displaystyle p$};
\draw (291,151) node [anchor=north west][inner sep=0.75pt]   [align=left] {$\displaystyle R_{C}$};
\draw (180,248) node [anchor=north west][inner sep=0.75pt]   [align=left] {$\displaystyle Ex^{t}_{l}$};
\draw (433,247) node [anchor=north west][inner sep=0.75pt]   [align=left] {$\displaystyle Ex^{t}_{r}$};

\end{tikzpicture}}
    \caption{An illustration of Fact \ref{f:ruleextend}. $w$ has $l$ symbols to its left and $r$ symbols to its right contained within a run with period $p$. An occurrence $w_t$ must have this property as well in order to be $(l,r)$-extendable.} \label{fig:extcase1ex}
\end{figure}
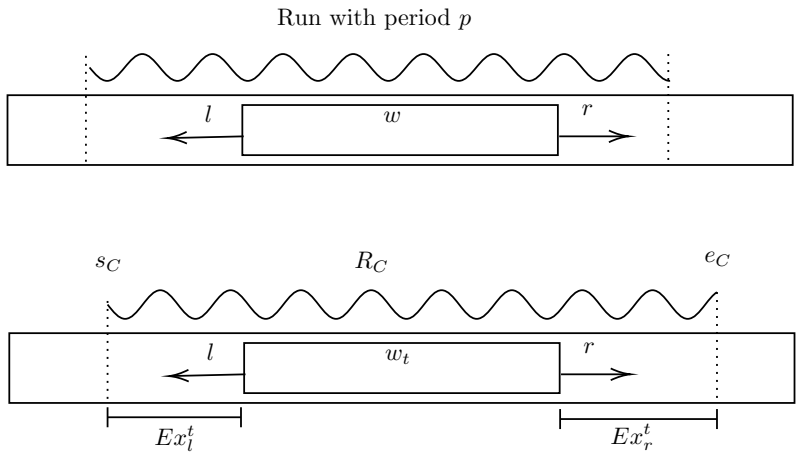

In words - $w_t$ is $(l,r)$-extendable iff $S[s_t - l \ldots e_t + r]$ is contained within $R_C$. See Figure \ref{fig:extcase1ex} for an illustration.

Let $q_R$, $q_L$, $r_R < p$ and $r_L < p$ be the unique non negative integers such that $r = q_R \cdot p + r_R$ and $l = q_L \cdot p + r_L$.
Consider a cluster $C = (a,b,p) \in POR_I$ with implied occurrences $w_t = S[s_t \ldots e_t]$ starting in indexes $s_t = a + t\cdot p$ for $t\in [0 \cdot |C|-1]$. We aim to provide a characterization of the $(r,l)$-extendable occurrences within $C$. For this purpose, we consider 3 types of occurrences represented by $C$:
\begin{enumerate}
    \item An occurrence $w_t$ is \textbf{central} if $r_p(w_t) \in [q_R + 1 \ldots |C| - q_L - 2]$
    \item An occurrence $w_t$ is \textbf{borderline} if $|C| \ge q_R + q_L +1$ and $r_p(w_t)\in \{ q_R, |C|-q_L - 1\}$
    \item An occurrence $w_t$ is \textbf{marginal} if it is neither central nor borderline.
\end{enumerate}

Let $R_C = S[s_C \ldots e_C]$. We present the notation $r_L(C) = s_0 - s_C$ and $r_R(C) = e_C - e_{|C|-1}$. $r_L(C)$ (resp. $r_R(C)$) is the remainder from the left (resp. right) of the run $R_C$ that is not contained within any of the occurrences in $C$.

\begin{example}
In the text $S=bbbababbababbababbababbababaaa$, there is a cluster $C$ of occurrences of $w=ababbababba$ with $R_C = S[2\ldots 27] = bbababbababbababbababbabab$. The prefix $bb$ of length $2$ of $R_c$ precedes the leftmost occurrence represented by $C$, therefore $r_L(C)=2$.The suffix $bab$ with length $3$ of $R_c$ follows the rightmost occurrence represented by $C$, therefore we have $r_R(C) = 3$
\end{example}

We make the following simple Claim.
\begin{claim}\label{c:extensiontorank}
An occurrence $w_t$ represented by a cluster $C$ has a part with length exactly $(|C| - 1 - r_p(w_t)) \cdot p + r_L(C)$ of $R_C$ to its left and a part with length $r_p(w_t) \cdot p + r_R(C)$ of $R_C$ to its right.
\end{claim}
\textbf{Proof:} By definition, $w_0$ has a part of length exactly $r_L(C)$ of $R_C$ to its left, which is consistent with our claim as $r_p(w_0) = |C|-1$. Every consecutive occurrence to the right has a period rank lower by $1$, and a part with length longer by $p$ of $R_C$ to its left (as it starts $p$ indexes to the right of the previous occurrence). This inductively proves the correctness of our claim for all the occurrences represented by $|C|$. The length of the extension to the right can be proven in a symmetric manner. \qedsymbol

The following holds due to Fact \ref{f:ruleextend}
\begin{lemma}\label{l:sufcontrib}
\begin{enumerate}
\item Every central occurrence is an $(l,r)$-extendable occurrence.
\item Every marginal occurrence is not an $(l,r)$-extendable occurrence.
\item A borderline occurrence with $r_p(w_t) = q_R < |C| - q_L - 1$ is an $(l,r)$-extendable occurrence iff $r_R(C) \ge r_R$
\item A borderline occurrence with $r_p(w_t) = |C|-q_L - 1 > q_r$ is an $(l,r)$-extendable occurrence iff $r_L(C) \ge r_L$
\item A borderline occurrence with $r_p(s_t) = |C|-q_L - 1 = q_r$ is an $(l,r)$-extendable occurrence iff $r_L(C) \ge r_L$ and $r_R(C) \ge r_R$.
\end{enumerate}
\end{lemma}

\textbf{Proof:} 
Applying Claim \ref{c:extensiontorank} and Fact \ref{f:ruleextend} on every case yields the corresponding result. \QED 

Note that if $|C| < q_L+q_R + 1$, all the occurrences are marginal occurrences. So $C$ does not contain any $(l,r)$-extendable occurrences.

We recall the partition $I=T_m,T_{m-1} \ldots T_1,T_0$ as in the previous section. Let $T_{r_i}$ be the sub array containing $i$ and $T_{r_j}$ be the sub array containing $j$. And let $i_{r_i}$ and $j_{r_j}$ be two indexes such that $T_{r_i}[i_{r_i}] = I[i]$ and $T_{r_j}[j_{r_j}] = I[j]$. Every sub array $T_r$ with $r\in [r_j + 1 \ldots r_i - 1]$ is fully contained within $I[i \ldots j]$. Therefore, every occurrence with rank $r\in [r_j + 1 \ldots r_i - 1]$ is in $I[ i \ldots j]$. 
 
We define $5$ subsets of the $(l,r)$-extendable occurrences that a cluster $C$ may contribute to $I[i\ldots j]$:
 
\begin{enumerate}
     \item $Cent(C)$ - Central occurrences $w_t$ with period rank $r_p(w_t) \in  [r_j + 1 \ldots r_i - 1]$.
     \item $Bor_R(C)$ - an $(l,r)$-extendable borderline occurrence $w_t$ with $r_p(w_t) = q_R \in [r_j + 1 \ldots r_i - 1]$
     \item $Bor_L(C)$ - an $(l,r)$-extendable borderline occurrence $w_t$ with $r_p(w_t) = |C| - q_L - 1 \in [r_j + 1 \ldots r_i - 1]$.
     \item $T_{r_i}(C)$ - an $(l,r)$-extendable occurrence $w_t$ with $r_p(w_t) = r_i$
     \item $T_{r_j}(C)$ - an $(l,r)$-extendable occurrence $w_t$ with $r_p(w_t) = r_j$
\end{enumerate}
 
Note that apart from $Cent(C)$, all the listed sets are either of size $0$ or of size $1$. Also note that the the union of these sets form the set of $(l,r)$-extendable occurrences implied by $C$ in $I[i\ldots j]$, so our problem is reduced to accumulating the sizes of these unions over all the clusters in $POR_I$.
 
We start by considering the size of $Cent(C)$. Notice that only clusters with $|C| \ge q_R + q_L + 3$ have central occurrences.

\begin{fact}\label{f:centralcontribution}
\begin{enumerate}
    \item If $|C| - q_L - 2 \ge r_i - 1$, then $|Cent(C)| = r_i - 1 - max(r_j , q_R )$.
    \item If $q_R + q_L + 3 \le |C|$ and $|C|- q_L -1 < r_i - 1$, then $|Cent(C)| = |C| - q_L - 2 - max(r_j , q_R )$
\end{enumerate}
\end{fact}
\textbf{Proof:} The Central occurrences of $C$ are those with ranks in $I_{cen} = [q_R + 1 \ldots |C| - q_L - 2]$. $Cents(C)$ is the intersections of the Central occurrences with the occurrences with rank in $I_{c} = [r_j + 1 \ldots r_i - 1]$ . This intersection is $[max(r_j +1, q_R + 1) \ldots \min(r_i - 1, |C| - q_L - 2)]$. The size of the interval yields the formula \QED.

Next, we consider the sizes of $Bor_R(C)$ and $Bor_L(C)$. Note that each of these sets can contain at most $1$ occurrence. Also note that only clusters with $|C| \ge q_R + q_L + 1$ have borderline occurrences. 

The following facts hold directly from Lemma \ref{l:sufcontrib} and the definitions of $Bor_R(C)$ and $Bor_L(C)$.
\begin{fact}\label{f:bordercontdist}
If $|C| \ge q_R + q_L + 2$:
\begin{enumerate}
    \item $Bor_R(C)$ and $Bor_L(C)$ are disjoint.
    \item We have $|Bor_R(C)| = 1$ iff $r_R(C) \ge r_R$ and $ q_R \in [r_j + 1 \ldots r_i - 1]$.
    \item We have $|Bor_L(C)| = 1$ iff $r_L(C) \ge r_L$ and $|C| - q_L - 2 \in [r_j + 1 \ldots r_i - 1]$.
\end{enumerate}
\end{fact}

\begin{fact}\label{f:bordercontsame}
If $|C| = q_R + q_L + 1$:
\begin{enumerate}
    \item $Bor_R(C)=Bor_L(C)$ 
    \item $|Bor_R(C)|=1$ iff $r_R(C) \ge r_R$, $r_L(C) \ge r_L$ and $ q_R \in [r_j + 1 \ldots r_i - 1]$.
\end{enumerate}

\end{fact}

Finally, we present a useful fact for evaluating the sizes of $T_{r_i}(C)$ and $T_{r_j}(C)$. These are also of size at most $1$. Recall that the occurrences in $T_{r_i}$ and $T_{r_j}$ are sorted by lexicographical order of the tails of their clusters. We denote as $C_i$ and $C_j$ be the clusters representing $I[i]$ and $I[j]$, respectively. The following Facts directly follow:

\begin{fact}\label{f:onetcont}
Let $T_z[i_z \ldots j_z]$ be a subarray of some $T_z$. Let $C_{i_z}$ and $C_{j_z}$ be the clusters representing $T_z[i_z]$ and $T_z[j_z]$, respectively. The cluster $C$ has an $(l,r)$ extendable occurrence in $T_z[i_z \ldots j_z]$ iff the occurrence $w_t$ implied by $C$ with $r_p(w_t) = z$ is $(l,r)$-extendable and $Tail(C_{i_z}) \le_L Tail(C) \le_L Tail(C_{j_z})$. 
\end{fact}

Equipped with facts \ref{f:centralcontribution} - \ref{f:onetcont}, we can express the sum of the contributions of all the clusters to $I[i\ldots j]$ as a set of $O(1)$ range queries. 
We recall the notations of $tr(C)$ - the lexicographic rank of $Tail(C)$ among the clusters in $POR_I$. We initialize a $4$-dimensional range queries data structures $Tlex$. For every cluster $C$, we insert the point $p_C = (|C|, r_L(C), r_R(C),tr(C))$ with $v(p_C) = |C|$ to $Tlex$. 

\begin{lemma}\label{l:sumcentc}
The sum $\Sigma_{Cent} = \Sigma_{C \in POR_I}|Cent(C)|$ can be evaluated in $O(\log^4(n))$ time given $Tlex$.
\end{lemma}
\textbf{Proof:}
We make the following range queries on $Tlex$:
\begin{enumerate}
    \item $sum_{mid} = SUM([q_L + q_R + 3 \ldots r_i + q_L] \times \ast \times \ast \times \ast)$
    \item $count_{mid} = COUNT([q_L + q_R + 3 \ldots r_i + q_L] \times \ast \times \ast \times \ast)$ 
    \item $count_{large} = COUNT([max(r_i + q_L + 1,q_R + q_L +3) \ldots \infinity{}] \times \ast \times \ast \times \ast)$
\end{enumerate}
The "$\ast$" symbol denotes the complete range of the corresponding coordinate and is equivalent to $[-\infinity \ldots \infinity]$. According to Fact \ref{f:centralcontribution}, Every cluster $C$ such that $p_C$ is in the interval counted in $count_{large}$ has $|Cent(C)| = r_i - 1 - max(r_j , q_R )$. So the overall contribution of these clusters to $\Sigma_{Cent}$ is $cont_{large} = count_{large} \cdot( r_i - 1 - max(r_j , q_R ))$

Every cluster $C$ such that the value of $p_C$ is accumulated in $sum_{mid}$ has $|Cent(C)| = |C| - q_L - 2 - max(r_j , q_R )$. The overall contribution of these clusters to $\Sigma_{Cent}$ is $cont_{mid} = sum_{mid} - count_{mid} \cdot (q_L + 2 + max(r_j , q_R ))$. We have $\Sigma_{Cent} = cont_{mid} + cont_{large}$ since the clusters whose contributions are considered in $const_{mid}$ are disjoint from the clusters considered in $const_{large}$, and every cluster that was not considered in either of the two is too small to have central occurrences. We used a constant number of $4$-dimensional range queries so the time complexity is $O(\log^4(n))$. \QED 

\begin{lemma}\label{l:sumbordc}
$\Sigma_{Bord} = \Sigma_{C \in POR_I}|Bor_L(C) \cup Bor_R(C)|$ can be evaluated in $O(\log^4(n))$ time given $Tlex$.
\end{lemma}
\textbf{Proof:} We make the following range queries on $Tlex$:
\begin{enumerate}
    \item $count_L = COUNT([max(q_R + q_L + 2,r_j + q_L + 2) \ldots r_i + q_L] \times [r_L \ldots \infinity{}] \times \ast \times \ast ])$
    \item $count_R =  COUNT([q_R + q_L + 2 \ldots \infinity{}] \times \ast \times [r_R \ldots \infinity{}] \times \ast ])$. If $q_R \notin [r_j + 1 \ldots r_i - 1]$, we set $count_R =0$
    \item $count_M = COUNT([q_R + q_L + 1 \ldots q_R + q_L + 1] \times [r_L \ldots \infinity{}] \times [r_R \ldots \infinity{}] \times \ast ])$. If $q_R \notin [r_j + 1 \ldots r_i - 1]$, we set $count_M =0$
\end{enumerate}

Note that according to Fact \ref{f:bordercontdist}, the point $p_C$ is counted in $count_L$ iff $|C| > q_R + q_L + 2$ and $|Bor_L(C)| = 1$, and the point $p_C$ is counted in $count_R$ iff $|C| > q_R + q_L + 2$ and $|Bor_R(C)| = 1$.  Fact \ref{f:bordercontdist} suggests that $Bor_L(C)$ and $Bor_L(C)$ are distinct when $|C| \ge q_L + q_R + 2$. So the overall contribution of clusters with size at least $q_R + q_L + 2$ to $\Sigma_{Bord}$ is exactly $count_L + count_R$. 

According to Fact \ref{f:bordercontsame}, a point $p_C$ is counted in $count_M$ iff $C$ has a single Borderline $(r,l)$-extendable occurrence $w_t$ with $r_p(w_t) = q_r$.  It follows that the contribution of clusters with size $q_L + q_R + 1$ to $\Sigma_{Bord}$ is $count_M$. To summarize, $\Sigma_{Bord} = count_L + count_R + count_M$. We use a constant number of $4$-dimensional range queries to evaluate $\Sigma_{Bord}$, so the time complexity is $O(\log^4(n))$. \QED

\begin{lemma}\label{l:sumtrc}
$\Sigma_{T} = \Sigma_{C\in POR_I} |T_{r_i}(C) \cup T_{r_j}(C)|$ can be evaluated in $O(\log^4(n))$ time given $Tlex$. 
\end{lemma}
Specifically, we prove the following: 
\begin{lemma}\label{l:lrextintert}
Given an interval $[i_z \ldots j_z]$ and a period rank $z$, the amount of $(l,r)$-extendable occurrences in $T_z[i_z \ldots j_z]$ can be evaluated in $O(\log^4(n))$ time given $Tlex$.
\end{lemma}
Lemma \ref{l:lrextintert} can be applied to either find $\Sigma_{C\in POR_I}|T_{r_i}(C)|$ and  $\Sigma_{C\in POR_I}|T_{r_j}(C)|$ independently if $r_i \neq r_j$ or $\Sigma_{C\in POR_I}|T_{r_i}(C)|$ ($= \Sigma_{C\in POR_I}|T_{r_j}(C)|$) if $r_i = r_j$.

\textbf{Proof:} Let $C_{i_z}$ and $C_{j_z}$ be the clusters containing $T_z[i_z]$ and $T_z[j_z]$ respectively. If $z < q_R$, all the occurrences with rank $z$ are Marginal, and therefore there are 0 $(l,r)$-extendable occurrences in $T_z$. Otherwise, we execute the following range queries on $Tlex$:
\begin{enumerate}
    \item $count_{cent} = COUNT([z + q_L + 2 \ldots \infinity{}]\times \ast \times \ast \times [tr(C_{i_z}) \ldots tr(C_{j_z})])$
    \item $count_{bordL} = COUNT([z + q_L +1 \ldots z +q_L + 1] \times [r_L \ldots \infinity{}] \times \ast \times [tr(C_{i_z}) \ldots tr(C_{j_z})])$
    \item $count_{bordR} = COUNT([q_R + q_L + 2 \ldots \infinity{}]\times \ast \times [r_R \ldots \infinity{}] \times [tr(C_{i_z}) \ldots tr(C_{j_z})])$
    \item $count_{bordM} = COUNT([q_R + q_L + 1 \ldots q_R + q_L + 1]\times [r_L \ldots \infinity{}] \times [r_R \ldots \infinity{}] \times [tr(C_{i_z}) \ldots tr(C_{j_z})])$
\end{enumerate}
In all the above queries, the constraint on the last coordinate ensures that we are only counting points that correspond to clusters $C$ with $Tail(C_{i_z}) <_L Tail(C) <_L Tail(C_{j_z})$. According to Fact \ref{f:onetcont}, we are interested in counting the clusters with this property, that also has an $(l,r)$-extendable occurrence with degree $z$.

\textbf{If $z = q_R$}, every occurrence in $T_z$ is a borderline occurrence. For arguments similar to the ones used in the evaluation of $\Sigma_{Bord}$, a point $p_C$ is counted in $count_{bordR}$ iff $C$ has an $(l,r)$-extendable occurrences with rank $q_R$ and $|C| \ge q_R + q_L + 2$. Similarly, $p_C$ is counted in $count_{bordM}$ iff $C$ has an $(l,r)$-extendable occurrence with rank $q_R$ and $|C| = q_R + q_L + 1$. Overall, there are $count_{bordR} + count_{bordM}$ $(l,r)$-extendable occurrences in $T_z[i_z \ldots j_z]$ .

\textbf{If $z \ge q_R+1$}, an occurrence in $T_z[i_z \ldots j_z]$ can be a borderline occurrence $w_t$ in a cluster $C$ with $r_p(w_t) = z = |C| - q_L - 1$ in this case, $|C| = z + q_L + 1 \ge q_R + q_L + 2$, this occurrence is $(l,r)$-extendable iff $r_L(C) \ge r_L$. These conditions are precisely described by the range counting query that was used to yield $count_{bordL}$. Additionally, we need to count the clusters that have a central occurrence in $T_z[i_z \ldots j_z]$. A cluster $C$ has a central occurrence with degree $z$ iff $z\in [q_R + 1 \ldots |C|-q_L - 2]$. In this case, it is given that $z > q_R + 1$, so our problem is reduced to counting the clusters with $|C| > q_L + z + 2$. This constrain is precisely described by the range query used to evaluate $count_{cent}$. It is clear that the clusters counted in $count_{bordL}$ and in $count_{cent}$ are disjoint, so $count_{bordL} + count_{cent}$ is the number of $(l,r)$-extendable occurrences in $T_z[i_z \ldots j_z]$. 

We check if $z = q_R$ and return $count_{bordR} + count_{bordM}$ or $count_{bordL} + count_{cent}$ accordingly. We used a constant number of $4$-dimensional range queries, so the time complexity is $O(\log^4(n))$. We also need to obtain $tr(C_{i_z})$ and $tr(C_{j_z})$. We can use Theorem \ref{t:sortclusters} to obtain $C_{i_z}$ and $C_{j_z}$ in $O(\log^3(n))$ time. We store the clusters such that every cluster is associated with the auxiliary information $tr(C)$. \QED

We conclude this subsection with the following:
\begin{theorem}\label{t:countextper}
The periodic occurrences representation $POR_w$ of a word $w=S[s\ldots e]$ with $per(w) = p$ can be preprocessed in time $\cOtilde(|POR_w|)$ to answer Expression Restricting Range Counting queries of the form $(l,r,i,j)$ in $O(\log^4(n))$ time, provided that $per(S[s - l \ldots e +r]) = p$. The generated data structure uses $\cOtilde(|POR_w|)$ space.
\end{theorem}
\textbf{Proof:}
We preprocess $POR_w$ for $A_w$ lookup queries by employing Theorem \ref{t:sortclusters}. We also construct $Tlex$.

To process a query $(l,r,i,j)$ with $per(S[s - l \ldots e +r]) = p$, we use the data structure of Theorem \ref{t:sortclusters} to obtain $T_{r_i}$, $T_{r_j}$, $i_{r_i}$, $j_{r_j}$ and the clusters $C_{i}$ and $C_j$ containing $i$ and $j$ respectively. We also evaluate $q_L$, $q_R$, $r_L$ and $r_R$. We now have all the data required to execute the range queries on $Tlex$ to evaluate $\Sigma_{Bord}$, $\Sigma_{T}$ and $\Sigma_{Cent}$ by employing lemmas \ref{l:sumcentc}-\ref{l:sumtrc}. It can be easily verified that the sets of $(l,r)$-extendable occurrences counted in $\Sigma_{Bord}$, in $\Sigma_{T}$ and in $\Sigma_{Cent}$ are disjoint. We output thee sum $\Sigma_{Bord}+\Sigma_{T}+\Sigma_{Cent}$.

\textbf{Preprocessing time and space:} The construction of the data structure for $A_w$ look-ups takes $\cOtilde(POR_w)$ time and space by Theorem \ref{t:sortclusters}. The arguments required to construct the points of $Tlex$ can be evaluated by calculating $|C|$, $r_L(C)$ and $r_R(C)$ in $O(1)$ for every cluster $|C|$, and sorting the tails to get $tr(C)$ for every cluster $C$. We then insert every one of the $|POR_w|$ points to $Tlex$ in $O(\log^4(n))$. The overall time is $\cOtilde(|POR_w|)$, and so is the space of the generated data structure.

\textbf{Query time:} The query consists of two lookup queries in $A_w$, a constant number of arithmetic operations to evaluate $q_R$, $q_L$, $r_R$ and $r_L$, and a single applications of each of lemmas \ref{l:sumcentc}-\ref{l:sumtrc}. The time is bounded by $O(\log^4(n))$. \QED
 
\subsubsection{Case 2: An A-periodic Extended Word}
In this subsection, we examine the case in which $w' =S[s - l \ldots e+r]$ does not have a period $per(w) = p$.
We show that in this case, every cluster contributes at most a single $(l,r)$-extendable occurrence to $A_w[i\ldots j]$.
Let $R = S[s-E_l \ldots e + E_r]$ be the run with period $p$ containing $w$. Let $q_L,q_R,r_R,r_L$ defined as in the previous section, and similarly let $q_{E_L}$, $q_{E_r}$, $r_{E_l} < p$ and $r_{E_r}<p$ be non negative integers such that $E_l= q_{E_l} \cdot p + r_{E_l}$ and $E_r= q_{E_r} \cdot p + r_{E_r}$.

Note that if $w'$ does not have a period $p$, it must be the case that either $r > E_r$ or $l > E_l$. Let $C_w$ be the periodic cluster containing $w$. The following facts are the key for efficiently executing Restricted Range Counting Queries:

\begin{fact}\label{l:aperleftex}
If $l > E_l$ and $r \le E_r$, only an occurrence $w_t$ implied by cluster $C$ with period rank $r_p(w_t) = |C|-q_{E_l} - 1$ can be $(l,r)$-extendable. This $w_t$ is $(l,r)$-extendable iff one of these conditions hold:
\begin{enumerate}
    \item $LCS(Head(C),Head(C_w)) \ge l - p\cdot q_{E_l}$ and $|C| \ge q_{E_l} + q_R + 2$
    \item $LCS(Head(C),Head(C_w)) \ge l - p\cdot q_{E_l}$ , $|C| = q_{E_l} + q_R + 1$ and $r_R(C) \ge r_R$
\end{enumerate}

\end{fact}

\begin{fact}\label{l:aperrightex}
If $l \le E_l$ and $r > E_r$, only an occurrence $w_t$ implied by cluster $C$ with period rank $r_p(w_t) = q_{E_r}$ can be $(l,r)$-extendable.
This $w_t$ is $(l,r)$-extendable iff one of these conditions hold: 
\begin{enumerate}
    \item $LCP(Tail(C),Tail(C_w)) \ge r - p\cdot q_{E_r}$ and $|C| \ge q_L + q_{E_r} + 2$
    \item $LCP(Tail(C),Tail(C_w)) \ge r - p\cdot q_{E_r}$ , $|C| = q_L + q_{E_r} + 1$ and $r_L(C) \ge r_L$
\end{enumerate}
\end{fact}

\begin{fact}\label{l:apermidex}
If $l > E_l$ and $r > E_r$, only an occurrence $w_t$ implied by cluster $C$ with period rank $r_p(w_t) = q_{E_r}$ can be $(l,r)$-extendable. 
This $w_t$ is $(l,r)$-extendable iff $|C| = q_{E_l} + q_{E_r} + 1$,  $LCP(Tail(C),Tail(C_w)) \ge r - p\cdot q_{E_r}$ and $LCS(Head(C),Head(C_w)) \ge l - p\cdot q_{E_l}$
\end{fact}

\begin{figure}[htpb!] 
    \centering
    \scalebox{0.9}{\input{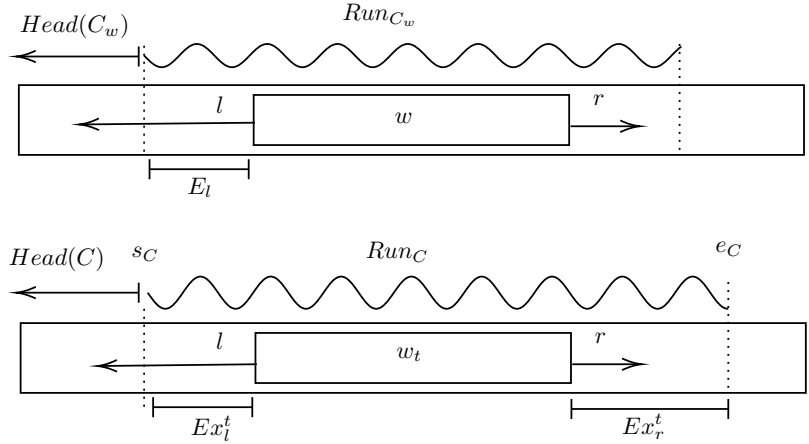}}
    \caption{An illustration of Fact \ref{l:aperleftex}. $w$ has $r$ symbols to its right contained within $Run_{C_w}$. To its left, the first $Ex_l$ symbols are within $Run_{C_w}$ and the following $l-Ex_l$ symbols are a suffix of $Head(C_w)$. An occurrence $w_t$ implied by cluster $C$ must have a similar relation to $Run_C$ in order to be $(l,r)$-extendable. Additionally, $Head(C)$ must have a sufficiently long agreement with $Head(C_w)$. In this demonstration, $Head(C)$ starts to the left of $Run(C)$. Note that it does not have to be the case, and a portion of $Head(C)$ may be contained within $Run(C)$.}\label{fig:extcase2ex}
\end{figure}

The intuition behind facts \ref{l:aperleftex} - \ref{l:apermidex} is the following. An occurrence $w_0=S[i_0 \ldots j_0]$ of $w$ has period $p$. For $w_0$ to be $(l,r)$-extendable, it must agree with $w'$ on the extension of the period $p$ around $w_0$ (at least within $w_0 = S[i_0 - l \ldots j_0 + r]$). Furthermore, there must be a sufficiently long agreement between the non periodic 'left overs'. For an illustration, see Figure \ref{fig:extcase2ex}. We provide a proof for Fact \ref{l:aperleftex}, facts \ref{l:aperrightex} and \ref{l:apermidex} can be proven by using similar arguments.

\textbf{Proof:} Let $w_t=S[s_t\ldots e_t]$ be an occurrence implied by $C$, and let the run with period $p$ containing $C$ be $R = S[s_C \ldots e_C]$. Recall that the size of the extension of $R$ to the left of $w_t$ is $(|C| - 1 - r_p(w_t)) \cdot p + r_L(C)$ (Claim \ref{c:extensiontorank}).

We start by proving the following claim.
\begin{claim}
An $(l,r)$-extendable occurrence $w_t$ implied by $C$ must have $r_p(w_t) = |C|-q_{E_l} - 1$ .
\end{claim}
\textbf{Proof:}
If $r_p(w_t) < |C|-q_{E_l} - 1$, the extension of $R$ to the left from $w_t$ is at least $(q_{E_l} + 1) \cdot p > q_{E_l} \cdot p + r_{E_l} = E_l$ and therefore, (according to Lemma \ref{l:clusterlceprog}) $lcs(s_t - 1,s - 1) = E_l < l$ so $w_t$ is not $(l,r)$-extendable.  

If $r_p(w_t) > |C|-q_{E_l} - 1$, the extension of $R$ to the left from $w_t$ is at most $(q_{E_l} - 1)\cdot p < q_{E_l}\cdot p + r_{E_l} = E_l$. Therefore, $lcs(s-1,s_t - 1) < E_l<l$ and $w_t$ is not $(l,r)$-extendable. We get that for $w_t$ to be $(l,r)$ extendable, it must have $r_p(w_t) = |C|-q_{E_l} - 1$.  \QED

We proceed with the following claim:
\begin{claim}
An occurrence $w_t$ with $r_p(w_t) = |C|-q_{E_l} - 1$ has $lcs(s-1,s_t-1) \ge l$ iff $lcs(Head(C_w),Head(C)) \ge l - p \cdot q_{E_l}$
\end{claim}
\textbf{Proof:} If $r_p(w_t) = |C|-q_{E_l} - 1$, $w_t$ has a (not necessarily maximal) interval with period $p$ of size $q_{E_l} \cdot p$ to its left, proceeded by $Head(C)$. $w$ has the same structure - it has a periodic interval of size at least $q_{E_l} \cdot p$ to its left proceeded by $Head(C_w)$. Comparing from right to left, the suffixes $S^{s_t - 1}$ and $S^{s-1}$ match within the periodic interval. Therefore we have $S[s_t - p \cdot q_{E_l} \ldots s_t - 1] =S[s - p \cdot q_{E_l} \ldots s - 1 ]$. From this point, we are comparing between the heads of $C$ and $C_w$. So $lcs(s_t - 1, s - 1) = lcs(Head(C_w),Head(C)) + p \cdot q_{E_l}$ and we need  $lcs(Head(C_w),Head(C)) \ge l - p \cdot q_{E_l}$ for $w_t$ to have a sufficiently long agreement the left with $w$. \QED

Let $w_t$ be an occurrence represented by a cluster $C$ with $lcs(Head(C_w),Head(C)) \ge l - p \cdot q_{E_l}$. According to the above claim, $w_t$ is sufficiently extendable to the left to be an $(l,r)$ extendable occurrence. We consider the following cases for the size of $C$.

If $|C| < q_{E_l} + q_R + 1$, the extension of $w_t$ with $r_p(w_t) = |C| - q_{E_l} - 1$ to the right is at most $(q_R - 1) \cdot p < q_R + r_R = r$. Since $E_r \ge r$, Lemma \ref{l:clusterlceprog} yields $lcp(e+1,e_t+1) < r$ and $w_t$ is not $(l,r)$-extendable.

If $|C| \ge q_{E_l} + q_R + 2$, the extension of $R$ to the right of $w_t$ with $r_p(w_t) = |C| - q_{E_l} - 1$ is at least $(q_R + 1) \cdot p > q_R + r_R = r$. Since $E_r \ge r$, the period $p$ extends at least $r$ indexes both to the right of $w_t$ and to the right of $w$. It follows that $lcp(e+1,e_t+1) \ge r$ and $w_t$ is sufficiently extendable from the right making it an $(l,r)$-extendable occurrence.

If $|C| = q_{E_l} + q_R + 1$, $w_t$ has an extension of $Run_C$ of size exactly $q_R \cdot p + r_R(C)$ to its right. Since $E_r \ge r$ We need the period $p$ to extend at least $r = q_R\cdot p + r_R$ indexes to the right of $w_t$ for it to be sufficiently extendable to the right. Therefore, in this case $w_t$ is $(l,r)$-extendable iff $r_R(C) \ge r_R$. \QED

Equipped with facts \ref{l:aperleftex} - \ref{l:apermidex}, we are ready to formulate the problem as a set of range queries.
We construct a data structure for $6$-dimensional range queries $Tlex$. For every cluster $C$, we insert the point:

$p_C = (|C|,r_L(C),r_R(C),lcs(Head(C),Head(C_w)),lcp(Tail(C),Tail(C_w)),tr(C))$

More precisely, we will present our queries \textbf{as if} they are executed on the aforementioned $6$-dimensional data structure. We will never use more than $4$ non-trivial ranges ("$\ast$") in a single query, so it is sufficient to store a set of $4$-dimensional data structures, each containing a different subsets of the coordinates of the points $p_C$ that is required for a certain query.

Let our input Extension Restricting Range Query be $(l,r,i,j)$. We recall the notations of $T_{r_i}$,$T_{r_j}$, $i_{r_i}$ and $j_{r_j}$ such that $T_{r_i}[i_{r_i}] = I[i]$ and $T_{r_j}[j_{r_j}] = I[j]$, as well as $C_i$ and $C_j$ the clusters respectively implying the occurrences starting in $I[i]$ and in $I[j]$. We distinguish between three cases with each case corresponding to one of the facts \ref{l:aperleftex} - \ref{l:apermidex}.

\textbf{Case 1:} $l > E_l$ and $r \le E_r$. We start by counting the $(l,r)$-extendable occurrences in $T_z$ for $z\in [r_j + 1 \ldots r_i - 1]$. To do that, we execute the following range queries on $Tlex$-
\begin{enumerate}
    \item $count_{mid1} = COUNT([max(q_{E_l} + r_j +  2, q_{E_l} + q_R + 2) \ldots q_{E_L} + r_i] \times \ast \times \ast \times [l - q_{E_L} \cdot p) \ldots \infinity{}] \times \ast \times \ast)$
    \item $count_{mid2} = COUNT( q_{E_l} + q_R + 1 \ldots q_{E_l} + q_R + 1] \times \ast \times [r_R \ldots \infinity] \times [l - q_{E_L} \cdot p) \ldots \infinity{}] \times \ast \times \ast)$. We set $count_{mid2} = 0$ if  $q_R \notin [r_j + 1 \ldots r_i - 1]$.
\end{enumerate}

$count_{mid1}$ counts occurrences from clusters $C$ with $|C| \ge q_{E_l} + q_R + 2$ and $LCS(Head(C_w,C)) \ge l - q_{E_l} \cdot p$ - these are $(l,r)$-extendable occurrences that correspond to the first condition in Fact \ref{l:aperleftex}. The additional constraint on $|C|$ ensures that $|C| - q_{E_l} -1 \in [r_j + 1 \ldots r_i - 1]$. Therefore, $count_{mid1}$ counts the amount of $(l,r)$-extendable occurrences $w_t$ corresponding to the first condition of Fact \ref{l:aperleftex} with $r_p(w_t) = |C| - q_{E_l} - 1 \in [r_j + 1 \ldots r_i - 1]$.

$count_{mid2}$ counts $(l,r)$-extendable occurrences that correspond to the second condition. Every $(l,r)$-extendable occurrence $w_t$ in this case has rank $r_p(w_t) = |C| - q_{E_l} - 1 = q_R$. Since we nullify $count_{mid2}$ if $q_R \notin [r_j + 1 \ldots r_i -1]$, we get that $count_{mid2}$ counts the $(l,r)$-extendable occurrences that corresponds to the second condition and are within the relevant interval of ranks.

Overall, $count_{mid1} + count_{mid2}$ is the number of $(l,r)$-extendable occurrences in $T_z$ for $z \in [r_j + 1 \ldots r_i -1]$.

We are left with the tasks of counting the $(l,r)$-extendable occurrences in the parts of $T_{r_i}$ and  $T_{r_j}$ that are contained in $I[i \ldots j]$. As in the periodic case, this is easily reducible to counting the $(l,r)$-extendable occurrences within a given interval $T_z[i_z\ldots j_z]$ of a sub-array $T_z$.

As in the periodic case, we construct a range query to count the $(l,r)$-extendable occurrences within an interval $T_{z}[i_z \ldots j_z]$ in $T_z$. Let $C_i$ and $C_j$ be the clusters implying $T_{z}[i_z]$ and $T_z[j_z]$, respectively. We execute the following range query on $Tlex$:

\begin{enumerate}
    \item $count_1 = COUNT([z + q_L + 1 \ldots z + q_L + 1 ] \times \ast \times \ast \times [L - q_{E_L} \cdot p \ldots \infinity] \times \ast \times [tr(C_i) \ldots tr(C_j)])$
    \item $count_2 = COUNT([z + q_L +1 \ldots z + q_L + 1 ] \times \ast \times [r_R \ldots \infinity] \times  [L - q_{E_L} \cdot p \ldots \infinity] \times \ast \times [tr(C_i) \ldots tr(C_j)])$
\end{enumerate}

We proceed to make the following claims regarding $count_1$ and $count_2$.

\begin{claim}\label{c:apertzsize}
A cluster $C$ implies an $(l,r)$-extendable occurrence in $T_z$ iff $|C| = q_{E_l} + z + 1$ and it implies an $(l,r)$-extendable occurrence. 
\end{claim}
\textbf{Proof:}
We start by proving $\xrightarrow[]{}$. Let $w_t$ be an $(l,r)$-extendable occurrence in $T_z$ implied by cluster $C$. According to Lemma \ref{l:aperleftex}, $w_t$ must have $r_p(w_t) = |C| - q_{E_l} - 1$. Since $w_t$ is in $T_z$, it must have $r_p(w_t) = z$. By transitivity, we get $|C| -q_{E_l} - 1 =z$ which implies $|C| =q_{E_l} +z + 1$. 

We proceed to prove $\xleftarrow[]{}$. Let $w_t$ be an $(l,r)$-extendable occurrence implied by $C$, with $|C|=q_{E_l} + z +1$. Lemma \ref{l:aperleftex} suggests that $r_p(w_t) = z$ \QED

\begin{claim}\label{c:apertint1}
If $z \ge q_R +1$, The number of $(l,r)$-extendable occurrences in $T_{z}[i_z \ldots j_z]$ is $count_1$.
\end{claim}
\textbf{Proof:} The constraint on the first coordinate ensures that the clusters counted by $count_1$ are with size $q_{E_l} + z +1$.

If $q_{E_l} + z + 1 \ge q_{E_l}  + q_R+ 2$, the first case of Lemma \ref{l:aperleftex} applies and every cluster $C$ with $|C| = q_{E_l} + z + 1$ contains an $(l,r)$-extendable occurrence iff $lcs(Head(C),Head(C_w)) \ge L - q_{E_L} \cdot p$. the constraint on the $4$th coordinate ensures that we are only counting clusters that contain an $(l,r)$-extendable occurrence.

So far, we have shown that points satisfying the constraints on the first and on the 
$4$th coordinates corresponds to the set of clusters with size $q_{E_l} + z + 1$ and containing an $(l,r)$-extendable occurrence. Claim \ref{c:apertzsize} suggests that this is exactly the set of clusters that contains an $(l,r)$-extendable occurrence in $T_z$. 

Finally, the constraint on $tr(C)$ ensures that we only count clusters $C$ with $tr(C_i) \le tr(C)\le tr(C_j)$. Since the occurrences $T_z$ are sorted by the lexicographic order of the tails of their clusters, that constraint ensures that we only count clusters with an $(l,r)$-extendable occurrence in $T_z[i_z \ldots j_z]$ \QED

\begin{claim}\label{c:apertint2}
If $z = q_R$, the number of $(l,r)$-extendable occurrences in $T_z[i_z \ldots j_z]$ is $count_2$.
\end{claim}
\textbf{Proof:} The constraint on the first coordinate ensures that the clusters counted by $count_1$ are with size $q_{E_l} + z +1$.

If $z = q_R$, the second case of Lemma \ref{l:aperleftex} applies and every cluster $C$ with $|C| = q_{E_l} + z + 1$ contains an $(l,r)$-extendable occurrence with degree $z$ iff $lcs(Head(C),Head(C_w)) \ge L - q_{E_L} \cdot p$ and $r_R(C) \ge r_R(C_w)$. The constraints on the third and on the forth coordinates ensures that we only count clusters that imply an $(l,r)$-extendable occurrence. 

So far, we have shown that points satisfying the constraints on the first, the third and the 
$4$th coordinates corresponds to the set of clusters with size $q_{E_l} + z + 1$ that contain an $(l,r)$-extendable occurrence. Claim \ref{c:apertzsize} suggests that this is exactly the set of clusters that contains an $(l,r)$-extendable occurrence in $T_z$. 

As in the proof of Claim \ref{c:apertint1}, the constraint on $tr(C)$ ensures that we only count clusters with an $(l,r)$-occurrence within the interval $T_z[i_z \ldots j_z]$ in $T_z$. \QED

\begin{claim}\label{c:apertint3}
If $z < q_R $, there are no $(l,r)$-extendable occurrences in $T_z[i_z \ldots j_z]$.
\end{claim}
\textbf{Proof:} Assume to the contrary that $T_z$ contains an $(l,r)$-extendable occurrence $w'$ implied by a cluster $C'$. Claim \ref{c:apertzsize} suggests that that $r_p(w_t) = |C'| - q_{E_l} - 1$. Since $w_t$ is in $T_z$, we get $z = |C'| - q_{E_l} - 1$. From our assumption that $z < q_R $ we get $|C'| < q_{E_l} + q_R + 1$. This is a contradiction to Lemma \ref{l:aperleftex}, in which both cases require $|C'| \ge  q_{E_l} + q_R + 1$. \QED

With claims \ref{c:apertint1}-\ref{c:apertint3}, it is straightforward to construct an algorithm for counting $(l,r)$-extendable occurrence within $T_z[i_Z \ldots j_z]$. We compare $z$ to $q_R $. If $z \ge q_R + 1$- we compute $count_1$ and return it (Claim \ref{c:apertint1}). If $z = q_R$ we compute $count_2$ and return it (Claim \ref{c:apertint2}). If $z<q_R$ we simply return $0$ (Claim \ref{c:apertint3}). We execute at most a single range query in the process of reporting the number of $(l,r)$-extendable occurrences within $T_z[i_z \ldots j_z]$. The time complexity is therefore $O(\log^4(n))$
\\\\
\textbf{Case 2:} $l \le E_l$ and $r > E_r$. This case is corresponding to Lemma \ref{l:aperrightex} and is handled symmetrically to case 1.
\\\\

\textbf{Case 3:} $l > E_l$ and $r > E_r$. Again, we start by counting the $(l,r)$-extendable occurrences in $T_z$ for $z\in [r_j + 1 \ldots r_i - 1]$. Note that in this case, according to Fact \ref{l:apermidex}, the only acceptable rank for an $(l,r)$-extendable occurrence is $q_{E_r}$, and the only acceptable size for a cluster containing an $(l,r)$ extendable occurrence is $q_{E_l} + q_{E_r} + 1$. if $q_{E_r} \in  [r_j + 1 \ldots r_i - 1]$, we query:

$count = COUNT( q_{E_l} + q_{E_r} +  1 \ldots q_{E_l} + q_{E_r} +  1] \times \ast \times \ast \times [l - q_{E_l} \cdot p \ldots \infinity{}] \times [r-q_{E_r} \cdot p \ldots \infinity{}] \times \ast)$.

According to Fact \ref{l:apermidex}, these are all the $(l,r)$-extendable occurrences of $w$. Since $T_{q_{E_r}}$ is completely contained within $I[i \ldots j]$, all of these occurrences are in $I[i \ldots j]$ and we output $count$.

If $q_{E_r} \in \{ r_i, r_j \} ,$ we have $T_{q_{E_r}}$ only partially contained in $I[i \ldots j]$.  We need to count $(l,r)$-extendable occurrences within a certain interval $T_{q_{E_r}}[ i_z \ldots j_z]$ with $C_{i_z}$ and $C_{j_z}$ the clusters implying the occurrences starting in $T_{q_{E_r}}[i_z]$ and in $T_{q_{E_r}}[j_z]$ respectively. We query: 

$count = COUNT([ q_{E_l} + q_{E_r} +  1 \ldots q_{E_l} + q_{E_r} +  1] \times \ast \times \ast \times [l - q_{E_l} \cdot p \ldots \infinity{}] \times [r-q_{E_r} \cdot p \ldots \infinity{}] \times [tr(C_{i_z}) \ldots tr(C_{j_z}))$.

$count$ counts only the $(l,r)$-extendable occurrences within $T_{q_{E_r}}[ i_z \ldots j_z]$ since the occurrences in $T_{q_{E_r}}$ are sorted by the lexicographic order of their clusters' tails.

The following theorem concludes the result of this subsection:
\begin{theorem}\label{t:countextaper}
The periodic occurrences representation $POR_w$ of a word $w=S[s\ldots e]$ with $per(w) = p$ can be preprocessed in time $\cOtilde(|POR_w|)$ to answer Extension Restricting Range Counting queries of the form $(l,r,i,j)$ in $O(\log^4(n))$ time, provided that $per(S[s - l \ldots e +r]) \neq p$.
\end{theorem}
\textbf{Proof:} We preprocess $POR_w$ for look-ups in $A_w[i]$ using Theorem \ref{t:sortclusters}. We process every cluster $C$ to obtain $|C|$, $r_L(C)$, $r_R(C)$, $lcs(Head(C),Head(C_w))$ and $lcp(Tail(C),Tail(C_w))$. We sort the tails of the clusters to obtain $tr(C)$ for every cluster $C$. Now we have the required data to construct the point $p_C$ for every cluster. We insert subsets of size at most $4$ of the coordinates of $p_C$ to several $4$-dimensional range queries data structures corresponding to the coordinates with non-trivial intervals in the queries described in this section. We also evaluate $E_r$ and $E_l$ the extensions of the run with period $p$ to the right and to the left of $w$, respectively.

Upon a query $(l,r,i,j)$, we classify the query into one of the 3 cases described in this section. We execute a look-up query to $I[i]$ and $I[j]$ to obtain $T_{r_i}$,$T_{r_j}$, $i_{r_i}$, $j_{r_j}$, $C_i$ and $C_j$. We now have all the information required to execute the range queries.

\textbf{Preprocessing time:} The time for constructing a data structure for $I[i]$ look ups is $\cOtilde(|POR_w|)$. Every cluster in $POR_w$ is processed with a constant number of $LCE$ queries and basic arithmetic operation, which are all polylogarithmic operations. Sorting the tails of the clusters takes $\cOtilde(|POR_w|)$. Finally, inserting all the (partial) points to $4$-dimensional range queries data structures takes polylogarithmic time per point. The overall complexity is $\cOtilde(|POR_w|)$.

\textbf{Query time:} The query time consists of a constant amount of comparison operations to classify the query, two look-up queries for $I$ and a constant amount of $4$-dimensional range queries. The complexity is dominated by $O(\log^4(n))$. 

Recall that a similar data structure needs to be constructed for the subarray $D$ of $A_w$ containing occurrences from decreasing clusters.
\QED

Putting it together with Theorem \ref{t:countextper}, we get the following:

\begin{theorem}\label{t:countext}
The periodic occurrences representation $POR_w$ of a word $w=S[s\ldots e]$ can be preprocessed in time $\cOtilde(|POR_w|)$ to answer Extension Restricting Range Counting queries in $O(\log^4(n))$ time.
\end{theorem}
\textbf{Proof:} given a query $(l,r,i,j)$, we can check if $l \le E_l$ and $r \le E_r$. If both are true, we employ Theorem \ref{t:countextper}. Otherwise, we employ Theorem \ref{t:countextaper}. The classification test takes additional constant time. \QED

We can use the above result to finally prove Theorem \ref{t:extSelect}

\textbf{Proof:} We preprocess $POR_w$ for $A_w$ look-up queries and for Extension Restricting Range Counting queries with theorems \ref{t:sortclusters} and \ref{t:countext}.

Upon a query for $A_w^{(l,r)}[i]$, we execute a binary search for the index $i'$ such that $A_w[i'] = A_w^{(l,r)}[i]$ as follows: start by executing the Extension Restricting Range Counting query $(l,r, 1 , \frac{|A_w|}{2})$. Denote the output of the query as $L$. If $L \ge i$ we deduce that $A_w^{(l,r)}[i]$ is in $A_w[1 \ldots \frac{|A_w|}{2}]$. Otherwise, we deduce that $A_w^{(l,r)}[i]$ is in $A_w[\frac{|A_w|}{2} + 1 \ldots |A_w|]$. We shrink our search range and proceed recursively until our search interval is of size $1$. The sole index in the final interval is $i'$. We conclude by returning $A_w[i']$

The preprocessing time and the space consumed by the data structure are both bounded by $\cOtilde(|POR_w|)$ directly from Theorem \ref{t:sortclusters} and Theorem \ref{t:countext}. The query time consists of $\log(n)$ binary search iterations ($|A_w|\le n$). In every iteration of the binary search we execute an Extensions Restricting Range Counting query costing $O(\log^4(n))$ time each. We conclude with a single look-up query in $A_w$. The time complexity is dominated by $O(\log^5(n))$. \QED

\section{Maintaining Indexed Integers Under Batched Updates}\label{s:updates}
In this section, we provide proofs for Claim \ref{c:intupdates} and Theorem \ref{t:stairsp}.

\subsection{Proof of Claim \ref{c:intupdates}}

\intupdates*

Claim \ref{c:intupdates} may be considered as folklore and can be solved via the following well known, simple construction. $C$ can be represented as a full binary tree with $n$ leaves and weighted edges. The $k$'th leaf from the right represents $C[k]$ and the sum of the edges from the root to that leaf equals to $C[k]$. Upon update $(i,j,x)$, a set of $O(\log(n))$ edges can be identified such that every leaf $l$ in position $[i \ldots j]$ from the left has exactly one of these edges in the route from $l$ to the root, and every other leaf does not have any of these edges in its route to the root. Therefore, increasing the weights of these edges by $x$ will have the desired effect.

This set of edges can be identified by constructing the paths from $i$ and from $j$ to the root. The lowest common ancestor $lca(i,j)$ of the leaves $i$ and $j$ is the lowest node to appear in both paths. For every node $u$ in the sub - path from $lca(i,j)$ to $i$ (not including $lca(i,j)$), if the right edge emerging from $u$ is not a part of the path - add it to the set. For every node $v$ in the path from $lca(i,j)$ to $j$ - if the left node emerging from $v$ is not in the path - add it to the set. In both cases- include the edge entering the leaf in the set. It can be easily verified that this construction yields a set with the required property and can be found in $O(\log(n))$ time.

\subsection{Proof of Theorem \ref{t:stairsp}}

\stairsp*

We denote the set of maintained counters as $C[1\ldots n]$. We assume that all the counters are initially set to $C[i] = 0$. Any other initialization of $C$ can be easily supported by keeping the initial values in a static array $A$ and a stairs updates data structure $D$ with initialized values $D[i]=0$. Upon a lookup query for $C[i]$, we report $C[i] = A[i] + D[i]$. 

Our data structure for restricted decreasing stairs updates works as follows:
We maintain two range queries data structure. A 2-dimensional data structure $O$ and a 3-dimensional data structure $R$. Note that $p$ is fixed among all the updates, so it does not need to be specified in the query.

Upon a query for applying a decreasing stairs update $(i , j)$ we insert the point $p_O = (i,j)$ with the value $v(p_O) = \lceil \frac{j}{p} \rceil$ to $O$, and the point $p_R=(i,j, ( j \bmod p ) )$ (with an arbitrary value assigned to $v(p_R)$) to $R$.  If $(j \bmod p) = 0$ we set the third coordinate of $p_R$ to be $p$ instead.

Upon a query for reporting the value of $C[x]$, we evaluate the unique positive integers $q_x$ and $r_x$ such that $x = q_x \cdot p + r_x$ with $r_x < p$. We execute the following range queries:
\begin{enumerate}
    \item Query $O$ for $S = SUM([1 \ldots x] \times [x \ldots n])$
    \item Query $O$ for $C = COUNT([1 \ldots x] \times [x \ldots n])$
    \item Query $R$ for $D = COUNT([1 \ldots x] \times [x \ldots n] \times [1 \ldots r_x - 1])$. If $r_x = 0$, we set $D=0$
\end{enumerate}
We return $S - C \cdot q_x - D$.

\begin{lemma}
$C[x] = S - C \cdot q_x - D$
\end{lemma}
\textbf{Proof:} We denote $con_j(k)$ to be the value that a decreasing stairs update $(i,j,p)$ adds to index $k$, assuming that $k \in [i \ldots j]$. If $k > j$, the value of $con_j(k)$ is undefined and allowed to be an arbitrary number. Note that for $k\in [i \ldots j]$, the value of $con_j(k)$ is completely independent from $i$. Therefore, it can be equivalently defined as the value that the update $(1,j,p)$ adds to index $k \le j$. Consider an update $(i,j,p)$ that was applied to $C$ before the query for the value $C[x]$. Let $r_j,q_j$ be the two unique non negative integers such that $j = q_j \cdot p + r_j$ and $r_j \in [1 \ldots p]$. 

Note that if $r_x \le r_j$, $con_j(r_x) = \lceil \frac{j}{p} \rceil$, and otherwise, $con_j(r_x) = \lceil \frac{j}{p} \rceil - 1$ That is due to the fact that the leftmost step in the update $(1,j,p)$ is the  $ \lceil \frac{j}{p} \rceil$'th step and it is of size $r_j$. Another property of $con_j(k)$ is that for every two integers $q,k$ we have $con_j(q \cdot p + k) = con_j(k) - q$. Putting the above observations together, we get $con_j(x) = con_j(q_x \cdot p + r_x) = con_j(r_x) - q_x$. For $r_x \le r_j$, that is $con_j(x) = \lceil \frac{j}{p} \rceil - q_x$ and for $r_x > r_j$ that is $con_j(x) = \lceil \frac{j}{p} \rceil - q_x - 1$.

Let $U$ be the set of updates $(i,j,p)$ with $x\in [i \ldots j]$ that were applied to $C$ prior to the query. The definition of $con_j(x)$ directly implies that $C[x] = \Sigma_{(i,j,p) \in U} con_j(x)$. Let $U_1 = \{(i,j,p) \in U \textit{ and } r_x \le r_j\}$ and $U_2 = U \setminus U_1$. So we have $C[x] = \Sigma_{(i,j,p) \in U_1} con_j(x) + \Sigma_{(i,j,p) \in U_2} con_j(x) = \Sigma_{(i,j,p) \in U_1}(\lceil \frac{j}{p} \rceil - q_x) +  \Sigma_{(i,j,p) \in U_2}( \lceil \frac{j}{p} \rceil - q_x - 1) = \Sigma_{(i,j,p) \in U} \lceil \frac{j}{p} \rceil - |U| \cdot q_x - |U_2|$. One can easily confirm that $S = \Sigma_{(i,j,p) \in U} \lceil \frac{j}{p} \rceil$, $C = |U|$ and $D = |U_2|$. \QED

With that, the proof of Theorem \ref{t:stairsp} is completed. Applying a decreasing stairs update and querying for the value of $C[x]$ is done using a constant number of 2-dimensional and 3-dimensional range queries, so the time complexity for both query and update is $O(\log^3(U))$. The space complexity is $O(U \cdot \log^2(U))$ with $U$ being the amount of updates. The initialization of the data structure requires initializing two empty dynamic range queries data structures.

\subsection{Variants}
The following symmetric variant of decreasing stairs updates can also be supported by a construction symmetric to ours:
\begin{definition}
An increasing stairs update is given as $U = (i,j,p)$. Applying $U$ to a set of indexed integers $C[ 1 \ldots n]$ increases $C[a]$ for $a \in [i + (k-1) \cdot p \ldots \min(i + k \cdot p - 1,j)]$ by $k$ for every $1 \le k \le \lceil \frac{j-i+1}{k} \rceil$.
\end{definition}

Increasing stairs updates are equivalent to decreasing stairs updates with inverted indexes. Namely, let $C'[1\ldots n]$ be an array with $C'[x] = C[n-x - 1]$. Applying an increasing stairs update $u=(i,j,p)$ to $C[i \ldots j]$ is equivalent to applying a decreasing stairs update $((n-j-1,n-i-1,p)$ to $C'$. Trivially, both types of queries can be supported by maintaining two data structures, $I$ and $D$ for increasing and decreasing stairs updates respectively, and reporting $C[i] = D[i] + I[i]$. 

One can also consider a \textit{negative} stairs update.
\begin{definition}
An increasing (resp. decreasing) negative stairs update is given as $U = (i,j,p)$. Applying $U$ to a set of indexed integers $C[ 1 \ldots n]$ \textbf{decreases} $C[a]$ for $a \in [i + (k-1) \cdot p \ldots \min(i + k \cdot p - 1,j)]$ (resp. $a\in [max(j - k\cdot p +1,i) \ldots j - (k-1)\cdot p )]$ ) by $k$ for every $1 \le k \le \lceil \frac{j-i+1}{k} \rceil$.
\end{definition}

In words, a negative stairs updates decreases the values of indexes on the $k$th step by $k$ rather than increasing it by $k$. This kind of updates can also be supported by maintaining two stairs data structures, $P$ and $N$, for stairs updates and negative stairs updates respectively. Positive updates are applied to $P$, and negative updates are applied to $N$ as if they were positive. Upon a query index $i$, we output $C[i] = P[i] - N[i]$.

\subsection{Updates reduction}
We say that a set of updates $U=u_1,u_2\ldots u_t$ can be reduced to another set of updates $U'$ if for every array $C$, applying all the updates in $U$ on an array $C$ will result in the same modified array $C'$ as applying all the updates in $U'$ on $C$.

In some settings, we can reduce a sequence of interval increment updates to a small set of stairs updates and interval increment updates.

\begin{figure}[htpb!] 
    \centering
    \scalebox{0.9}{\tikzset{every picture/.style={line width=0.75pt}} 

\begin{tikzpicture}[x=0.75pt,y=0.75pt,yscale=-1,xscale=1]

\draw    (170.5,69.67) -- (354.5,68.67) ;
\draw [shift={(354.5,68.67)}, rotate = 539.69] [color={rgb, 255:red, 0; green, 0; blue, 0 }  ][line width=0.75]    (0,5.59) -- (0,-5.59)   ;
\draw [shift={(170.5,69.67)}, rotate = 539.69] [color={rgb, 255:red, 0; green, 0; blue, 0 }  ][line width=0.75]    (0,5.59) -- (0,-5.59)   ;
\draw  [dash pattern={on 0.84pt off 2.51pt}]  (361,68.98) -- (390.5,68.69) ;
\draw [shift={(392.5,68.67)}, rotate = 539.4300000000001] [color={rgb, 255:red, 0; green, 0; blue, 0 }  ][line width=0.75]    (10.93,-4.9) .. controls (6.95,-2.3) and (3.31,-0.67) .. (0,0) .. controls (3.31,0.67) and (6.95,2.3) .. (10.93,4.9)   ;
\draw [shift={(359,69)}, rotate = 359.43] [color={rgb, 255:red, 0; green, 0; blue, 0 }  ][line width=0.75]    (10.93,-3.29) .. controls (6.95,-1.4) and (3.31,-0.3) .. (0,0) .. controls (3.31,0.3) and (6.95,1.4) .. (10.93,3.29)   ;
\draw   (296,144.83) -- (307.88,144.83) -- (307.88,125.83) -- (331.63,125.83) -- (331.63,144.83) -- (343.5,144.83) -- (319.75,157.5) -- cycle ;
\draw  [dash pattern={on 0.84pt off 2.51pt}]  (440,179.98) -- (469.5,179.69) ;
\draw [shift={(471.5,179.67)}, rotate = 539.4300000000001] [color={rgb, 255:red, 0; green, 0; blue, 0 }  ][line width=0.75]    (10.93,-4.9) .. controls (6.95,-2.3) and (3.31,-0.67) .. (0,0) .. controls (3.31,0.67) and (6.95,2.3) .. (10.93,4.9)   ;
\draw [shift={(438,180)}, rotate = 359.43] [color={rgb, 255:red, 0; green, 0; blue, 0 }  ][line width=0.75]    (10.93,-3.29) .. controls (6.95,-1.4) and (3.31,-0.3) .. (0,0) .. controls (3.31,0.3) and (6.95,1.4) .. (10.93,3.29)   ;
\draw [line width=1.5]    (119.5,104.67) -- (490.5,103.67) ;
\draw [shift={(490.5,103.67)}, rotate = 539.85] [color={rgb, 255:red, 0; green, 0; blue, 0 }  ][line width=1.5]    (0,6.71) -- (0,-6.71)   ;
\draw [shift={(119.5,104.67)}, rotate = 539.85] [color={rgb, 255:red, 0; green, 0; blue, 0 }  ][line width=1.5]    (0,6.71) -- (0,-6.71)   ;
\draw [line width=1.5]    (119.5,211.67) -- (490.5,210.67) ;
\draw [shift={(490.5,210.67)}, rotate = 539.85] [color={rgb, 255:red, 0; green, 0; blue, 0 }  ][line width=1.5]    (0,6.71) -- (0,-6.71)   ;
\draw [shift={(119.5,211.67)}, rotate = 539.85] [color={rgb, 255:red, 0; green, 0; blue, 0 }  ][line width=1.5]    (0,6.71) -- (0,-6.71)   ;
\draw    (434.5,200.67) -- (471.5,200.67) ;
\draw [shift={(471.5,200.67)}, rotate = 180] [color={rgb, 255:red, 0; green, 0; blue, 0 }  ][line width=0.75]    (0,5.59) -- (0,-5.59)   ;
\draw [shift={(434.5,200.67)}, rotate = 180] [color={rgb, 255:red, 0; green, 0; blue, 0 }  ][line width=0.75]    (0,5.59) -- (0,-5.59)   ;
\draw    (210.5,78.67) -- (394.5,77.67) ;
\draw [shift={(394.5,77.67)}, rotate = 539.69] [color={rgb, 255:red, 0; green, 0; blue, 0 }  ][line width=0.75]    (0,5.59) -- (0,-5.59)   ;
\draw [shift={(210.5,78.67)}, rotate = 539.69] [color={rgb, 255:red, 0; green, 0; blue, 0 }  ][line width=0.75]    (0,5.59) -- (0,-5.59)   ;
\draw    (250.5,86.67) -- (434.5,85.67) ;
\draw [shift={(434.5,85.67)}, rotate = 539.69] [color={rgb, 255:red, 0; green, 0; blue, 0 }  ][line width=0.75]    (0,5.59) -- (0,-5.59)   ;
\draw [shift={(250.5,86.67)}, rotate = 539.69] [color={rgb, 255:red, 0; green, 0; blue, 0 }  ][line width=0.75]    (0,5.59) -- (0,-5.59)   ;
\draw    (287.5,94.67) -- (471.5,93.67) ;
\draw [shift={(471.5,93.67)}, rotate = 539.69] [color={rgb, 255:red, 0; green, 0; blue, 0 }  ][line width=0.75]    (0,5.59) -- (0,-5.59)   ;
\draw [shift={(287.5,94.67)}, rotate = 539.69] [color={rgb, 255:red, 0; green, 0; blue, 0 }  ][line width=0.75]    (0,5.59) -- (0,-5.59)   ;
\draw [line width=0.75]  [dash pattern={on 4.5pt off 4.5pt}]  (287.5,98.67) -- (288.5,196) ;
\draw [line width=0.75]  [dash pattern={on 4.5pt off 4.5pt}]  (354.5,76.67) -- (355.5,196) ;
\draw    (395.5,200.67) -- (432.5,200.67) ;
\draw [shift={(432.5,200.67)}, rotate = 180] [color={rgb, 255:red, 0; green, 0; blue, 0 }  ][line width=0.75]    (0,5.59) -- (0,-5.59)   ;
\draw [shift={(395.5,200.67)}, rotate = 180] [color={rgb, 255:red, 0; green, 0; blue, 0 }  ][line width=0.75]    (0,5.59) -- (0,-5.59)   ;
\draw    (356.5,200.67) -- (393.5,200.67) ;
\draw [shift={(393.5,200.67)}, rotate = 180] [color={rgb, 255:red, 0; green, 0; blue, 0 }  ][line width=0.75]    (0,5.59) -- (0,-5.59)   ;
\draw [shift={(356.5,200.67)}, rotate = 180] [color={rgb, 255:red, 0; green, 0; blue, 0 }  ][line width=0.75]    (0,5.59) -- (0,-5.59)   ;
\draw    (288.5,200.67) -- (354.5,200.67) ;
\draw [shift={(354.5,200.67)}, rotate = 180] [color={rgb, 255:red, 0; green, 0; blue, 0 }  ][line width=0.75]    (0,5.59) -- (0,-5.59)   ;
\draw [shift={(288.5,200.67)}, rotate = 180] [color={rgb, 255:red, 0; green, 0; blue, 0 }  ][line width=0.75]    (0,5.59) -- (0,-5.59)   ;
\draw    (250.5,200.67) -- (287.5,200.67) ;
\draw [shift={(287.5,200.67)}, rotate = 180] [color={rgb, 255:red, 0; green, 0; blue, 0 }  ][line width=0.75]    (0,5.59) -- (0,-5.59)   ;
\draw [shift={(250.5,200.67)}, rotate = 180] [color={rgb, 255:red, 0; green, 0; blue, 0 }  ][line width=0.75]    (0,5.59) -- (0,-5.59)   ;
\draw    (210.5,200.67) -- (247.5,200.67) ;
\draw [shift={(247.5,200.67)}, rotate = 180] [color={rgb, 255:red, 0; green, 0; blue, 0 }  ][line width=0.75]    (0,5.59) -- (0,-5.59)   ;
\draw [shift={(210.5,200.67)}, rotate = 180] [color={rgb, 255:red, 0; green, 0; blue, 0 }  ][line width=0.75]    (0,5.59) -- (0,-5.59)   ;
\draw    (170.5,200.67) -- (207.5,200.67) ;
\draw [shift={(207.5,200.67)}, rotate = 180] [color={rgb, 255:red, 0; green, 0; blue, 0 }  ][line width=0.75]    (0,5.59) -- (0,-5.59)   ;
\draw [shift={(170.5,200.67)}, rotate = 180] [color={rgb, 255:red, 0; green, 0; blue, 0 }  ][line width=0.75]    (0,5.59) -- (0,-5.59)   ;

\draw (160,110) node [anchor=north west][inner sep=0.75pt]  [font=\footnotesize] [align=left] {$\displaystyle i_0$};
\draw (466,106) node [anchor=north west][inner sep=0.75pt]  [font=\scriptsize] [align=left] {$\displaystyle j_k$};
\draw (372,50) node [anchor=north west][inner sep=0.75pt]  [font=\footnotesize] [align=left] {$\displaystyle p$};
\draw (152.5,64.67) node [anchor=north west][inner sep=0.75pt]  [font=\scriptsize] [align=left] {+1};

\draw (451,161) node [anchor=north west][inner sep=0.75pt]  [font=\footnotesize] [align=left] {$\displaystyle p$};
\draw (366.5,186.67) node [anchor=north west][inner sep=0.75pt]  [font=\scriptsize] [align=left] {+3};
\draw (312.5,187.67) node [anchor=north west][inner sep=0.75pt]  [font=\scriptsize] [align=left] {+4};
\draw (446,186) node [anchor=north west][inner sep=0.75pt]  [font=\scriptsize] [align=left] {+1};
\draw (407,186) node [anchor=north west][inner sep=0.75pt]  [font=\scriptsize] [align=left] {+2};
\draw (192.5,73.67) node [anchor=north west][inner sep=0.75pt]  [font=\scriptsize] [align=left] {+1};
\draw (232.5,81.67) node [anchor=north west][inner sep=0.75pt]  [font=\scriptsize] [align=left] {+1};
\draw (269.5,89.67) node [anchor=north west][inner sep=0.75pt]  [font=\scriptsize] [align=left] {+1};
\draw (259.5,186.67) node [anchor=north west][inner sep=0.75pt]  [font=\scriptsize] [align=left] {+3};
\draw (225,186) node [anchor=north west][inner sep=0.75pt]  [font=\scriptsize] [align=left] {+2};
\draw (182,186) node [anchor=north west][inner sep=0.75pt]  [font=\scriptsize] [align=left] {+1};
\draw (161,217) node [anchor=north west][inner sep=0.75pt]  [font=\footnotesize] [align=left] {$\displaystyle i_0$};
\draw (467,216) node [anchor=north west][inner sep=0.75pt]  [font=\scriptsize] [align=left] {$\displaystyle j_k$};

\end{tikzpicture}}
    \caption{An illustration of how applying a sequence of interval updates with both ends being an arithmetic progression with difference $p$ is equivalent to two stairs updates and an interval increment update. This set of updates can be viewed as a 'sliding window'. The application of all the interval updates can be represented as an increasing stairs update with the stairs corresponding to the starting (left) indexes of the 'sliding window' and a negative increasing stairs update with the stairs corresponding to the ending indexes of the windows.}\label{fig:slidexp}
\end{figure}
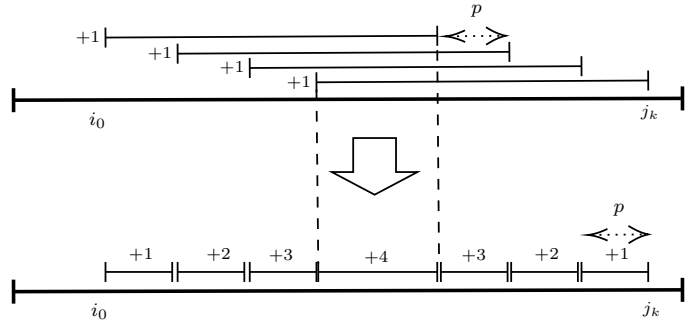

\begin{lemma} \label{l:intToStairs}
Let $p$ be an integer and $U= u_1,u_2 \ldots u_{|U|-1}$ be a sequence of interval increment updates with $u_t = (i_t,j_t,1)$. If both $i_t$ and $j_t$ are either fixed, or an arithmetic progression with difference $p$, the sequence $U$ can be reduced to a set $U'$ of interval increment updates and stairs updates with stair width $|p|$ such that $|U'| \in O(1)$. $U'$ can be evaluated in constant time.
\end{lemma}
\textbf{Proof:} we distinguish between the following cases
\begin{enumerate}
    \item Both $i_t$ and $j_t$ are fixed. In this case, $U$ can be reduced to a single interval update $u'=(i,j,|U|)$.
    \item $i$ is fixed and $j_t = j_0 + t \cdot p$ with $p>0$. In this case, $U$ can be reduced to two updates $U' =\{ u',s\}$ with $s$ being a decreasing stairs update $s = (j_0,j_{|U|-1},p)$ and $u'$ the interval increment update $u' = (i,j_0,|U|)$. The case in which $p < 0 $ can be handled the same way by considering $j_t$ in a reversed order.
    \item $i_t = i_0 + t \cdot p$ with $p >0$ and $j$ is a constant. $U$ can be reduced to two updates $U' = \{ u',s\}$ with $s$ being an increasing stairs update $s=(i_0,i_{|U|-1},p)$ and the interval update $u'=(i_{|U|-1},j,|U|)$. The case where $p< 0$ can be handled the same way by considering $i_t$ in a reversed order.
    \item $i_t = i_0 + t \cdot p$ and $j_t = j_0 + t \cdot p$ be two arithmetic progressions with $p > 0$. $U$ can be reduced to a set of $3$ updates $U' = \{ s_1,s_2,u' \}$ with an increasing stairs update $s_1= (i_0, i_{|U|-1},p)$, an increasing negative stairs update $s_2=(j_0, j_{|U|-1},p)$ and an interval increment update $u'_1 = (i_{|U|-1},j_{|U|-1}, |U|)$.
\end{enumerate}
It can be easily verified that in every case, applying the updates in $U'$ to a set of indexed integers is equivalent to applying $U$ to the same set. For a visualization, See Figures \ref{fig:stairsexp} and \ref{fig:slidexp}.
In every case, the set $U'$ contains at most three updates and can be evaluated in constant time in a straightforward way given the representation of $i_t$ and $j_t$ as arithmetic progressions (or fixed values, depending on the case). \QED

\begin{figure}[htpb!] 
    \centering
    \scalebox{0.9}{\tikzset{every picture/.style={line width=0.75pt}} 

\begin{tikzpicture}[x=0.75pt,y=0.75pt,yscale=-1,xscale=1]

\draw    (188.5,94.67) -- (441.5,93.67) ;
\draw [shift={(441.5,93.67)}, rotate = 539.77] [color={rgb, 255:red, 0; green, 0; blue, 0 }  ][line width=0.75]    (0,5.59) -- (0,-5.59)   ;
\draw [shift={(188.5,94.67)}, rotate = 539.77] [color={rgb, 255:red, 0; green, 0; blue, 0 }  ][line width=0.75]    (0,5.59) -- (0,-5.59)   ;
\draw    (188.5,81.67) -- (406.5,80.67) ;
\draw [shift={(406.5,80.67)}, rotate = 539.74] [color={rgb, 255:red, 0; green, 0; blue, 0 }  ][line width=0.75]    (0,5.59) -- (0,-5.59)   ;
\draw [shift={(188.5,81.67)}, rotate = 539.74] [color={rgb, 255:red, 0; green, 0; blue, 0 }  ][line width=0.75]    (0,5.59) -- (0,-5.59)   ;
\draw    (188.5,69.67) -- (372.5,68.67) ;
\draw [shift={(372.5,68.67)}, rotate = 539.69] [color={rgb, 255:red, 0; green, 0; blue, 0 }  ][line width=0.75]    (0,5.59) -- (0,-5.59)   ;
\draw [shift={(188.5,69.67)}, rotate = 539.69] [color={rgb, 255:red, 0; green, 0; blue, 0 }  ][line width=0.75]    (0,5.59) -- (0,-5.59)   ;
\draw    (188.5,57.67) -- (338.5,56.67) ;
\draw [shift={(338.5,56.67)}, rotate = 539.62] [color={rgb, 255:red, 0; green, 0; blue, 0 }  ][line width=0.75]    (0,5.59) -- (0,-5.59)   ;
\draw [shift={(188.5,57.67)}, rotate = 539.62] [color={rgb, 255:red, 0; green, 0; blue, 0 }  ][line width=0.75]    (0,5.59) -- (0,-5.59)   ;
\draw    (188.5,45.67) -- (305.5,44.67) ;
\draw [shift={(305.5,44.67)}, rotate = 539.51] [color={rgb, 255:red, 0; green, 0; blue, 0 }  ][line width=0.75]    (0,5.59) -- (0,-5.59)   ;
\draw [shift={(188.5,45.67)}, rotate = 539.51] [color={rgb, 255:red, 0; green, 0; blue, 0 }  ][line width=0.75]    (0,5.59) -- (0,-5.59)   ;
\draw  [dash pattern={on 0.84pt off 2.51pt}]  (411,81.98) -- (440.5,81.69) ;
\draw [shift={(442.5,81.67)}, rotate = 539.4300000000001] [color={rgb, 255:red, 0; green, 0; blue, 0 }  ][line width=0.75]    (10.93,-4.9) .. controls (6.95,-2.3) and (3.31,-0.67) .. (0,0) .. controls (3.31,0.67) and (6.95,2.3) .. (10.93,4.9)   ;
\draw [shift={(409,82)}, rotate = 359.43] [color={rgb, 255:red, 0; green, 0; blue, 0 }  ][line width=0.75]    (10.93,-3.29) .. controls (6.95,-1.4) and (3.31,-0.3) .. (0,0) .. controls (3.31,0.3) and (6.95,1.4) .. (10.93,3.29)   ;
\draw   (273,153) -- (290.5,153) -- (290.5,134) -- (325.5,134) -- (325.5,153) -- (343,153) -- (308,165.67) -- cycle ;
\draw    (188.5,200.67) -- (301.5,200.67) ;
\draw [shift={(301.5,200.67)}, rotate = 180] [color={rgb, 255:red, 0; green, 0; blue, 0 }  ][line width=0.75]    (0,5.59) -- (0,-5.59)   ;
\draw [shift={(188.5,200.67)}, rotate = 180] [color={rgb, 255:red, 0; green, 0; blue, 0 }  ][line width=0.75]    (0,5.59) -- (0,-5.59)   ;
\draw  [dash pattern={on 0.84pt off 2.51pt}]  (413,179.98) -- (442.5,179.69) ;
\draw [shift={(444.5,179.67)}, rotate = 539.4300000000001] [color={rgb, 255:red, 0; green, 0; blue, 0 }  ][line width=0.75]    (10.93,-4.9) .. controls (6.95,-2.3) and (3.31,-0.67) .. (0,0) .. controls (3.31,0.67) and (6.95,2.3) .. (10.93,4.9)   ;
\draw [shift={(411,180)}, rotate = 359.43] [color={rgb, 255:red, 0; green, 0; blue, 0 }  ][line width=0.75]    (10.93,-3.29) .. controls (6.95,-1.4) and (3.31,-0.3) .. (0,0) .. controls (3.31,0.3) and (6.95,1.4) .. (10.93,3.29)   ;
\draw    (303.5,200.67) -- (337.5,200.67) ;
\draw [shift={(337.5,200.67)}, rotate = 180] [color={rgb, 255:red, 0; green, 0; blue, 0 }  ][line width=0.75]    (0,5.59) -- (0,-5.59)   ;
\draw [shift={(303.5,200.67)}, rotate = 180] [color={rgb, 255:red, 0; green, 0; blue, 0 }  ][line width=0.75]    (0,5.59) -- (0,-5.59)   ;
\draw [line width=1.5]    (119.5,104.67) -- (490.5,103.67) ;
\draw [shift={(490.5,103.67)}, rotate = 539.85] [color={rgb, 255:red, 0; green, 0; blue, 0 }  ][line width=1.5]    (0,6.71) -- (0,-6.71)   ;
\draw [shift={(119.5,104.67)}, rotate = 539.85] [color={rgb, 255:red, 0; green, 0; blue, 0 }  ][line width=1.5]    (0,6.71) -- (0,-6.71)   ;
\draw [line width=1.5]    (119.5,211.67) -- (490.5,210.67) ;
\draw [shift={(490.5,210.67)}, rotate = 539.85] [color={rgb, 255:red, 0; green, 0; blue, 0 }  ][line width=1.5]    (0,6.71) -- (0,-6.71)   ;
\draw [shift={(119.5,211.67)}, rotate = 539.85] [color={rgb, 255:red, 0; green, 0; blue, 0 }  ][line width=1.5]    (0,6.71) -- (0,-6.71)   ;
\draw    (339.5,200.67) -- (373.5,200.67) ;
\draw [shift={(373.5,200.67)}, rotate = 180] [color={rgb, 255:red, 0; green, 0; blue, 0 }  ][line width=0.75]    (0,5.59) -- (0,-5.59)   ;
\draw [shift={(339.5,200.67)}, rotate = 180] [color={rgb, 255:red, 0; green, 0; blue, 0 }  ][line width=0.75]    (0,5.59) -- (0,-5.59)   ;
\draw    (375.5,200.67) -- (409.5,200.67) ;
\draw [shift={(409.5,200.67)}, rotate = 180] [color={rgb, 255:red, 0; green, 0; blue, 0 }  ][line width=0.75]    (0,5.59) -- (0,-5.59)   ;
\draw [shift={(375.5,200.67)}, rotate = 180] [color={rgb, 255:red, 0; green, 0; blue, 0 }  ][line width=0.75]    (0,5.59) -- (0,-5.59)   ;
\draw    (411.5,200.67) -- (445.5,200.67) ;
\draw [shift={(445.5,200.67)}, rotate = 180] [color={rgb, 255:red, 0; green, 0; blue, 0 }  ][line width=0.75]    (0,5.59) -- (0,-5.59)   ;
\draw [shift={(411.5,200.67)}, rotate = 180] [color={rgb, 255:red, 0; green, 0; blue, 0 }  ][line width=0.75]    (0,5.59) -- (0,-5.59)   ;

\draw (183,110) node [anchor=north west][inner sep=0.75pt]  [font=\footnotesize] [align=left] {$\displaystyle i$};
\draw (432,104) node [anchor=north west][inner sep=0.75pt]   [align=left] {$\displaystyle j_k$};
\draw (422,63) node [anchor=north west][inner sep=0.75pt]  [font=\footnotesize] [align=left] {$\displaystyle p$};
\draw (171,90) node [anchor=north west][inner sep=0.75pt]  [font=\scriptsize] [align=left] {+1};
\draw (170.5,77.67) node [anchor=north west][inner sep=0.75pt]  [font=\scriptsize] [align=left] {+1};
\draw (170.5,64.67) node [anchor=north west][inner sep=0.75pt]  [font=\scriptsize] [align=left] {+1};
\draw (170.5,52.67) node [anchor=north west][inner sep=0.75pt]  [font=\scriptsize] [align=left] {+1};
\draw (170.5,39.67) node [anchor=north west][inner sep=0.75pt]  [font=\scriptsize] [align=left] {+1};
\draw (183,217) node [anchor=north west][inner sep=0.75pt]  [font=\footnotesize] [align=left] {$\displaystyle i$};
\draw (434,212) node [anchor=north west][inner sep=0.75pt]   [align=left] {$\displaystyle j_k$};
\draw (424,161) node [anchor=north west][inner sep=0.75pt]  [font=\footnotesize] [align=left] {$\displaystyle p$};
\draw (349.5,186.67) node [anchor=north west][inner sep=0.75pt]  [font=\scriptsize] [align=left] {+3};
\draw (312.5,187.67) node [anchor=north west][inner sep=0.75pt]  [font=\scriptsize] [align=left] {+4};
\draw (419,186) node [anchor=north west][inner sep=0.75pt]  [font=\scriptsize] [align=left] {+1};
\draw (386,186) node [anchor=north west][inner sep=0.75pt]  [font=\scriptsize] [align=left] {+2};
\draw (235.5,185.67) node [anchor=north west][inner sep=0.75pt]  [font=\scriptsize] [align=left] {+5};

\end{tikzpicture}}
    \caption{An illustration of how applying all the interval increment updates $U$ with $i_t$ being an arithmetic progression and a fixed $j_t$ has the same effect as applying a stairs update (and a 'reminder' interval increment update)} \label{fig:stairsexp}
\end{figure}
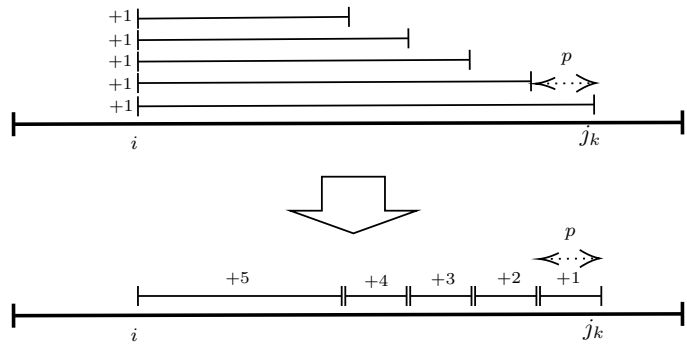

\section*{Acknowledgements}
We warmly thank Eylon Yogev for assistance with the presentation and the organization of the results presented in this manuscript.
\bibliographystyle{acm}

\bibliography{paper}

\end{document}